\documentclass[
  journal=pasa, 
  manuscript=article, 
  year=2023,
  volume=??,
]{cup-journal}

\usepackage{amsmath}
\usepackage[nopatch]{microtype}
\usepackage{booktabs}

\usepackage{CJKutf8}    
\usepackage{tikz,xcolor,hyperref} 
\usepackage{amssymb}    
\usepackage{longtable}
\usepackage{multicol}
\usepackage{float}
\usepackage{lscape}

\definecolor{lime}{HTML}{A6CE39}
\DeclareRobustCommand{\orcidicon}{%
    \begin{tikzpicture}
    \draw[lime, fill=lime] (0,0) 
    circle [radius=0.16] 
    node[white] {{\fontfamily{qag}\selectfont \tiny ID}};
    \draw[white, fill=white] (-0.0625,0.095) 
    circle [radius=0.007];
    \end{tikzpicture}
    \hspace{-1mm}
}

\newcommand{\orcidChrisO}{\href{https://orcid.org/0000-0003-0017-349X}{\orcidicon}}
\newcommand{\orcidChrisW}{\href{https://orcid.org/0000-0002-4569-016X}{\orcidicon}}
\newcommand{\orcidMike}{\href{https://orcid.org/0000-0001-7801-1410}{\orcidicon}}
\newcommand{\orcidSeoWon}{\href{https://orcid.org/0000-0002-3118-8275}{\orcidicon}}
\newcommand{\orcidJohn}{\href{https://orcid.org/0000-0003-2858-9657}{\orcidicon}}
\newcommand{\orcidMarc}{\href{https://orcid.org/0000-0003-3882-418X}{\orcidicon}}
\newcommand{\orcidGary}{\href{https://orcid.org/0000-0001-7019-649X}{\orcidicon}}

\newcolumntype{C}[1]{>{\centering\let\newline\\\arraybackslash\hspace{0pt}}m{#1}}

\newcommand\arcsec{$^{\prime\prime}$}

\title{SkyMapper Southern Survey: Data Release 4}

\author{{Christopher A.\ Onken}~\orcidChrisO}
\affiliation{
Research School of Astronomy and Astrophysics, Australian National University, Canberra ACT 2611, Australia}
\alsoaffiliation{
Centre for Gravitational Astrophysics, Australian National University, Canberra ACT 2600, Australia}
\email[Christopher A.\ Onken]{christopher.onken@anu.edu.au}

\author{~Christian Wolf~\orcidChrisW}
\affiliation{
Research School of Astronomy and Astrophysics, Australian National University, Canberra ACT 2611, Australia}
\alsoaffiliation{
Centre for Gravitational Astrophysics, Australian National University, Canberra ACT 2600, Australia}

\author{~Michael S.\ Bessell~\orcidMike}
\affiliation{
Research School of Astronomy and Astrophysics, Australian National University, Canberra ACT 2611, Australia}

\author{~Seo-Won Chang~\orcidSeoWon}
\affiliation{
Astronomy Program, Department of Physics and Astronomy, Seoul National University (SNU), Seoul 08826, Republic of Korea}
\alsoaffiliation{
SNU Astronomy Research Center, Seoul National University (SNU), 1 Gwanak-rho, Gwanak-gu, Seoul 08826, Republic of Korea}

\author{~Lance C.\ Luvaul}
\affiliation{
Research School of Astronomy and Astrophysics, Australian National University, Canberra ACT 2611, Australia}

\author{~John L.\ Tonry~\orcidJohn}
\affiliation{
Institute for Astronomy, University of Hawaii, 2680 Woodlawn Drive, Honolulu, HI 96822, USA}

\author{~Marc C.\ White~\orcidMarc}
\affiliation{
Research School of Astronomy and Astrophysics, Australian National University, Canberra ACT 2611, Australia}

\author{~Gary S.\ Da~Costa~\orcidGary}
\affiliation{
Research School of Astronomy and Astrophysics, Australian National University, Canberra ACT 2611, Australia}

\keywords{optical astronomy; sky surveys; catalogs} 

\begin{document}

\begin{abstract}
We present the fourth data release (DR4) of the SkyMapper Southern Survey (SMSS), the last major step in our hemispheric survey with six optical filters: $u, v, g, r, i, z$. 
SMSS DR4 covers 26,000~deg$^{2}$ from over 400,000 images acquired by the 1.3~m SkyMapper telescope between 2014-03 and 2021-09. 
The 6-band sky coverage extends from the South Celestial Pole to $\delta=+16^{\circ}$, with some images reaching $\delta\sim +28^{\circ}$.
In contrast to previous DRs, we include all good-quality images from the facility taken during that time span, not only those explicitly taken for the public Survey. From the image dataset, we produce a catalogue of nearly 13 billion detections made from $\sim700$ million unique astrophysical objects. 
The typical 10$\sigma$ depths for each field range between 18.5 and 20.5~mag, depending on the filter, but certain sky regions include longer exposures that reach as deep as 22~mag in some filters.
As with previous SMSS catalogues, we have cross-matched with a host of other imaging and spectroscopic datasets to facilitate additional science outcomes. SMSS DR4 is now available to the worldwide astronomical community.
\end{abstract}

\section{Introduction}

As originally proposed in 2002, the Stromlo Southern Sky Survey (S4) aimed to create the first digital map of the southern sky, providing a database of a billion objects. While it was always expected that the database would be used by the entire community, the four key areas driving the S4 design at the Australian National University (ANU) were: "studying the creation of our solar system through a census of distant asteroids, exploring how stars and planets form by observing nearby young stars, probing the shape and extent of the Galaxy's dark matter halo, and discovering when the first stars in the Universe formed." 

In January 2003, bushfires destroyed the original S4 telescope at Mount Stromlo Observatory: the 1.27~m "Great Melbourne Telescope", which had recently completed a seven-year survey of the Galactic Bulge and Magellanic Clouds in search of microlensing by MAssive Compact Halo Objects, i.e., the MACHO Project \citep{2000ApJ...542..281A}. Pursuit of the original S4 science goals required a new facility, and with the new telescope came a new survey name and new survey plan.

The 1.3~m SkyMapper telescope, located at Siding Spring Observatory, has been conducting the SkyMapper Southern Survey (SMSS) since early 2014. The telescope has a 5.7~deg$^{2}$ field-of-view and a 32-CCD mosaic camera (10$\times$ the field-of-view of the original facility, with a slightly smaller pixel scale). The SMSS includes multiple visits of varying depth in six optical filters: $u, v, g, r, i,$ and $z$ \citep{2011PASP..123..789B}. Full 6-band coverage of the survey now extends from the South Celestial Pole to $\delta=+16^{\circ}$, with some fields of partial coverage reaching as far North as $\delta\sim +28^{\circ}$.

SMSS Data Release 1 \cite[DR1, and its photometric recalibration, DR1.1;][]{2018PASA...35...10W} presented a shallow pass over the hemisphere, with 10$\sigma$ point-source depths\footnote{All magnitudes are given on the AB scale \citep{1983ApJ...266..713O} unless otherwise indicated.} of $\sim18$~mag. In DR2 \citep{2019PASA...36...33O}, we began introducing longer images that extend the depth by $1.5-3$~mag. With DR3, we nearly doubled the number of images and detections, as the coverage of the deeper component of the survey expanded.

SMSS data have been used for a diverse array of scientific investigations, ranging from Earth-impacting asteroids closer than the Moon \cite[namely, the last ex-atmospheric measurements of 2018~LA on UT 2 June 2018;][]{2021M&PS...56..844J} to the most luminous quasars at redshift $\sim5$ \cite[including the most UV-luminous quasar known, SMSS~J2157-3602, and the most complete survey of the bright end of the $z\sim 5$ quasar luminosity function;][]{2018PASA...35...24W,2022MNRAS.511..572O}. However, the greatest impact of SMSS, in publications and citations, has been in the field of Extremely Metal Poor star searches, one of the principal science goals underpinning the SkyMapper project and the design of its unique filter set \cite[e.g.,][]{
2019MNRAS.488L.109N,2019MNRAS.489.5900D,2021MNRAS.507.4102Y,2023MNRAS.524..577O}.

\begin{figure*}[ht!]
    \begin{center}
    \includegraphics[width=\columnwidth]{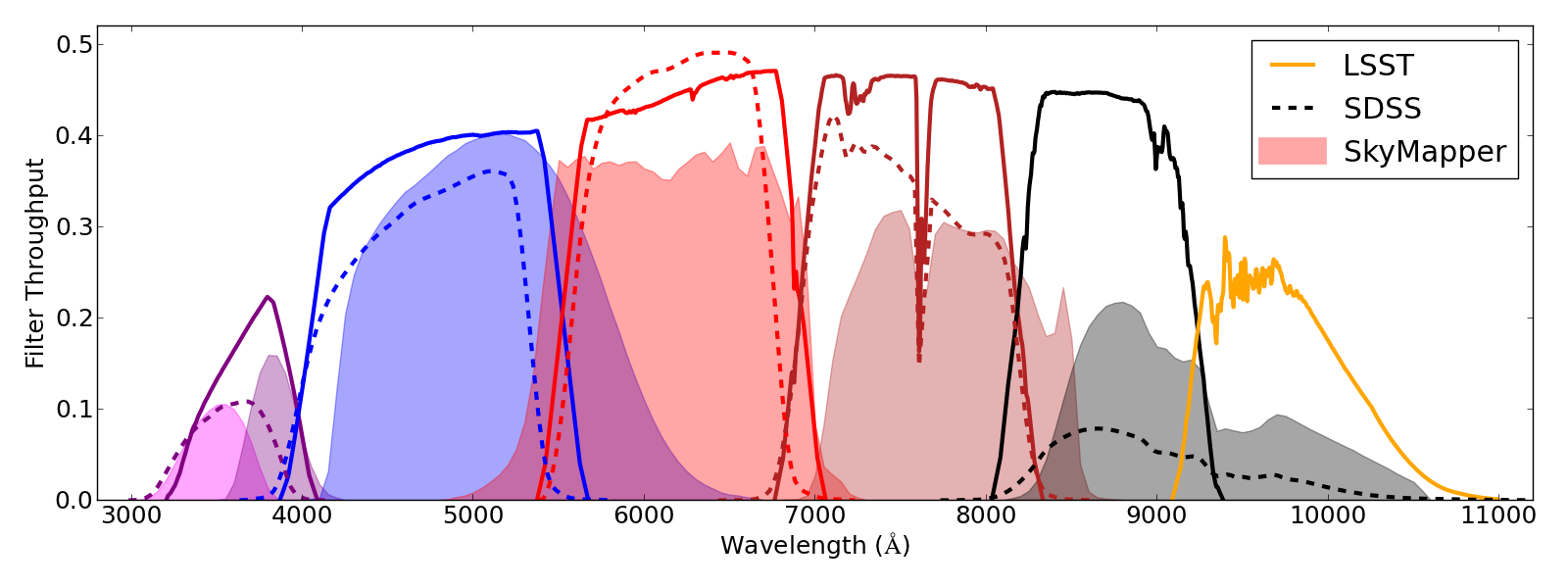}
    \caption{SkyMapper bandpasses: throughput curves are shown for the six SMSS passbands $uvgriz$ relative to SDSS $ugriz$ and LSST $ugrizy$. The SkyMapper curves describe the end-to-end throughput including atmosphere, all optical components and the detector, at airmass 1.3 in good weather with a recently cleaned main mirror; note, that the $u$-band sensitivity varies among the mosaic CCDs. These curves were calibrated from the count rates of standard stars in a range of survey images. According to the \href{https://www.sdss3.org/instruments/camera.php}{SDSS documentation}, the SDSS passbands do not show the total throughput from atmosphere to detector. The LSST passbands have been modelled to include the full system throughput (optical elements, detectors, and an airmass 1.0 atmosphere with aerosols; version 1.5 from \href{https://github.com/lsst/throughputs/blob/main/baseline/README.md}{this GitHub page}).}
    \label{fig:filters}
    \end{center}
\end{figure*}

Here, we present SMSS~DR4, which, compared to DR2, nearly doubles the time baseline from 4~years to 7.5~years, more than triples the number of images (to over 400,000), expands the sky coverage by 5,000~deg$^{2}$ (to over 26,000~deg$^{2}$), and improves the astrometry and photometry of the dataset. DR4 is immediately available to the worldwide community. 

Section~\ref{sec:overview} provides an overview of the SkyMapper facility and operations. We describe the SMSS design, nightly operations, and data release history in Section~\ref{sec:smss}. In Section~\ref{sec:processing}, we describe the SMSS data and its DR4 processing, highlighting the differences from prior releases. Section~\ref{sec:distill} describes the distillation process to go from photometric detection lists to astrophysical object lists. Section~\ref{sec:properties} details the properties of the DR4 dataset, from the selection of the input images to the final catalogue. In Section~\ref{sec:data}, we describe how to access the DR4 dataset, as well as the data format.
Section~\ref{sec:future} describes our plans for augmentation of DR4, as well as future aspirations. In Section~\ref{sec:summary}, we summarise the data release.

\section{SkyMapper Overview}
\label{sec:overview}

The SkyMapper telescope operates at Siding Spring Observatory (SSO), near Coonabarabran, New South Wales, and was inaugurated in 2009. In the subsections below, we describe the facility itself, some of its operational constraints, the history of the telescope operations, and the absolute calibration of its passbands.

\subsection{The Facility}
\label{sec:facility}

The SkyMapper telescope is a modified Cassegrain design, with a 3-element corrector lens assembly, providing a system $f$-number of $f$/4.79 and a delivered field-of-view 3.4~deg in diameter \citep{2006SPIE.6267E..0ER}. The primary mirror has a diameter of 1.35m, with an unobstructed aperture of 1.30m, and features a protective coating that may be washed (rather than having to re-aluminise the mirror). The telescope has an alt-az design with an image rotator, and sits inside an 11.5m tall, 3-level enclosure. Both the telescope and dome were designed and constructed by Electro Optic Systems (EOS). The mean geographical coordinates of the telescope focus (in the WGS-84 frame) have been measured by GPS to be (Latitude, Longitude) = ($-31.272147 \pm 0.000012, 149.061416 \pm 0.000014$)~deg, with a Height Above Ellipsoid of $1165.5 \pm 1.3$~m. The location has been registered with the International Astronomical Union (IAU) Minor Planet Center\footnote{See \url{https://minorplanetcenter.net}.} and given observatory code Q55.

\begin{table}[ht]
\begin{threeparttable}
\caption{Properties of the SkyMapper filter set: name, central wavelength $\lambda_\mathrm{cen}$ and full width at half maximum (FWHM), peak system efficiency $Q_\mathrm{peak}$ as in Figure~\ref{fig:filters}, zeropoint loss ($\Delta$ZP) per airmass (AM), median PSF FWHM among Quality 1 and 2 Shallow and Main Survey images, and magnitude limit of a typical Q1/2 100~sec exposure (peak of source count histogram). Note that the $u$-band has a 0.7~per~cent red leak; the $\lambda_\mathrm{cen}$ and $\mathrm{FWHM}$ refer to the main bandpass.}
\label{tab:filters}
\centering
\begin{tabular}{lccccc}
\toprule
\headrow Filter & $\lambda_{\rm cen}$/FWHM & $Q_{\rm peak}$ & $\Delta$ZP/AM & PSF FWHM & 100sec limit \\
\headrow & (nm) &    & (mag~AM${-1}$) & (arcsec) & ABmag \\
\midrule
$u$ & 350/\phantom{0}43 & 0.11 & 0.65 & 3.15 & 18.1 \\  
$v$ & 384/\phantom{0}31  & 0.16 & 0.41 & 3.00 & 18.3 \\ 
$g$ & 510/157 & 0.40 & 0.19 & 2.80 & 19.5 \\ 
$r$ & 617/158 & 0.39 & 0.12 & 2.63 & 19.2 \\  
$i$ & 779/141 & 0.32 & 0.08 & 2.54 & 18.4 \\ 
$z$ & 916/\phantom{0}84  & 0.22 & 0.08 & 2.49 & 17.8 \\  
\bottomrule
\end{tabular}
\end{threeparttable}
\end{table}

The custom-designed SkyMapper filter set \citep{2011PASP..123..789B} is shown in Figure~\ref{fig:filters} alongside those of the Sloan Digital Sky Survey (SDSS) and the Vera C.\ Rubin Observatory/Large Synoptic Sky Survey (VRO/LSST). The filters are  $31\times31$~cm in size, and are primarily of coloured-glass construction ($uvgz$), although the $r$ and $i$ filters utilise dielectric coatings ($r$ on both wavelength extremes, and $i$ on the long-wavelength side). The $u$-band is known to have a red leak centred at 717~nm with $\approx0.7$~per~cent throughput relative to the transmission peak in the main bandpass at 350~nm (together with the CCD response, the effective throughput of the leak 
is twice as large). Similarly, the $v$-band was found to have a red leak at 690~nm, but 10 times lower amplitude than that of $u$-band \cite[$\sim0.05$~per~cent;][]{2013_Rocci_PhD} and has not been included in the passband models.

The six filters\footnote{An additional filter, centred around H$\alpha$, has also been fabricated, although its use has been limited and its calibration has not been extensively studied, thus no H$\alpha$ images are included in SMSS DR4.} are housed in a slide system, with filter pairs residing in one of three levels. The inactive filters are moved to either side of the optical path and provide extra baffling of stray light. The filters are locked into place with pneumatic pins, which are activated with a dry-air system that also supplies a flow of low-humidity air over the camera window and detector controllers to mitigate condensation. Filter-free observations are possible, but no such images are included in SMSS DR4. Absolute calibration of the passbands is described in Section~\ref{sec:abscal}.

The camera shutter was manufactured by a group from the Argelander Institute for Astronomy of the University of Bonn (now the private company, Bonn-Shutter GmbH), and consists of two moving blades that work together to expose and then obscure the detectors with high precision ($\pm$200 $\mu$s variation in the effective exposure time across the field of view). While providing a high uniformity in the exposure time, the time at which the exposure begins then becomes a function of location within the mosaic\footnote{The blades travel horizontally, so the exposure timing is simply related to the mosaic x-axis coordinate.}. The travel time is 658~milliseconds, and each successive exposure sees the direction of blade travel reversed.

At the focal plane of the telescope is the ANU-built mosaic camera \citep{2006SPIE.6269E..27G}. The 32 CCDs in the SkyMapper mosaic (e2v CCD44-82-1-D03 deep-depletion, back-illuminated devices) are each 2048$\times$4096 pixels, giving a total of 268 million on-sky pixels with a plate scale of $\approx0.497$~arcsec~pix$^{-1}$. Variations in the pixel area are seen across the field-of-view, with two, opposing corners exhibiting areas approximately 2~per~cent smaller than in the mosaic centre (the other pair of corners show a smaller change).

\begin{figure}
    \begin{center}
    \vspace{-3.5mm} 
    \includegraphics[width=\columnwidth]{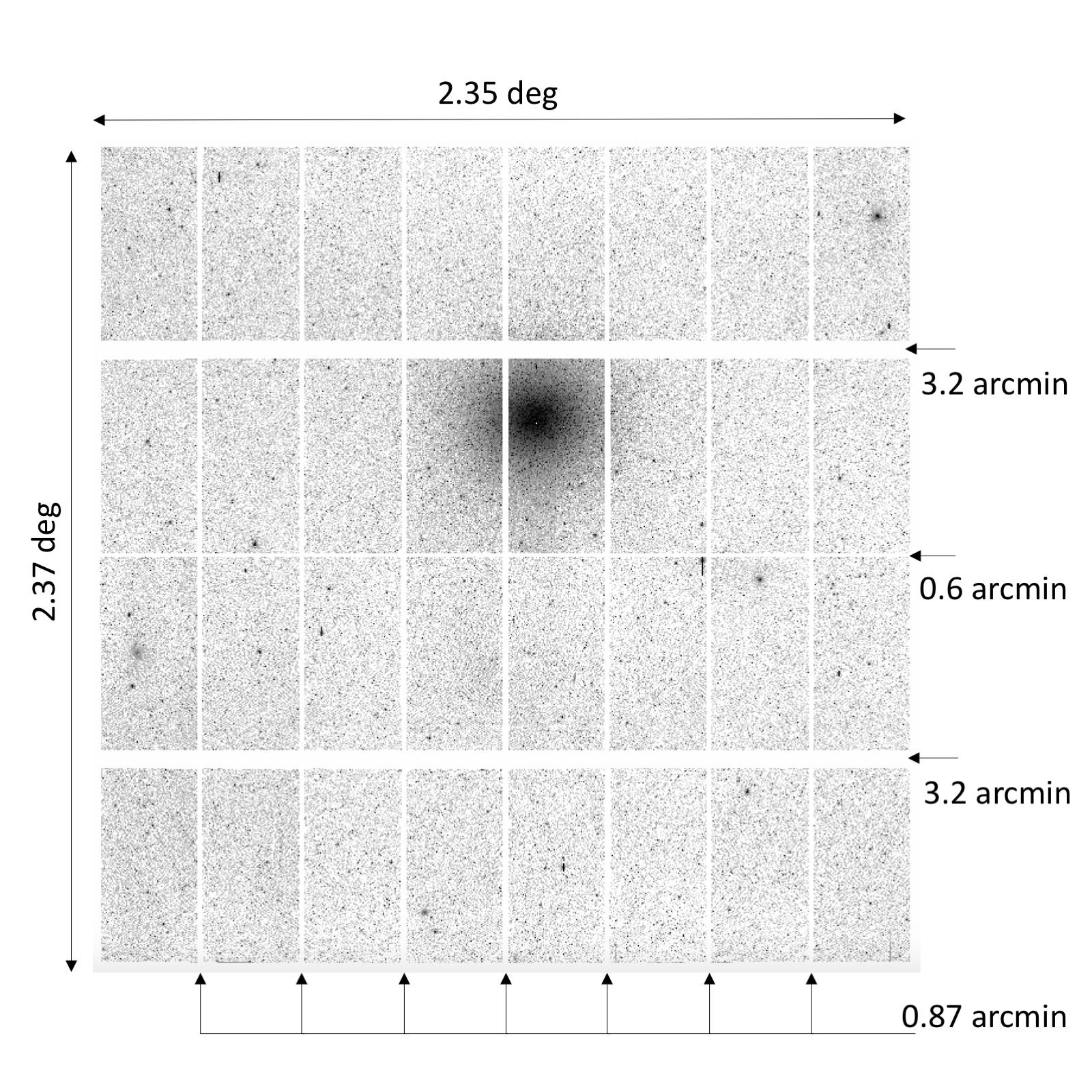}
    \caption{SkyMapper detector mosaic, with sky coverage and gaps between CCDs indicated. Each CCD is approximately $17\times34$~arcmin in size. The background is a 100-s $i$-band image of the region around the Milky Way globular cluster, Omega Centauri.}
    \label{fig:mosaic}
    \end{center}
\end{figure}

Allowing for the gaps between detectors illustrated in Figure~\ref{fig:mosaic}, the camera delivers a 90~per~cent fill factor over $2.35 \times 2.37$~deg. The mosaic is read out by 64 amplifiers that are driven by a Scalar Topology Architecture of Redundant Gigabit Readout Array Signal Processors (STARGRASP) controller system developed by the Institute for Astronomy at the University of Hawai'i \citep{2008SPIE.7014E..0DO}. Using the methodology of \citet{2021PASA...38....6R}, measurements of the detector gain from pre-flash images\footnote{The pre-flash images use an array of light emitting diodes inside the camera to impose an illumination pattern on the mosaic without opening the camera shutter, giving rise to a highly consistent count level well above the level of the bias.} taken in 2015 were found to be $0.74$~ADU~e$^{-1}$, with a root-mean-square (RMS) scatter of $0.03$~ADU~e$^{-1}$. A repeat of the measurement in 2023 found $0.76\pm0.03$~ADU~e$^{-1}$. Thus, we assume a universal gain value of $\approx0.75$~ADU~e$^{-1}$. Each amplifier includes 50 pre-scan and 50 post-scan pixels, and each 1124$\times$4096 readout is stored in a separate extension of a single multi-extension FITS (MEF) file per exposure. 

\subsection{Facility constraints}
\label{sec:constraints}

\begin{figure}
    \begin{center}
    \includegraphics[width=\columnwidth]{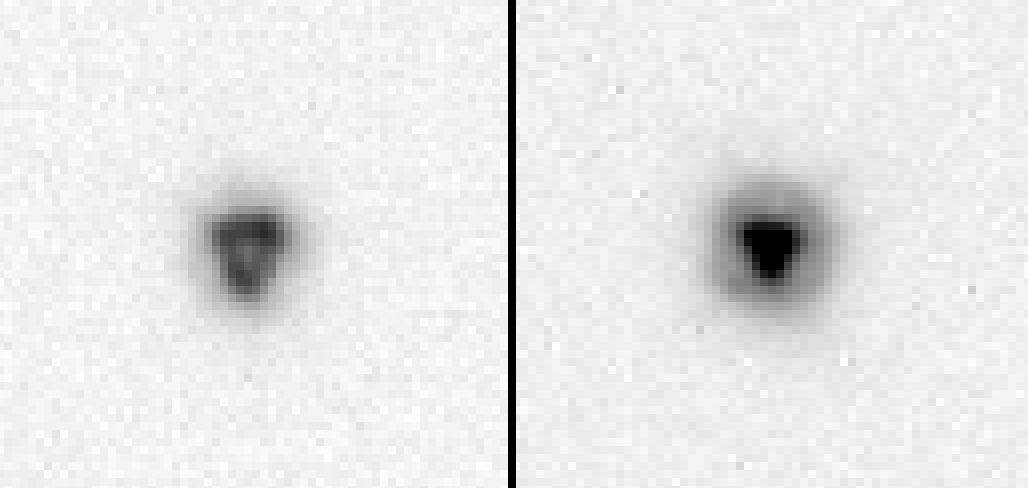}
    \caption{Trefoil distortions in the point spread function (PSF), comparing the mosaic corner (left) with the mosaic centre (right). Each cutout is $30\times30$~arcsec and comes from the same 10-s $i$-band image. The greyscale uses a logarithmic stretch to highlight the extended emission of the PSF.}
    \label{fig:PSF}
    \end{center}
\end{figure}

\begin{figure}[ht]
    \begin{center}
    \includegraphics[width=\columnwidth]{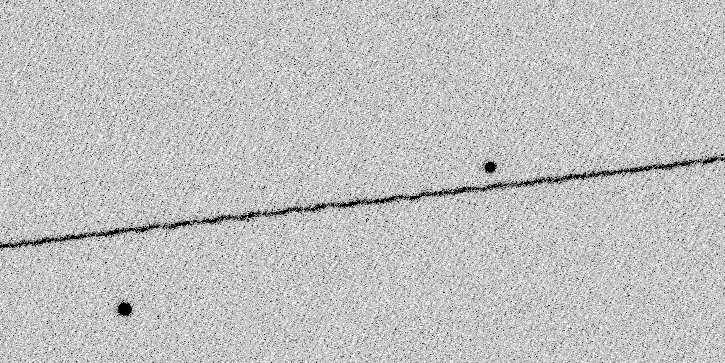}
    \caption{Satellite trail showing the residual vibrations in the system, which contribute to a floor in the PSF FWHM distribution. The portion shown is $10\times5$~arcmin, with North up and East left, and comes from a 100-s $u$-band exposure (\texttt{IMAGE\_ID}=20141019163153) with a mean FWHM of 2.3~arcsec.}
    \label{fig:trails}
    \end{center}
\end{figure}

The images obtained by SkyMapper are affected by curvature in the focal plane, with significant point spread function (PSF) variations with radius \cite[for detail, see Fig.~4 in the DR1 paper:][]{2018PASA...35...24W}. 
The focal position has been selected to balance the image quality across the mosaic, but the corners exhibit a trefoil shape that is particularly visible in good seeing conditions (Figure~\ref{fig:PSF}). The degraded PSF quality compared to expectations is likely due to the mechanical pressure applied to the primary mirror at three locations around the perimeter. These clamps were installed in 2013 to mitigate the vibrations of the primary mirror, which had contributed to 
an effective seeing as bad as 8~arcsec across the whole mosaic.
Residual vibrations in the secondary mirror can be seen in the satellite trails present in some DR4 images (Figure~\ref{fig:trails}), and sets a floor in the seeing statistics of $\sim1$~arcsec.

The vibrations in the system are driven by the closed-cycle gaseous helium cooling system used to maintain the CCDs at their operating temperature of 155~K. The two single-stage Gifford-McMahon coldheads on either side of the camera use free-floating displacers that cannot be mechanically coupled. This leads to vibrational impulses from the two coldheads that are constantly changing in relative phase, and which drive vibrations through the telescope structure. A series of mechanical modifications to stiffen the telescope supports and damp the vibrations were enacted between 2010 and 2013, which eventually improved the image quality sufficiently to begin the Survey in March 2014.

The cable drape that connects the fixed components of the dome to the rotating telescope and camera imposes limitations on the on-sky position angle (PA) that may be reached. Thus, exposures that approach the cable drape limits may be forced to rotate by $180^{\circ}$ at the same RA/Dec position, imposing an overhead of $\sim$40~s.
The images for the public Survey are typically acquired with a PA of $0^{\circ}$ (meaning North is aligned with increasing y-axis pixel). In SMSS DR4, such images constitute 92~per~cent of the dataset, with 75~per~cent of the remainder having ${\rm PA}=180^{\circ}$ (giving the same mosaic footprint on the sky). 
The strong preference for ${\rm PA}=0^{\circ}$ results from that being the default setting, if available. Images with ${\rm PA}=180^{\circ}$ are required on the Eastern side of the meridian, but these were disallowed for most Survey images from 2016-12-20 onwards. It was found that images just East of the meridian exhibited worse PSFs than those just West of it, with the FWHM in $g$-band, e.g., increasing from an average of 2.6~arcsec to 3.2~arcsec. 

\vspace{0.5cm}

\subsection{Operational History}
\label{sec:history}

{\it "The best-laid schemes o' mice an' men / Gang aft agley"}\\
- Robert Burns, 1785

The original description of the SkyMapper facility \citep{2007PASA...24....1K} was published before the telescope had achieved first light at the factory. In this section, we describe the operational history of the facility, particularly any modifications to plans previously published.

SkyMapper was officially opened in May 2009, but underwent a long commissioning period. Much of the effort during that time was devoted to improving the image quality, which was significantly worse than expected for the site. The main culprit was found to be the vibrations noted above (Sec.~\ref{sec:constraints}), which were mitigated through several rounds of mechanical engineering interventions to stiffen the structure and damp the vibrations of the primary and secondary mirrors.

Another unanticipated factor degrading the quality of SkyMapper images was residue on the telescope axis encoders arising from infestations of the dome by ladybird beetles (taxonomic family {\it Coccinellidae}). The contamination of the optical encoders caused repeatable tracking errors. These were largely resolved by delicate cleaning of the encoder tapes, but the difficulty (for humans) in accessing the encoders and lasting damage to the surfaces by the insects has continued to impact SkyMapper tracking and is thought to be responsible for the gradual accumulation of pointing errors ($\sim$1~arcmin~d$^{-1}$, but reset by homing the telescope, which became regular practice).

In light of the image quality issues early on, the Shack-Hartmann wavefront sensing system was of little utility and has not been used in subsequent operations. The off-axis auto-guider was also never implemented.

Additional delays to the start of Survey operations were incurred because of a major bushfire which hit SSO in January 2013. In addition to the loss of 53 homes and the SSO Lodge, the Wambelong fire burned more than 95~per~cent of the Warrumbungle National Park that surrounds SSO, and 55,000 hectares overall. SkyMapper was subjected to intrusion of ash into the dome, but suffered no lasting physical damage. However, because of the potentially corrosive nature of the ash, an extensive cleaning process was undertaken, including the careful washing and baking out of most circuit boards in the dome. Ultimately, 10 weeks were spent on the bushfire cleanup process.

One of the elements which was exposed to the ash of the Wambelong fire was SkyDICE \cite[the SkyMapper Direct Illumination Calibration Experiment;][]{2013_Rocci_PhD, 2015A&A...581A..45R}. This set of calibrated photodiodes was intended to provide absolute measurements of the system throughput at 23 wavelengths (with LED emission widths of 20-50~nm) across the SkyMapper filter set, while also sampling any spatial variations thereof. Following the fire, the module was not put into regular deployment, and in 2016, the system was disassembled and shipped back to the team at Laboratoire de Physique Nucl\'{e}aire et de Hautes-\'{E}nergies in Paris. Lack of available personnel precluded its recalibration and return to SSO.

Despite these teething issues with the telescope, scientific observations prior to the commencement of the Survey had already demonstrated the power of the bandpass design for selecting low metallicity stars. These discoveries included the most iron-deficient star known at that time \citep{2014Natur.506..463K}, and the first large collection of metal-poor stars in the Galactic Bulge \citep{2015Natur.527..484H}.

Ahead of the start of the Survey in early 2014, the strategy was reassessed to take account of the intervening scientific progress. Priority was given to the Shallow Survey (see Sec.~\ref{sec:design}) for the first year, to obtain a uniform dataset across the full RA range for the scientific community. In addition, the Shallow Survey exposure times were increased for both the bluest filters and the reddest filters, to deliver a similar depth of $\sim18$~ABmag.

In contrast, the exposure time for the Main Survey was reduced from 110~s to 100~s. While this came at the cost of a small amount of depth, it was much less than the effects of the image quality being worse than expected, and helped to improve the rate of progress for the Survey. When the Main Survey became fully activated in April 2015, the cadence of images was significantly relaxed compared to the original schedule -- the loss of depth having reduced the importance of the sampling of RR~Lyrae light curves, and the start of {\it Gaia} observations having made Trans-Neptunian Object proper motions less compelling to measure with SkyMapper. As a result, the telescope was able to place a stronger emphasis on observations close to the meridian.

When the Shallow Survey was prioritised early in the Survey, it also involved a reconsideration of the photometric calibration strategy. No longer was the Shallow Survey only undertaken in photometric conditions -- ultimately, photometricity was not actively measured in real-time. Thus, the fields with spectrophotometric standard stars (see Sec.~\ref{sec:design}) were not used to anchor the overall Survey calibration, but external all-sky data sources were relied upon, culminating with the {\it Gaia}-based solution described below (Sec.~\ref{sec:zp}).

The early months of the Survey operations also revealed other properties of the data, which motivated certain alterations of the hardware configuration and nightly activities. For example, Sec.~4.3.3 of \citet{2018PASA...35...10W} describes the changes to the detector voltages that were required to remove spatially-varying curvilinear features in the images. However, the images obtained prior to that correction in July 2014 retain those artefacts. Similarly, the approach used to remove large-scale flux gradients from the twilight flatfields (see Sec.~\ref{sec:operations}) was only developed near the end of 2014, and the data taken prior to November 2014 will be less well calibrated because of the flatfields having been obtained at fixed position angle.

\begin{figure}
    \begin{center}
    \includegraphics[angle=270,width=\columnwidth]{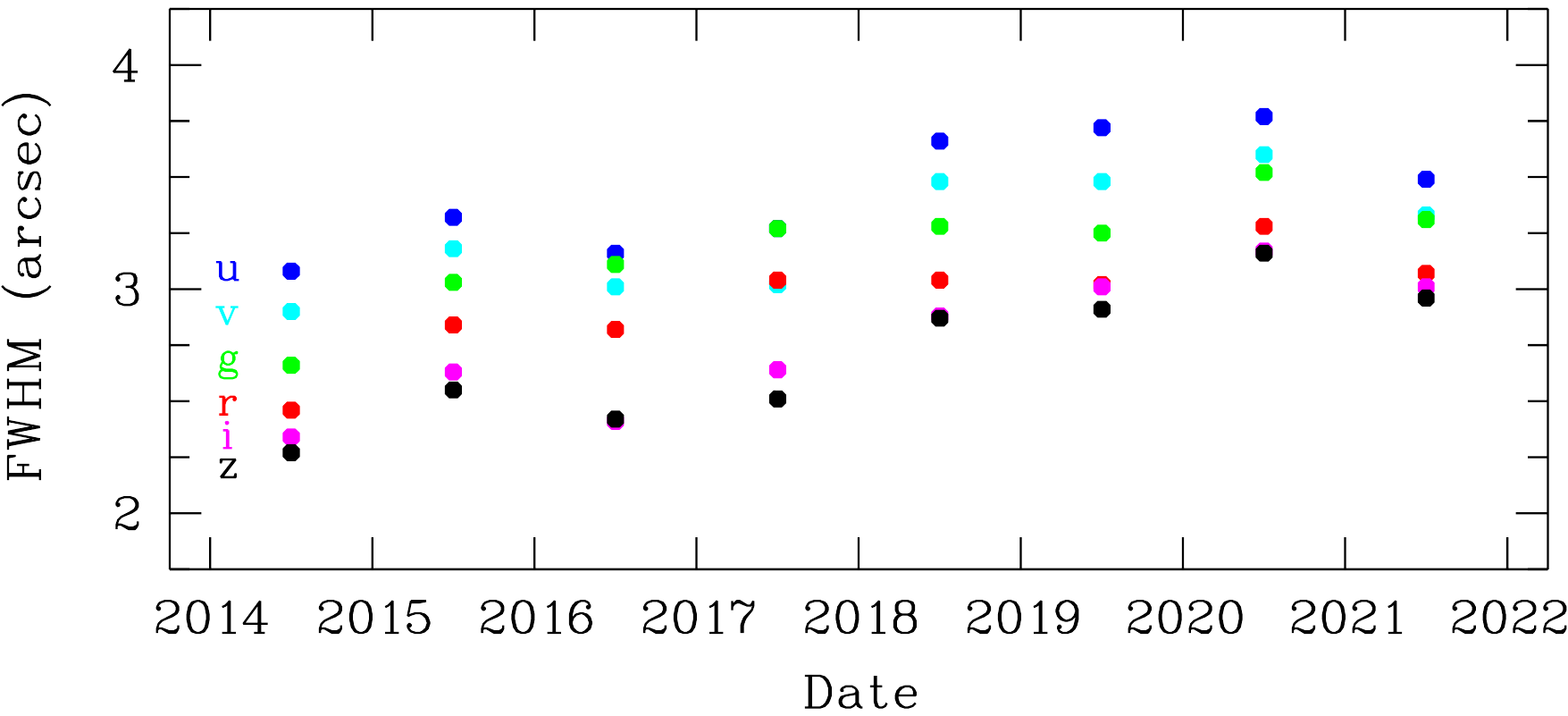}
    \caption{Time series of PSF FWHM per-filter in DR4, averaged per calendar year: an increase during the second half of the survey period can be seen. The reason for the variation are likely imperfect focus settings.}
    \label{fig:Seeingtrends}
    \end{center}
\end{figure}

The dome cooling systems installed in the SkyMapper enclosure proved unable to pre-cool the internal air to the expected nighttime ambient temperatures. Therefore, rather than performing the planned focus runs at the start of each night, a more dynamic focus system was enacted. Suitable focus control was found to be achievable by simple correction for the thermal expansion of the telescope truss along with an airmass-dependent flexure compensation with the secondary mirror (much improved after updates to the coefficients in Feb 2014). The focus equation is linear in temperature, with a slope of $-46~\mu$m~K$^{-1}$. 

In 2020, a gradual degradation of the mean PSF over the years of the Survey was noticed, leading to a mild revision of the focus equation. Pairs of out-of-focus "doughnut" images bracketing the predicted focus setting by $\pm 400~\mu$m are obtained during twilight, and the ratio of their diameters indicates the offset between real and predicted focus. 

Over the Survey years, the average PSF FWHM has evolved in a pattern that is common to all passbands (see Figure~\ref{fig:Seeingtrends}). The focus offset reconstructed from doughnut images evolve broadly similarly to the mean PSF FWHM, indicating that an imperfect focus equation is the cause of the evolution. The worst image quality was recorded in early 2020, when the hot summer led to catastrophic bushfires across Eastern Australia; this period also shows the strongest reconstructed focus offsets, of up to $50\mu$m, and equivalent to misjudging the relevant temperature for the telescope structure by $\sim 1$~degree. The seeing records from the 3.9~m Anglo-Australian Telescope (AAT), also located at SSO, shows relatively stable seeing behaviour over the DR4 period. However, the median AAT seeing from the start of 2020 to the end of the DR4 date range increased from 1.5 to 1.75~arcsec (C.~Ramage, priv. comm.).

Faults in the detector cooling system have been the primary source of extended periods of maintenance downtime since the start of Survey operations: each of the three helium compressor failures (in 2015, 2016, and 2017) has taken roughly 8 weeks to resolve. In addition, faults with some of the detector controllers have resulted in periods of partial mosaic operation\footnote{Incomplete mosaic operation motivated the decision to limit the scope of images for DR4 to those taken before October 2021.}. SSO was closed from 25 March due to COVID-19, but SkyMapper was able to restart operations from 07 May 2020.

\subsection{Absolute passband calibration}\label{sec:abscal}

We determined the end-to-end throughput of the SkyMapper passbands from DR4 data. The throughput of the glass filters is expected to change little in wavelength dependence over time. However, the atmospheric throughput fluctuates from day to day and even during the night. The reflectivity of the telescope mirrors tends to degrade over time between the less-than-annual mirror washing cycles; time series data of the image zeropoints show an average loss of reflectivity by $\sim 1$~per~cent per month (see Section~\ref{sec:properties}). We chose to evaluate the throughput from images of the southern spectrophotometric standard stars Feige~110, GD~50, GD~108 and LDS~749B, taken in good weather soon after a mirror washing restored high system throughput. The photon flux of these stars arriving outside of the Earth's atmosphere is known from CALSPEC\footnote{See \url{https://www.stsci.edu/hst/instrumentation/reference-data-for-calibration-and-tools/astronomical-catalogs/calspec}} \citep{2014PASP..126..711B,2020AJ....160...21B} and can be compared to the electron count recorded by the CCD camera. 

We start from laboratory measurements of the filter transmission curves \citep{2011PASP..123..789B}. 
In the $u$-band, the quantum efficiency varies between the detectors in the mosaic, so we use a mean CCD efficiency for the synthetic photometry of the standard stars. Given that we imaged the standard stars only on a subset of CCDs, our throughput estimation is only a rough average for the mosaic in $u$-band. The observations had an average airmass of 1.2, which we use to predict the wavelength dependence of the atmospheric transmission. We base our expectations on a reflective aperture area of 0.95~m$^2$ resulting from a primary mirror with 1.30~m unobstructed aperture diameter and an obstruction from a secondary mirror with 0.69~m diameter. A CCD gain of 0.75~ADU~e$^{-1}$ is used. The resulting end-to-end throughput curves are shown in Figure~\ref{fig:filters}. At the blue end, they are comparable to SDSS, while the red-sensitive CCDs in SkyMapper provide better sensitivity at longer wavelengths.

\citet{2011PASP..123..789B} stated a need to recalibrate the filter transmission curves with evidence from on-sky measurements in the converging beam of the telescope, which is expected to make a difference, especially for the $r$- and $i$-bands that have dielectric coatings. In Section~\ref{calspec_comp} we discuss what we can learn from the DR4 data and give an outlook to potential future calibration improvements.

\begin{table}
\begin{threeparttable}
\caption{CALSPEC standard star fields.}
\label{tab:calspec}
\centering
\begin{tabular}{lcl}
\toprule
\headrow Name & RA / Dec (J2000) & DR4 N$_{\rm visits}$ $u, v, g, r, i, z$ \\ 
\midrule
HD009051 & \phantom{0}22.3079 / $-$24.2257 & 966, 990, 1002, 1005, 998, 1000 \\
HD031128 & \phantom{0}73.1554 / $-$26.9493 & 1181, 1186, 1177, 1200, 1176, 1187 \\
HD074000 & 130.3250 / $-$16.2293 & 806, 812, 802, 820, 817, 811 \\
HD111980 & 193.4275 / $-$18.4076 & 1002, 1008, 999, 1007, 1015, 1017 \\
HD160617 & 265.8200 / $-$40.2054 & 2090, 2086, 2070, 2057, 2034, 2022 \\
H200654  & 316.7583 / $-$49.8487 & 1152, 1148, 1168, 1169, 1171, 1170 \\
GJ754.1A & 290.2604 / \phantom{0}$-$7.5521 & 603, 617, 621, 612, 618, 620 \\
\bottomrule
\end{tabular}
\end{threeparttable}
\end{table}

\section{SkyMapper Southern Survey}
\label{sec:smss}

In the subsections below, we describe the overall design of the SMSS, the typical nightly operations, and the history of SMSS data releases.

\subsection{Survey Design}
\label{sec:design}

In the first year of SMSS operations, the telescope was mainly focused on completing an initial, rapid pass around the sky in all filters. With exposure times between 5 and 40~seconds, the Shallow Survey component of the SMSS achieved a depth of \mbox{$\sim18$~mag} in all 6 filters. The Shallow Survey observations of each field were obtained sequentially in order of increasing filter wavelength, with any interruptions to the 4-minute image set causing the full sequence to be repeated. This dataset was then processed into SMSS DR1 \citep{2018PASA...35...10W}.

After significant sky coverage was obtained, the Shallow Survey was restricted to days around Full Moon (since the short exposure times leave the background levels at modest levels). This change was enacted on UT~2017-03-05.

Early in the Survey, a set of seven fields containing CALSPEC spectrophotometric standard stars \citep{2014PASP..126..711B,2020AJ....160...21B} were observed multiple times per night, again in order of increasing filter wavelength. These stars (Table~\ref{tab:calspec}) were originally intended to form the basis for the photometric calibration of the entire SMSS, but the variable observing conditions and the eventual availability of all-sky photometric datasets of high uniformity and precision (culminating in the {\it Gaia} low-resolution spectroscopy described in Sec.~\ref{sec:zp}), meant that the Standard fields were never used for that purpose. Because of the shorter exposure times (from 3~seconds in $g$ and $r$ to 20~seconds in $u$), these seven fields are nearly the only regions in which sources brighter than $\sim9$~mag are unsaturated. The nightly visits (weather and season permitting) resulted in $\sim$1000 visits per filter for each Standard field. However, after the Shallow Survey was limited to bright Moon phases, the Standard field observations were similarly restricted (from UT~2017-03-05), and then were halted altogether from UT~2021-05-01 onwards, because they consumed $\sim6$~per~cent of the observing time in a clear night.

The next major SMSS component is the Main Survey, wherein the images had a standard exposure time of 100~seconds. These were acquired in two modes: image pairs ($u + v$, $g + r$, or $i + z$; taken together to help protect against uncorrected cosmic rays implying spurious brightening) and colour sequences (10-image collections acquired over a span of 20 minutes in the filter order: $uvgruvizuv$). Pooling the $u + v$ exposures into a single visit was intended to enhance the depth for these two lowest-sensitivity filters. However, it does turn out that they are also useful for observing short-term variability, e.g., in compact eclipsing binaries \citep{2022MNRAS.515.3370L} and Blue Large-amplitude Pulsators \citep{2023arXiv231108775C}.

Late in the Survey operations, a portion of the $u + v$ image pairs were extended to 300~seconds each (with exposures South of $\delta$=-75$^{\circ}$ further lengthened to 400~s, while that region of sky was also restricted to seeing conditions better than 1.9~arcsec from April 2018 onwards). The longer exposures in $u + v$ were intended as a trade-off for a reduced number of visits (and thereby reduced overheads), and were still observationally suitable because the standard 100-s exposures in those filters were read-noise limited. However, the long exposures constitute less than 0.5~per~cent of the Main Survey images in those filters.

In the early years of the Survey, the bad seeing time (with a threshold that evolved between 2 and 3~arcsec) was principally used by the SkyMapper Transient (SMT) survey \citep{2017PASA...34...30S,2019IAUS..339....3M}, which searched $\sim2000$~deg$^{2}$ for supernovae and other transients. In addition, a fraction of the good-seeing time was made available to Australia-based applicants, totalling over 500~hours between 2014 and 2019.

\subsection{Nightly Operations}
\label{sec:operations}

In this section, we describe a typical night's operations plan for the SkyMapper facility. The telescope is fully robotic, with no human involvement expected during standard operations. The details have evolved over the course of the Survey, but the description below reflects the current framework.

Each afternoon, a crontab process launches the Scheduler software, a Perl framework that controls the telescope's activities until an automatic shutdown following the morning twilight. In preparation for the night, it pre-selects available survey fields while excluding those containing bright planets. For each SkyMapper image, the Scheduler prepares an observation definition that is provided to the high-level interface software, the Telescope Automation and Remote Observing System \cite[TAROS;][]{2005ASPC..347..563W}, as implemented for SkyMapper \citep{2008SPIE.7019E..2RV}. TAROS coordinates the activities between low-level systems, including the Configurable Instrument Control and Data Acquisition software \cite[CICADA;][]{1997ASPC..125..385Y, 1999ASPC..172..115Y} that interfaces to the camera hardware (filter selector, camera shutter, detector controllers, etc.), and the software that interfaces with the EOS telescope and dome control systems, which run on a separate pair of computers.

A typical night obtains a set of evening bias frames before sunset, and if the weather is suitable for observing, the dome is then opened after the Sun is down in order to obtain twilight flatfields. Working through a sequence of filters of increasing sky-level sensitivity ($u, v, z, i, r, g,$ and the filter-free clear aperture), the telescope takes 3 images at one PA, then rotates 180$^{\circ}$ to obtain another 3 images. (The rotation allows for the trivial correction for the large-scale gradient in sky illumination over the wide field-of-view.) The next filter begins from the same PA, then rotates back to the original PA for the second set of 3 images, and so on through all the filters until all of the individual exposure times exceed 60 seconds. The starting position is selected to be near an Hour Angle of $-1$~h and a Declination of $-35^{\circ}$, while avoiding the Moon, Galactic Plane, and any bright planets (from Venus to Saturn, inclusive). Each observation is executed while tracking the sky, with 30~arcsec dithers in RA and Dec between exposures.

During astronomical twilight (Sun angles between 12 and 18~deg below the horizon), the sky in the redder filters has become faint enough to allow useful astronomical observations. Thus, we allow 100~s Main Survey exposures in $i$ and $z$ to be taken before full nighttime darkness is achieved.

In full darkness, the Scheduler then cycles through the available image types (which may depend upon the Moon's phase and current position relative to the horizon) until it finds a suitable observation to execute. The top priority is given to the Target-of-Opportunity (ToO) programs, including the follow-up of gravitational wave alerts \citep{2021PASA...38...24C}. Next, any other non-Survey images are considered within the UT time boundaries defined by the user per exposure. Then, Shallow Survey, and Main Survey images are considered in turn. For Survey images, each available field is given a weight that incorporates its current Hour Angle, position within a sequence/pair, and other priority levels. The field with the highest weight is translated into a TAROS observing block, which passes the observation definition to the hardware.

For most image types, TAROS is configured to hold two observation definitions, so it can reconfigure the system as soon as it records the completed exposure of the first image. This allows the multiple system components to be reconfigured during the $\approx22$~sec overhead time between images (consisting of approximately 13~sec of readout and 9~sec of additional system overheads). Additional parameters affecting the time between images are the $\sim40$~sec time for the instrument rotator to execute a 180$^{\circ}$ rotation, and the slew speeds of 4~deg~s$^{-1}$ in azimuth and 2~deg~s$^{-1}$ in elevation.

After each image has been written to disk, the QuickLook analysis process is run on two of the image's central amplifiers, one each from CCDs on opposite sides of the mosaic centre. 
Basic image parameters are recorded in the Scheduler's postgreSQL database, including an estimate of the seeing. After normalising between filters (the SkyMapper seeing improves notably towards longer wavelengths) and to an airmass of 1 (using an empirically derived trend that seeing degrades as airmass to the power of 0.8), the last 30 images from within the past 30~minutes have their QuickLook seeing estimates medianed, which serves to establish the current seeing estimate used by the Scheduler in its next observation decision.

At the close of each night, twilight flatfields are obtained in the opposite filter order as in the evening (now beginning near an Hour Angle of $+1$~h), and a final set of 10 bias frames are taken. After observing has concluded, images are transferred from the telescope to the National Computational Infrastructure (NCI) on the ANU campus in Canberra.
The images are stored there until ready to be processed with NCI's high-performance computing system. The ToO images and other high-priority data are typically processed in near-real-time, through a separate data pipeline that operates on the computer systems at Mount Stromlo Observatory, but are still copied to NCI for long-term archiving\footnote{Because the general data processing did not keep pace with the observations, the steady trickle of data from the telescope that had been anticipated proved unnecessary.}. However, for inclusion in DR4, all such images are processed from a raw state as described in Section~\ref{sec:processing} below.

\subsection{Previous Data Releases}
\label{sec:previous}

Over the course of the Survey, data releases of increasing sky coverage, data volume, and photometric quality have been made available. The series of DR parameters, including the current DR4, is given in Table~\ref{tab:prev_dr}.

\begin{table*}[ht]
\begin{threeparttable}
\caption{History of data releases of the SkyMapper Southern Survey.}
\label{tab:prev_dr}
\centering
\begin{tabular}{lrrrrcl}
\toprule
\headrow Data Release & Images & Sky Area & Objects & Detections & Release Date & Reference\\
&  & deg$^{2}$ & ($\times10^{6}$) & ($\times10^{9}$) &   &   \\
\midrule
Test DR & 138 & 60 & 0.1 & 0.001 & 2015-05 & \ldots \\
Early DR & 19,147 & 6,700 & 49 & 0.5 & 2016-05 & \ldots \\
DR1 & 66,840 & 20,200 & 285 & 2.1 & 2017-06 & \citet{2018PASA...35...10W} \\
DR2 & 121,494 & 21,000 & 285 & 4.7 & 2019-02 & \citet{2019PASA...36...33O} \\
DR3 & 208,860 & 24,000 & 619 & 8.1 & 2020-02 & \ldots \\
\bf{DR4} & \bf{417,223} & \bf{26,000} & $\mathbf{724}$ & $\mathbf{13.0}$ & \bf{2024-02} & \bf{This paper} \\
\bottomrule
\end{tabular}
\end{threeparttable}
\end{table*}

Experience with the instrument and dataset, as well as the availability of new auxiliary data from other surveys, has led to an evolution in the SMSS image processing \cite[cf.][]{2017ASPC..512..289W, 2017ASPC..512..393L, 2018PASA...35...10W, 2019PASA...36...33O}. All previous data releases are now nearly obsolete: they include data that is excluded from DR4 on the grounds of low quality; such data might be useful if coverage is desired at a specific time. However, for static sources and for reliable statistical studies of variability, the DR4 data set is the best reference. In the following Section, we describe the processing approaches adopted in SMSS DR4.

\section{SMSS DR4 Data Processing}
\label{sec:processing}

Here we describe the main steps in the SMSS image reduction and extraction of photometric parameters in DR4. The images were processed on {\it Gadi}, NCI's peak supercomputer, utilising approximately 550,000 CPU-hours (including time for images which failed subsequent quality cuts). The image properties and derived photometry are stored in a PostgreSQL database (version 11.20), which also manages the data reduction flow of the pipeline by recording the ongoing and completed steps for each image, along with a status code for each step to determine how the image is treated by subsequent steps.

\subsection{Electronic noise filtering, overscan subtraction, and cross-talk correction}
\label{sec:ingest}

The SkyMapper electronics are subject to variable levels of high-frequency sinusoidal noise, which we filter from the images. Each amplifier is Fourier-transformed 
and we search for significant power corresponding to wavelengths between 6 and 8 pixels in the x-axis direction. 

If any amplifier shows such evidence of sinusoidal variations, then all amplifiers for that image are run through a row-by-row fitting procedure. For each row, we ignore the overscan region and subtract a 30-pixel boxcar-smoothed copy of the row from itself in order to isolate variations of the intended frequency. We then 
perform a least-squares fit of a sine function to the row, allowing the wavelength, phase, and amplitude to vary, and taking the results from the FFT analysis as the starting value for the wavelength. The best-fitting sinusoid is then subtracted from the original row (including the overscan region).

Next, the data is analysed with {\sc Source Extractor} \cite[version 2.19.5;][]{1996A&AS..117..393B} in order to identify saturated pixels (taken as those above 58,000 counts), which are flagged in the pixel masks associated with each CCD. The description of all bits in the pixel mask is given in Table~\ref{tab:masks}.
The data from each CCD is then merged into a single FITS image from its two constituent amplifiers, while simultaneously subtracting the bias level using the post-scan region (which is more stable in its behaviour than the pre-scan) and trimming both overscan regions. 

\begin{table}[ht]
\begin{threeparttable}
\caption{Pixel mask bit values.}
\label{tab:masks}
\centering
\begin{tabular}{ll}
\toprule
\headrow Bit value & Description\\
\midrule
1 & Known bad pixels.\\
2 & Not used.\\
4 & Saturated (above 58,000 ADU in raw image).\\
8 & Affected by uncorrectable cross-talk.\\
16 & Masking of full amplifier due to electronic fault.\\
32 & Corrected cosmic rays from L.A.Cosmic.\\
64 & Affected by correlated bias fluctuations.\\
\bottomrule
\end{tabular}
\end{threeparttable}
\end{table}

\begin{figure}
\begin{center}
\includegraphics[width=\columnwidth]{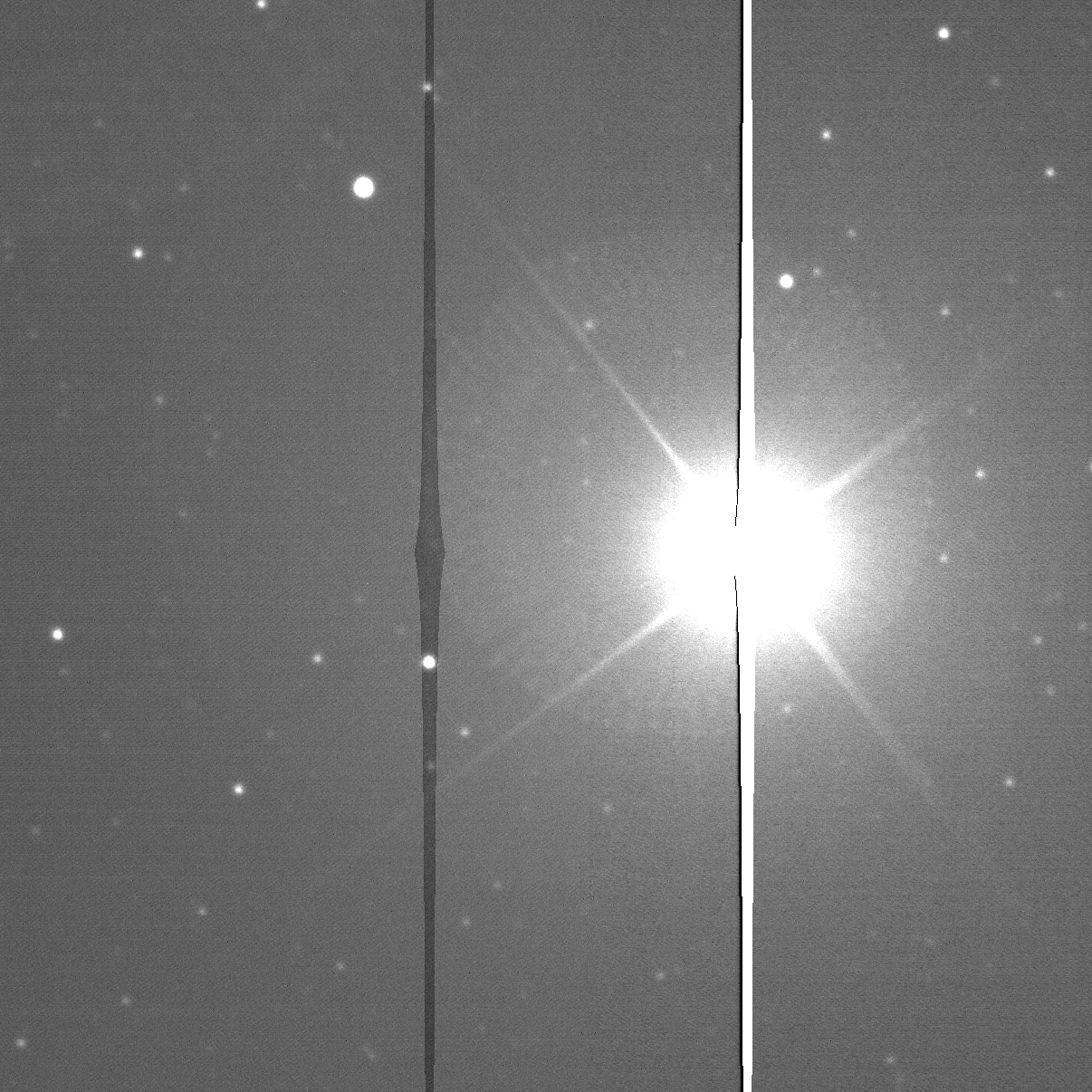}
\caption{SkyMapper image of Altair showing uncorrected cross-talk from saturated sources between adjacent amplifiers of a CCD. The image is a 20~sec $v$-band exposure. Counts in the adjacent amplifier are reduced, but those regions are flagged in the associated pixel masks, as are both the saturated pixels and the low-count ringing adjacent to saturation.
}\label{fig:cross-talk}
\end{center}
\end{figure}

We also correct for cross-talk between the two amplifiers of each CCD. The typical fractional amplitudes are ~$5\times10^{-4}$ and are subtracted from the neighbouring amplifier. Source pixels that are flagged in the previous step as saturated cannot have their cross-talk accurately corrected in the neighbour amplifier, and so the pixels in the latter are flagged as cross-talk-affected in the pixel masks (see Fig.~\ref{fig:cross-talk}). Pixels with full wells also induce amplifier ringing, where the next non-saturated pixel in the row has 0 counts and the subsequent two pixels have severely suppressed count levels. Such pixels are also flagged as saturated in the pixel mask. Additional cross-talk effects on other CCDs read out by the same controller are less than 5 ADU for a fully saturated pixel, and are not presently flagged.

\subsection{WCS solution}

Previous SMSS data releases have utilised the astrometric software of Astrometry.net \citep{2010AJ....139.1782L} to derive WCS solutions for each CCD by matching against the Fourth US Naval Observatory CCD Astrograph Catalog \citep[UCAC4;][]{2013AJ....145...44Z}. The resulting coordinate system provided a good match to the positions of stars presented in {\it Gaia} DR2 \citep{2018A&A...616A...1G}, with typical offset smaller than 0.2~arcsec \citep{2019PASA...36...33O}.

However, in some images, particularly in shorter $u$- and $v$-band exposures, there were insufficient stars matched to UCAC4 to produce a reliable coordinate system for certain CCDs, and the sources on those CCDs were then absent from the photometric catalogue. In DR3, 36~per~cent of images lost at least two CCDs, principally for lack of a WCS solution. This motivated a revised approach to recover those lost CCDs.

For DR4, we have adopted a mosaic-wide algorithm for determining the coordinate system. Based on a careful fitting of {\it Gaia} sources across the mosaic in a set of densely populated images, we improved the mapping of each CCD's location (offset and rotation) relative to the mosaic centre. 

First, to get the overall image boresight, we run {\sc Source Extractor} on all CCDs of an image, and select the 30 brightest stars in each of the central 8 CCDs. We map the CCD x/y positions into mosaic x/y positions and run the (up to 240) stars through Astrometry.net's {\sc solve-field} software (version 0.76), using the '5000-series' index files created from {\it Gaia} DR2 positions. The location of the mosaic centre and position angle of the mosaic system that is determined by that process is then used with the mosaic mapping to generate provisional RA/Dec positions for the full list of sources on each CCD. 

For each CCD, the brightest 300 sources are matched to the nearest {\it Gaia} DR2 source within 10~arcsec, where the {\it Gaia} source is required to have $G$<14~mag (Vega). The median shift in RA and Dec is determined and applied to all 300 sources when a second round of matching is performed, now with a maximum allowed offset of 5~arcsec. The {\it Gaia} positions are then adopted as the 'true' coordinates for those 300 stars. 

From the (up to) 9600 stars with {\it Gaia} DR2 coordinates over 32 CCDs, we then fit a mosaic-wide WCS solution, allowing polynomial corrections to the tangent projection of up to 3rd-order (but excluding radial terms). The resulting coordinate system is then overlayed on a dense grid of x/y positions for each CCD using the {\sc xy2sky} routine from the WCSTools package\footnote{See \url{http://tdc-www.harvard.edu/wcstools/index.html}.} \cite[version 3.8.7;][]{1996ASPC..101...96M,2019ASPC..523..281M}. The grid of CCD-based x/y points with associated RA/Dec coordinates is then re-fit for each CCD to yield the final WCS solution, adopting a TPV convention\footnote{See \url{https://fits.gsfc.nasa.gov/registry/tpvwcs/tpv.html}.}. This two-step fitting procedure ensures that each CCD has a coordinate system that is referenced to its individual CCD centre, and is well defined in relation to its native x/y axes, regardless of any small rotations relative to the overall mosaic. (In practice, the largest rotation of any CCD is less than 0.08$^{\circ}$, but this still translates to a shift of up to 3 pixels at the CCD corners.)

Finally, the WCS solution is saved in the header of each CCD image and its corresponding image mask. In Section~\ref{sec:prop_astrom}, we quantify the accuracy and precision of the resulting WCS solutions.

\subsection{Bias correction}
\label{sec:bias}

On each night of SkyMapper telescope operations, between 10 and 20 bias exposures are obtained. These are treated as in Section~\ref{sec:ingest} and then mean-combined with outlier clipping (to omit cosmic rays) to produce a 2D bias image. If no bias frames are available on a given night, those of adjacent nights are used. The 2D bias image is subtracted from the science frame.

Next, the pipeline addresses variable bias patterns in each row through the use of principal components analysis (PCA). The bias level during readout fluctuates in a manner that, for a given row, can be different from image to image, but which is effectively drawn from a small family of patterns. We first generate a set of principal components (PCs) by subtracting the 2D bias image from each of the input bias exposures, and then determining the top 10 PCs that describe the residual bias variations for each of the 4096 rows in that CCD (and performed separately for each of the two amplifiers). 

In applying the PCs to the science image, we first use {\sc Source Extractor} to determine background and object maps. The former is subtracted and the latter is slightly broadened (using a $\sigma=0.5$~pixels Gaussian kernel) before being used to mask data in the science image. Up to 10 of the PCs generated from the bias images are then fit to the background-subtracted and masked science image on a row-by-row basis, where the number of PCs varies in proportion to the unmasked pixel percentage, $p_{\rm unm}$, of each row (10 PCs for $p_{\rm unm}>25$~per~cent, 5 PCs for $p_{\rm unm}>10$~per~cent, only the first PC for $p_{\rm unm}>2$~per~cent, and no correction below that). The best-fit pattern for each row is then subtracted from the original science frame.

While this approach largely works well to model and remove the bias level, it is known to function less than optimally in the regions around extended sources. We return to this issue in Section~\ref{sec:extended}.

\subsection{Flatfield correction}

The large SkyMapper field-of-view creates challenges for uniformly illuminating internal screens for producing dome flats, so the SMSS relies on twilight flats, which are observed whenever the weather conditions allow. While some surveys utilise long-running mean flatfields for a first-pass calibration, as nightly variations in flatfield illumination are often larger than seasonal or long-term variations \citep[e.g.][]{2018ApJS..235...33D},
SkyMapper suffers from temporally varying sensitivity changes that are more impactful than localised effects arising from changes in the pattern of dust motes. 

The SkyMapper camera features an evolving pattern of sensitivity changes most profoundly observed around the edges of the mosaic. The pattern varies in a systematic way over the period of time following a warm-up of the camera, with changes occurring most rapidly soon after returning to the operating temperature, and then asymptoting to a persistent pattern over long timescales. The effect is wavelength-dependent, with $u$-band showing the strongest decreases in sensitivity at the mosaic edges relative to the centre, $g$-band showing very minor evolution, and $z$-band showing the inverse behaviour of increasing edge sensitivity compared to the mosaic centre. In $u$ band, the four corner CCDs in the mosaic may change their average sensitivity relative to central CCDs as rapidly as 1~per~cent per day before they converge to an aggregate sensitivity loss of 10 to 20~per~cent. To mitigate this effect, we gather twilight flatfields from $\pm 10$~days around each observing night in order to approximate the behaviour in the middle of that span. 

Twilight flats are obtained in two opposing position angles (PAs) during each twilight period, so that the large scale gradient in the sky emission can be cancelled out. Within the $\pm10$-day span (with hard cutoffs imposed for detector warm-ups or other configuration changes), the potential input frames are grouped by PA, and a tolerance-testing procedure is applied to remove flatfield affected by patchy clouds. Within each PA, the valid inputs are median-combined after rescaling the counts by the mean of the central 8 CCDs. For the opposing PAs within each twilight, the PA-medians are then mean-combined with equal weighting. Finally, the twilight-means are combined in a weighted mean, where the weights are taken as the number of contributing input frames, resulting in the master flatfield.

After dividing the science frame by the master flat, a small additive shift is applied between the two halves of a given CCD to ensure a smooth background level across the image.

\subsection{Fringing correction}

Similar to the bias PCA method described in Section~\ref{sec:bias}, to correct for fringing in the $i$- and $z$-band images, we generate a set of fringing PCs for each filter and each CCD, this time treating the whole CCD at once. The inputs for the PC creation were $\sim5,000$ Main Survey images in each filter processed as part of SMSS DR2. We employed 3 PCs for $i$-band images and 10 PCs for $z$-band images, fitting to a background-subtracted and object-masked science image, and subtracting the resulting fringe pattern. The fringe PCs remain the same as for DR2 and DR3, and more details can be found in \citet{2019PASA...36...33O}.

\subsection{Additional pixel masking and image compression}
\label{sec:masking}

During a portion of the survey operations (MJD = 57290 - 58323), ground loops in the detector electronics gave rise to correlated fluctuations in the bias level across the entire mosaic, which took the form of a spike in counts with adjacent count-depressed pixels on either side along the same row. To mask the affected pixels, we median-stack each of the four 8-CCD groups that shared a particular readout timing (after masking any detected astronomical sources). Sources with $\geq 7$-sigma positive fluctuations were flagged in the pixel masks (see Tab.~\ref{tab:masks}), as was one pixel on either side.

Cosmic rays are then identified using the {\sc lacosmicx} software package\footnote{https://github.com/cmccully/lacosmicx}, a {\sc Python} implementation of the L.A.Cosmic routine \citep{2001PASP..113.1420V}. Each amplifier is treated separately to allow for the use of specific read noise settings. The affected pixels are replaced with values typical of the local background, but the modified pixels are also flagged in the pixel masks.

The calibration process transforms the original 16-bit integer image data into 32-bit floating point values. However, because of the significant readnoise in the SkyMapper images, we are able to round the data back to 16-bit integers without suffering from significant degradation in data fidelity. To avoid truncation of the noise around low sky count levels, the allowed integer range is $-100$ to 65435 (by setting the BZERO header keyword to 32668). This transformation then allows us to reduce the data storage footprint by losslessly compressing the images using CFITSIO's {\sc fpack} routine \citep{2009PASP..121..414P}. Compared to the 32-bit floating point images, the conversion and compression amounts to a factor of $\sim4$ reduction in disk space. The pixel masks, natively having 8-bit integer format, are also compressed, reducing the footprint by a factor of $\sim32$ because of the sparse nature of the flagged pixels.

\subsection{Photometric measurements}

We run {\sc Source Extractor} on each CCD, with a detection threshold of 1.5$\sigma$ and a Gaussian filtering function adapted to the median image FWHM. We measure photometry in a series of circular apertures (diameters of 2, 3, 4, 5, 6, 8, 10, 15, 20, and 30 arcsec). We provide {\sc Source Extractor} with the CCD-specific pixel mask as a Flag image and a version of the global bad pixel mask as a Weight image (to un-weight bad pixels).

When {\sc Source Extractor} measured the photometry, the input gain value was provided as 1, rather than the 0.75~ADU~e$^{-1}$ determined from the preflash images (see Sec.~\ref{sec:facility}). The consequence is a slight overestimate of the magnitude errors for each individual measurement, which in the limit of high source counts and low background, asymptotes to a value 15~per~cent too large. Because of the significant time that would be required to revise the gain value used by {\sc Source Extractor} in DR4, we leave the photometric errors unmodified. In practice, the impact for each {\it object} (those in the \texttt{master} table) is not nearly as significant, because the final photometric errors in each filter are derived from the outlier-clipped median absolute deviation (as described in Sec.~\ref{sec:mean_phot_per_obj}).

\subsubsection{Aperture corrections and PSF variation}

The sequence of aperture magnitudes provides a growth curve, which depends on object morphology. For point sources, the growth curve has a fixed shape, which can be used to infer total point source photometry from any aperture magnitude. However, the PSF shape drifts across the focal plane and affects the required aperture correction. Thus, we determine the PSF correction for each aperture in each image as a function of position using unsaturated bright stars. We also use this information also to estimate a PSF magnitude for each source from the 1D sequence of aperture magnitudes (see next section). 

In other surveys, PSF magnitudes are commonly obtained by PSF fitting to sources in 2D image data. Our process works on table data and is equally robust for isolated sources. However, blended sources are not correctly deblended by our PSF magnitude calculation; instead, we provide warning flags based on the brightness difference and distance to any neighbouring objects, which specify (per filter) whether the PSF magnitude is likely compromised (see the description of the \texttt{FLAGS\_PSF} column in Sec.~\ref{sec:mean_phot_per_obj}). 

For each CCD, we derive aperture corrections for each aperture smaller than 15\arcsec\ by fitting the flux ratio between the 15\arcsec-aperture and the aperture in question as a function ($x$, $y$) pixel coordinates. The 2D linear gradient is intended to mitigate the large-scale variations in the PSF shape across the mosaic (see Sec.~\ref{sec:constraints}). The fit is performed iteratively with 4 cycles of 2.5-$\sigma$ clipping.
In this process, we preference sources that have counterparts in the photometric zeropoint catalogue (described further below), but relax that requirement when the number of such matches is less than 8 per CCD. If fewer than 5 stars are available to fit the ($x$, $y$) plane, we apply just the median aperture correction to all sources in the CCD.

\subsubsection{PSF magnitude calculation}

During the preparation of DR4, it was discovered that the aperture corrections were not being correctly propagated to the associated magnitude errors in previous DRs, leading to underestimates of the uncertainties for both the aperture-corrected magnitudes (\texttt{E\_MAG\_APCnn} for aperture $nn$ between 02 and 10~arcsec) and the 1D PSF magnitude derived therefrom
(\texttt{E\_MAG\_PSF}, along with the flux versions of the latter, 
\texttt{E\_FLUX\_PSF}). Consequently, the per-object mean magnitudes (\texttt{\{f\}\_PSF} for filter \texttt{\{f\}}) and uncertainties (\texttt{E\_\{f\}\_PSF}) were incorrectly weighted in the \texttt{master} table, and the estimates of photometric variability (the per-epoch \texttt{CHI2VAR} and the per-object \texttt{\{f\}\_RCHI2VAR}) were overestimated in a manner that worsened for brighter stars. These deficiencies have been rectified by suitably propagating the aperture corrections and their uncertainties.\footnote{We make note here of the convention adopted throughout the paper, in which the available tables of the data release are indicated in lower-case script as, e.g., \texttt{master}, while columns within each table are indicated as upper-case script, e.g., \texttt{OBJECT\_ID}.}

The DR4 approach to constructing the one-dimensional point-spread function (PSF) magnitudes is also different from previous DRs. In DR4, we consider each {\it annulus} of aperture-corrected flux for the 7 smallest apertures (diameters of $2-10$\arcsec), rather than a curve-of-growth approach that uses the entire flux within each aperture. We calculate the PSF magnitude and its uncertainty from the weighted mean annulus-corrected magnitude and the error in the weighted mean, as well as the $\chi^2_{\rm red}$ value relative to the expectations for that CCD's aperture corrections. Fitting annuli to the PSF model --- other surveys often do this per-pixel from the image data, whereas we do it in the tabulated fluxes per-annulus by differencing the nested aperture fluxes --- allows us to more properly propagate the photometric errors than had been done in previous SMSS DRs.

\subsection{Photometric Zeropoint}
\label{sec:zp}

For DR4, we adopt an entirely new photometric zeropoint (ZP) catalogue, based on synthetic photometry derived from the low-resolving-power ($R\sim50$) {\it Gaia} DR3 spectroscopic data \citep{2023A&A...674A...1G, 2023A&A...674A...2D} and the SkyMapper photometric bandpasses \cite[][and also available through the Spanish Virtual Observatory's Filter Profile Service\footnote{\url{http://svo2.cab.inta-csic.es/theory/fps/index.php?mode=browse&gname=SkyMapper}}]{2011PASP..123..789B}. The {\it Gaia} team used the {\sc GaiaXPy} {\sc Python} package on the set of over 200 million BP/RP spectra to produce synthetic photometry in the SkyMapper filters \citep{2023A&A...674A..33G}. By default, the {\it Gaia} fluxes are convolved with the filter throughput and mean CCD response, as well as the typical atmospheric transmission for an airmass of 1 \cite[as given in the unprimed columns of Table~2 of][]{2011PASP..123..789B}. 

For DR2 and DR3, the SMSS utilised the ATLAS All-Sky Stellar Reference Catalog \cite[known as Refcat2;][]{2018ApJ...867..105T}, which brought together a variety of survey data to provide a well calibrated catalogue of $griz$ photometry between 6 and 19~mag. However, this left the calibration for the SMSS $u$- and $v$-bands relying on extrapolations to shorter wavelengths. Recalibrations of the $uv$ photometry have been derived based on stellar-colour regression models \citep{2021ApJ...907...68H}, but the available spectroscopic data did not sample the full footprint of the SMSS, and still extrapolated the short-wavelength properties based on their stellar classifications.

In contrast, the {\it Gaia} spectroscopic sensitivity to wavelengths as short as 330~nm anchors the $u$- and $v$-band data in a way that was not possible for the calibration of previous DRs. However, initial testing with the synthetic photometry for $u$-band revealed that the uncertainties in the short-wavelength {\it Gaia} sensitivity led to undesirably large errors in the predicted $u$-band data for stars with high-quality CALSPEC data. For example, the predicted photometry for CALSPEC stars with ($B_P-R_P$)<0~mag was too faint by $\approx 0.25$~mag.

As a result, the {\it Gaia} team kindly reprocessed their spectra with a modified $u$-band throughput model having a cutoff at 340~nm instead of 330~nm. While this left an even larger fraction of the $u$-band throughput shortward of the {\it Gaia} cutoff\footnote{The $u$-band's throughput, relative to its peak at 350~nm, is 80~per~cent at 340~nm, 50~per~cent at 330~nm, and 20~per~cent at 320~nm.}, it substantially reduced the resulting photometric scatter of the comparison stars. We determined a residual colour-term as a difference between the synthetic photometry of CALSPEC stars using our full $u$ bandpass and the Gaia prediction that included the cutoff (both included the red leak). The $u$ band magnitudes in the zeropoint catalogue were then corrected with the residual, represented by the following 4-part relation:
\begin{equation}
u_{\rm pred} = u_{\rm 340} + \left \{
\begin{array}{ll}
-0.045 + 0.125 \times \alpha & \mbox{if } \alpha \leq -0.2 \\
-0.015 + 0.250 \times \alpha & \mbox{if } -0.2 \leq \alpha \leq 0.2 \\
0.000 + 0.175 \times \alpha & \mbox{if } \phantom{-}0.2 \leq \alpha \leq 0.6 \\
0.097 + 0.8 \times (\alpha - 0.5)^{2} & \mbox{if } \phantom{-}0.6 \leq \alpha
\end{array}
\right.
\end{equation}
where $\alpha =$ \mbox{$(B_{P}-R_{P})$} $-$ \mbox{$E(B_{P}-R_{P})_{\rm gspphot}$}, the extinction-\linebreak corrected colour, which was only permitted over the interval $-0.6\leq \alpha \leq 1.0$, and the latter term is the colour excess inferred from {\it Gaia}'s best-fitting GSP-Phot Aeneas library\footnote{See the {\it Gaia} DR3 Documentation, section 11.3.3, "General Stellar Parametrizer from Photometry (GSP-Phot)".}.

The ZP catalogue was further restricted to {\it Gaia} sources having the following characteristics:
\begin{itemize}
    \item synthetic magnitude < 16~mag
    \item synthetic signal-to-noise > 20
    \item \texttt{RUWE} < 1.4
    \item \texttt{PHOT\_VARIABLE\_FLAG} $\neq$ "VARIABLE"
    \item \texttt{IPD\_FRAC\_MULTI\_PEAK} < 7
    \item \texttt{IPD\_FRAC\_ODD\_WIN} < 7
    \item $|C^{*} / \sigma_{C^{*}}|$ < 2
\end{itemize}
where the central 4 constraints refer to the standard columns from the \texttt{gaiadr3.gaia\_source} table, and $C^{*}$ is the colour-corrected indicator of extended source flux from \citet{2021A&A...649A...3R}, which we require to be less than 2-$\sigma$ from the 0-value of point sources. In addition, for $u$-band only, we apply the following condition: 
\begin{itemize}
    \item \phantom{~~~~~}$E(B_{P}-R_{P})_{\rm gspphot} \leq E(B-V)_{\rm SFD} + 0.1$
    \\ OR
    \item \phantom{~~~~~}($E(B_{P}-R_{P})_{\rm gspphot} \leq 1.2$)\\\phantom{~~~~~}AND\\\phantom{~~~~~}($E(B_{P}-R_{P})_{\rm gspphot} \leq 1.2\times E(B-V)_{\rm SFD}$),
\end{itemize}
where the SFD colour excess is drawn from the maps of \citet{1998ApJ...500..525S}. Based on the reddening coefficients of \citet{2019MNRAS.482.2770C}, we expect a mean relation of $E(Bp-Rp)= (2.905-1.75)\times 0.86 \times E(B-V)_\mathrm{SFD} \approx E(B-V)_\mathrm{SFD}$. After applying these conditions, we have an all-sky ZP catalogue with between 6.3 million and 95.9 million sources per filter (from $u$- to $z$-band, growing in number by a factor of $\sim1.8$ with each progressively redder filter).

Previous work by \citet{2021ApJ...907...68H, 2022ApJ...924..141H} used stellar-colour regression (SCR) techniques to predict SkyMapper photometry based on precisely measured stellar parameters from the third data release of the Galactic Archaeology with HERMES survey \cite[GALAH+;][]{2021MNRAS.506..150B}. Based on SMSS DR2 and sparsely sampled spectroscopic data from GALAH+, Huang et al.\ fit high-order polynomials to describe the inferred $u$- and $v$-band magnitude biases as functions of Galactic reddening \cite[using the $E(B-V)$ maps of][]{1998ApJ...500..525S} as well as spatial position. 

\begin{figure}
\begin{center}
\includegraphics[width=\columnwidth]{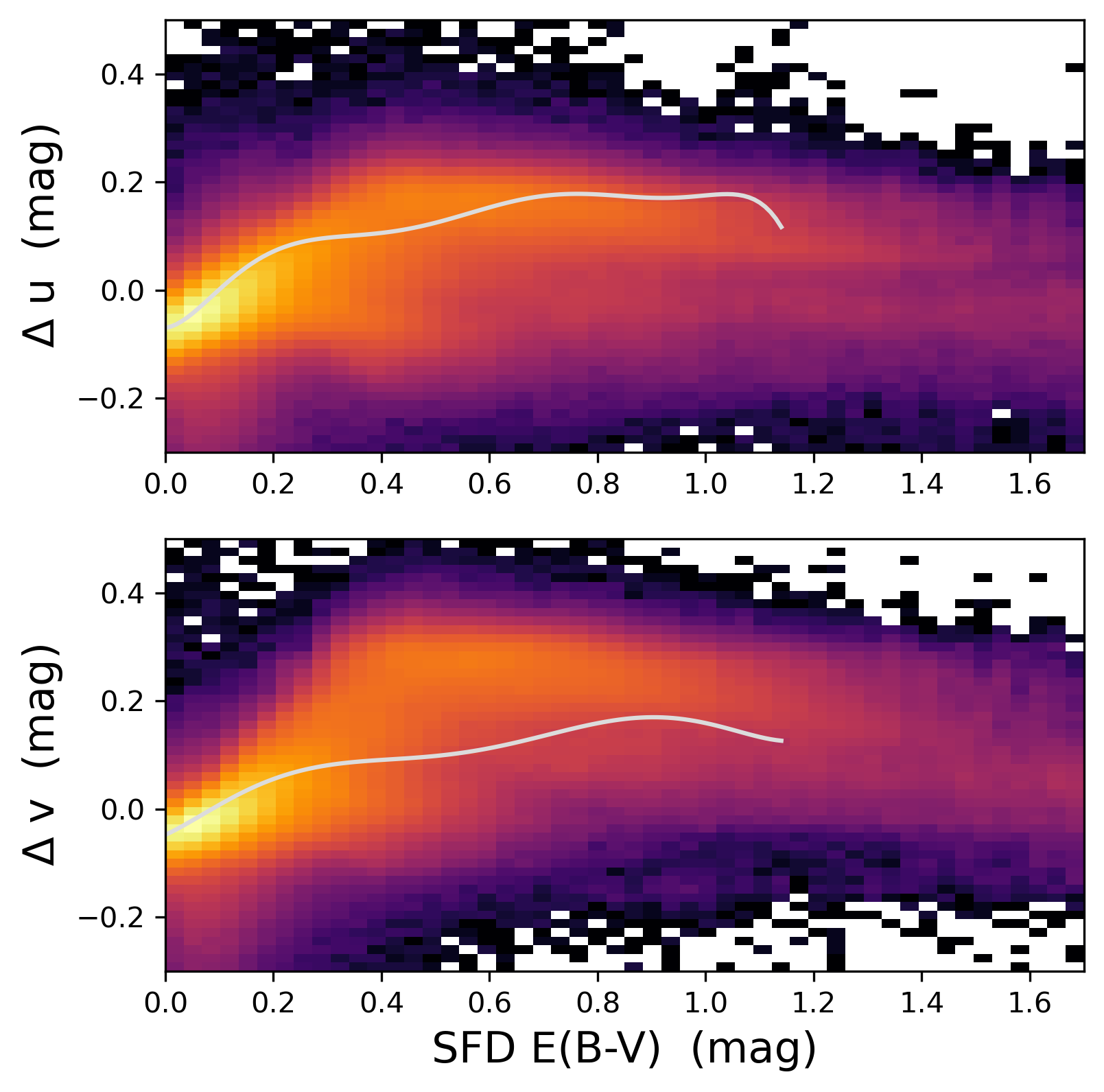}
\caption{Photometric differences (DR4 $-$ DR2) in $u$-band (top) and $v$-band (bottom) as functions of $E(B-V)$ drawn from the \citet{1998ApJ...500..525S} reddening maps. Lighter colours indicate logarithmically higher density of points. The polynomial corrections derived by \citet{2021ApJ...907...68H,2022ApJ...924..141H} are overplotted with white lines. The new ZP catalogue naturally corrects for the biases present in the SMSS DR2/DR3 photometric calibration and reproduces the trends found by \citet{2021ApJ...907...68H}. 
}\label{fig:zpcat_Huang_ebmv}
\end{center}
\end{figure}

By comparing the SMSS DR4 magnitudes to those of DR2\footnote{The SMSS DR3 photometry was calibrated in the same way as DR2.}, we can investigate whether the new ZP catalogue has corrected the features identified by \citet{2021ApJ...907...68H}. Figure~\ref{fig:zpcat_Huang_ebmv} shows a logarithmic density plot of SMSS DR4$-$DR2 photometry as a function of $E(B-V)$ for 7 million well measured stellar sources in $u$-band and $v$-band (restricted to magnitudes $12-16$, with photometric errors less than 0.05~mag, and \texttt{CLASS\_STAR} values above 0.9). We also plot the 7th- and 6th-order polynomials with the updated parameters of \citet{2022ApJ...924..141H} for the two filters, where we apply zeropoint shifts of ($u, v$) = ($-0.064, -0.042$)~mag to the polynomials to match them to the bulk of our low-reddening data; such a shift was previously unconstrained as the polynomial solution only represented trends between different reddening levels. The new ZP catalogue recalibrates the SMSS photometry in a manner that naturally reproduces the SCR trends with reddening, but which extends to higher reddening areas and accounts for localised variations.

\begin{figure*}
\begin{center}
\includegraphics[width=0.495\columnwidth]{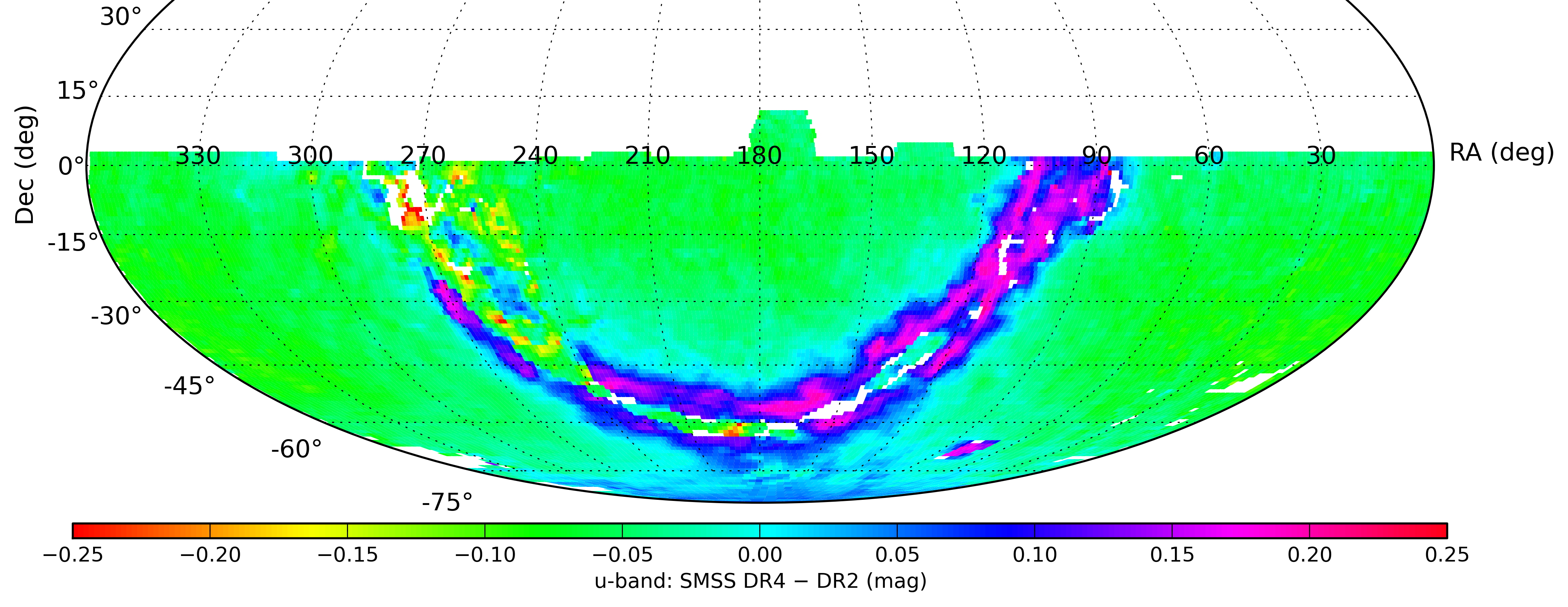}
\includegraphics[width=0.495\columnwidth]{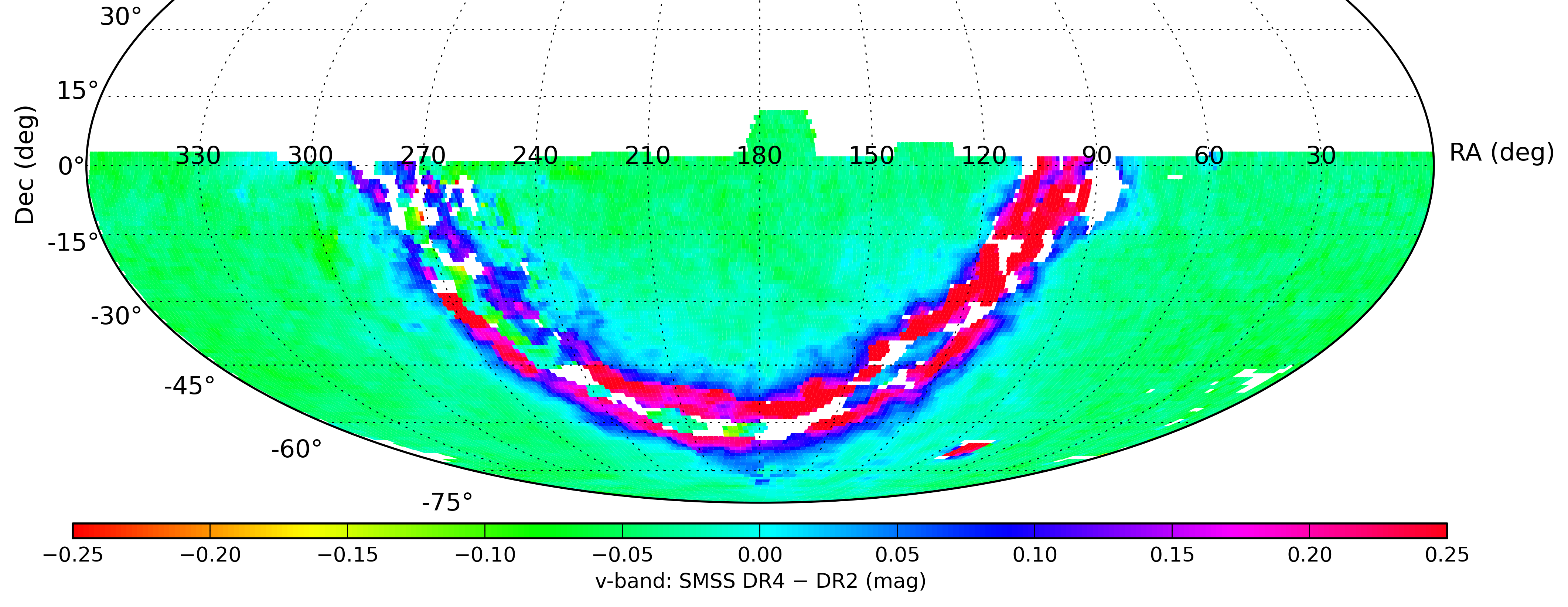}\\
\includegraphics[width=0.495\columnwidth]{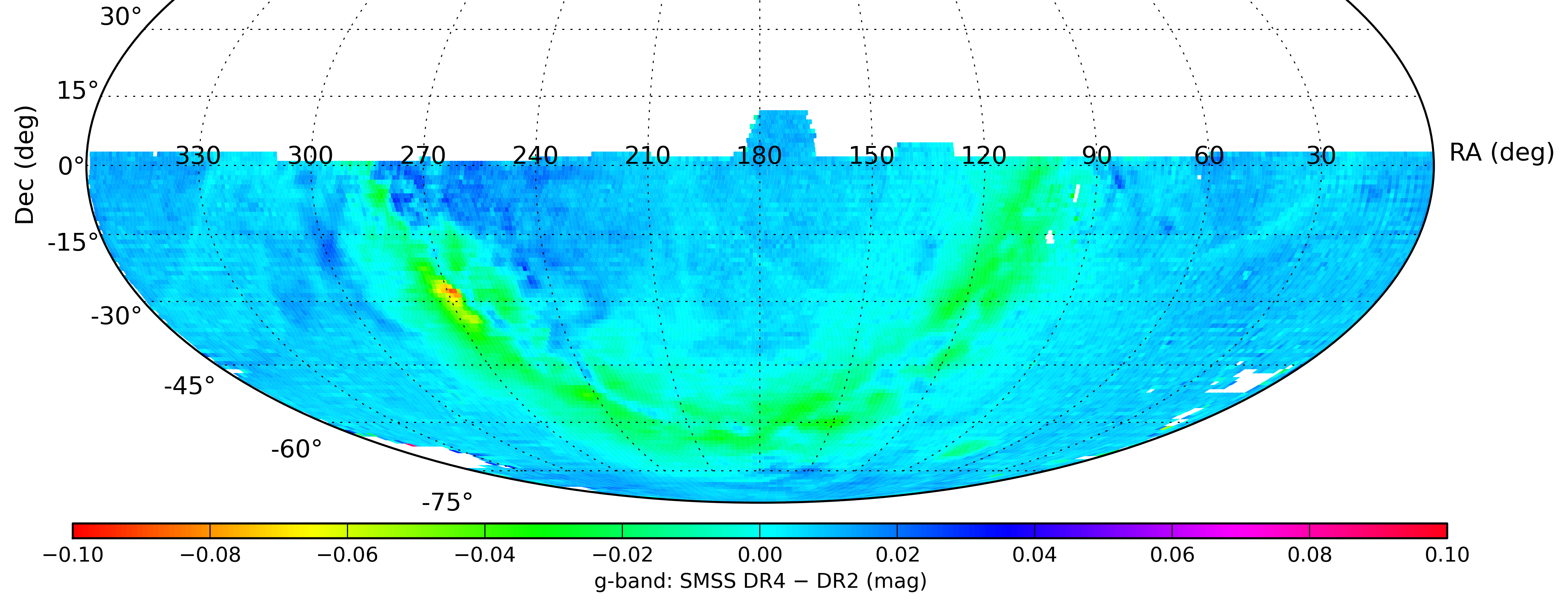}
\includegraphics[width=0.495\columnwidth]{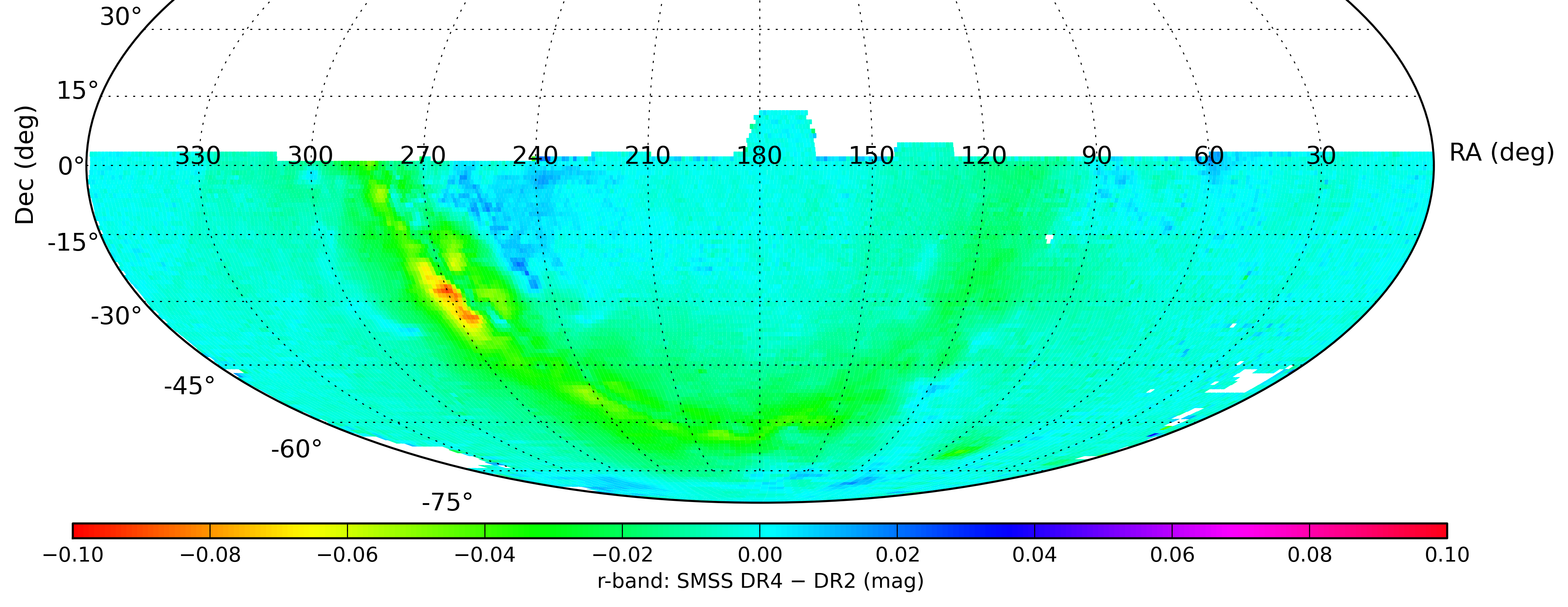}\\
\includegraphics[width=0.495\columnwidth]{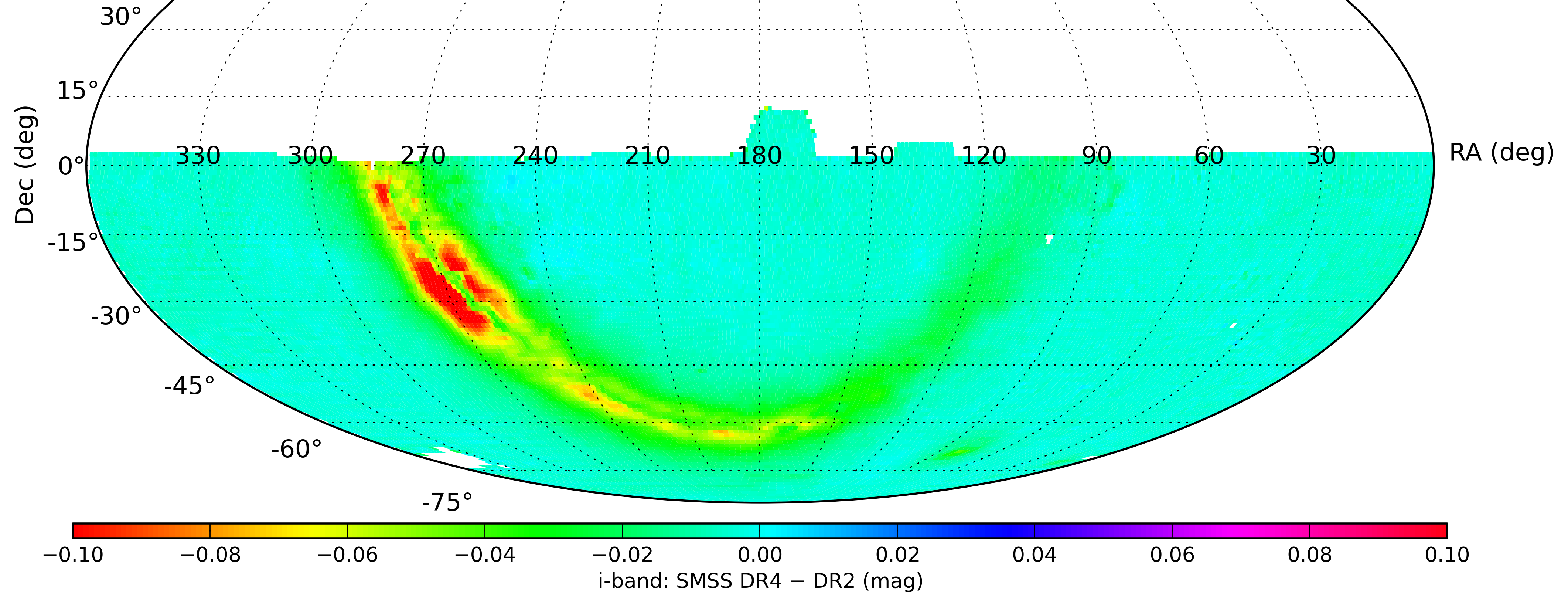}
\includegraphics[width=0.495\columnwidth]{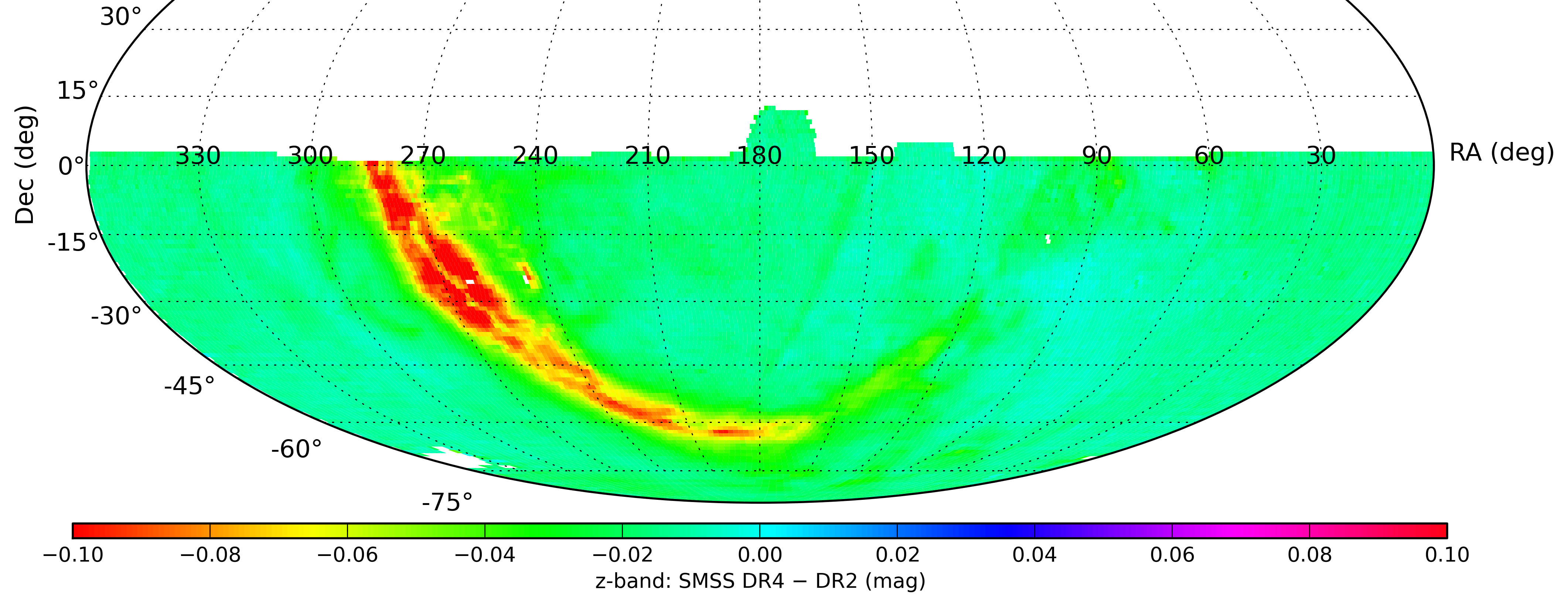}
\caption{Median photometric differences (DR4$-$DR2) per square degree for each of the 6 SMSS filters ($uvgriz$). Note that the colourbars cover the largest range for $u$ and $v$, and smaller ranges for the other four filters. The new ZP catalogue has resolved the known systematic biases present in the DR2/DR3 calibration.
}\label{fig:dr4_dr2}
\end{center}
\end{figure*}

Similarly, the photometric bias trends identified by \citep{2021ApJ...907...68H} as a function of sky position (($\alpha$, $\delta$) for $uv$, ($l$, $b$) for $gr$) are well matched by the updated ZP catalogue. The six panels of Figure~\ref{fig:dr4_dr2} present the sky maps of the photometry differences in each filter between SMSS DR4 and DR2. They show spatial patterns that trace Galactic structure as well as other regions of high source density --- precisely the sky areas in which the previous calibrations were known to be least reliable. 

The improvements in photometric calibration will greatly enhance the utility of the $u$- and $v$-band photometry from SkyMapper for a variety of scientific purposes. In Section~\ref{sec:prop_phot}, we compare the DR4 photometric results to various external datasets.

\subsection{Photometric calibration of images}

Using the instrumental PSF magnitudes derived above, we then derive the photometric calibration for each mosaic image by fitting a 2D plane of ZPs vs. ($x_{\rm mosaic}$, $y_{\rm mosaic}$) to account for atmospheric transmission gradients across the large FoV. In a photometric night, the most extreme transmission gradients are expected in $u$ band images taken near the South Celestial Pole at airmass $\sim 2$: across the image diagonal, the airmass may change by up to $\sim 0.3$, which will cause a throughput change of 0.2~mag (see also Table~\ref{tab:filters}). 

As with the aperture corrections, we perform 4 iterations of 2.5$\sigma$-clipping. If the number of remaining stars falls below 6, we discard the image from consideration for the DR. Mostly, these ZPs produce better fits with lower root-mean-square (RMS) scatter than ZPs that are flat across the image, but in rare cases the gradient procedure converges to a bad fit. Hence, we generally prefer the gradient solution but we choose the flat ZP instead whenever it produces a significantly better RMS ($\sigma$), i.e., when $\sigma_{\rm flat} < \sigma_{\rm gradient}-0.01$. The final ZP fit is applied to all source magnitudes in the image.

\section{Distilling the \texttt{master} catalogue}
\label{sec:distill}

Some applications of DR4 will involve a time-domain analysis and consider individual per-image detections of objects as described above and recorded in the \texttt{photometry} table. Many other applications would start better from a catalogue of unique astrophysical sources, here called the \texttt{master} table, and only use best estimates of per-object properties, or escalate into a time-domain analysis only after filtering the \texttt{master} table for relevant objects. In the \texttt{master} table, we record mean positions and mean per-filter magnitudes of objects by "distilling" them from the sample of individual detections, while assuming that the objects are not variable. In addition, we record quality flags, numbers of good-quality images in each filter, information on neighbouring objects, and cross-match IDs with external catalogues. Here, we describe our intentions for the properties of the \texttt{master} table and the distillation procedures.

\subsection{Distill platform}

In anticipation of this process, we transferred the $\sim13$~billion photometric measurements from the database server at NCI to a SkyMapper-dedicated server at ANU's Mount Stromlo Observatory\footnote{The images and pixel masks remained on disk at NCI, from where they will be made available through the data access tools described in Section~\ref{sec:access}.}. This 64-core machine, with over 500~GB of RAM, runs a similar PostgreSQL database (version 9.6.24) on which we have performed the final distillation of mean object properties for each SMSS DR to date. A master shell script launches a sequence of PostgreSQL scripts (parallelised where appropriate) that builds the \texttt{master} table via the steps below.

From the point-of-view of operational robustness, it is advantageous to move data into databases as soon as feasible and execute any further processing within the database \citep{4653207}. This allows one to take advantage of the fact that database engines are well-tested and robust enterprise-grade systems with built-in mechanisms to ensure integrity even during drastic failures such as power interruptions. However, our database instance is limited in its multiplex factor for parallel operations, and in a couple of instances, bottlenecks meant that some processes were forbiddingly slow compared to running them through standard Linux shell commands in the multi-processor environment at NCI. For this reason, a few steps in the DR4 process were shifted to the latter. End-to-end the distill process for DR4 has still taken six months.

\subsection{Selection of images and detections, flags}\label{sec:flags}

We first discard images that fail to meet any of the three quality levels in Section~\ref{sec:selection}, and further remove any individual CCDs for which the WCS solution yields a distance between opposite corners that differs from the expected value by more than 1~arcsec in either direction\footnote{Because the parameters of the WCS solutions are not numerically bounded, the best high-order fit to the brightest $\sim$10,000 stars in a mosaic image sometimes produces badly extrapolated coordinates, especially in the mosaic corners.}.

For each of the 6 filters, we restrict the photometry tables to those images and CCDs which remain, while simultaneously applying a number of bitmasks to the \texttt{FLAGS} column (where bits up to values of 128 are retained from {\sc Source Extractor}). In the following, we describe the meaning of the full list of extra flags, irrespective of whether the flags are determined per detection or, later in the distill process, per object:
\begin{itemize}
    \item 256 -- indicating the multiple detections in a single image that were spatially linked to a single \texttt{master} table object (applied in a later step of the distill process). This occurs as variations in image quality may sometimes render double sources separate and sometimes merged. The rate varies from 1 in 45\,000 detections in the larger-FWHM $u$-band images to 1 in 800 detections ($\sim$5 million out of $\sim4$ billion) in the smaller-FWHM $z$-band imagery.
    \item 512 -- indicating very faint detections that are considered dubious and a potential source of error. This flagging ensures that very faint detections in short exposures are ignored when the Main Survey or other deep observations can better define their distilled properties, while faint detections in deep images are still included into the master table (applied before distilling detections).
    \item 1024 -- indicating detections that appear much more concentrated than point sources, where the smallest aperture magnitude (2~arcsec) is significantly brighter than the largest aperture magnitude (15~arcsec), suggesting they may be affected by uncorrected cosmic rays or transient hot pixels (applied before distilling detections).
    \item 2048 -- indicating detections close to bright stars, which may suffer from scattered light or represent unreal sources. Using the stars from the ATLAS Refcat2 catalogue \citep{2018ApJ...867..105T}, we apply a flag to sources with $\log_{10} \left({\rm radius}\right) < -0.2 \times m$, where $m$ is the PS1-to-SkyMapper transformed magnitude in the filter in question, down to limits of (4, 5, 8.5, 8.5, 6, 5)~mag for ($u, v, g, r, i, z$). In contrast to previous DRs, such objects are now allowed to enter the \texttt{master} table, as many of them will have useful measurements. 
    \item 4096 -- indicating objects where all existing detections were affected by a bug in the distill code: the true flag values for individual images were unintentionally overwritten by NULL values, when their aperture correction failed, most often because the smallest (2\arcsec) aperture would have negative fluxes; these objects have no single detection without aperture-correction issues (3263075 objects, applied to the global \texttt{FLAGS} column in the \texttt{master} table after the distill).
    \item 8192 -- indicating detections from the very corners of the mosaic (beyond a radius of $\approx1.6$~deg), where the WCS solution was found to be systematically biased. Cross-matches to {\it Gaia} DR3 positions shows mean offsets larger than 1~pixel, with opposing corners showing consistent behaviour, and the other pair of corners being offset in the opposite sense. The trends appear to primarily be associated with the camera position angle (PA), where images acquired with PA=180$\deg$ reverse the behaviour in each corner. The flag setting suppresses the inclusion of such detections in distilled mean properties unless the measurements are all bad (applied before distilling detections).
    \item 16384 -- indicating objects that were never detected on any image other than Standard field images; these objects are typically extended and of low surface brightness so that their centroid positions are not well constrained in some of the shallowest Standard field images and thus show up as separate objects not matched properly to the correct astrophysical object seen more clearly in deeper images (applied to the global \texttt{FLAGS} column as well as the per-filter \texttt{\{f\}\_FLAGS} columns in the \texttt{master} table after the distill; note that the \texttt{FLAGS} column for such objects is {\it set} to 16384, rather than being OR-combined, and thus no longer reflects the OR-combined bits from all filters).
\end{itemize}

We also apply the final version of the photometric zeropoint -- either with the spatial gradient or flat across the image -- to all of the magnitude columns.

\subsection{Merging detections into unique astrophysical objects}

Within each filter, we then identify a set of "primary" sources by spatially matching the full list of detected sources against itself; in order to preference robust detections in good seeing, we use a maximum linking distance of 1.5~arcsec or $1.5\times$ the \texttt{FWHM} of the object, whichever is greater. From the detections in this neighbourhood we select as a {\it primary} counterpart the highest-ranked detection by first sorting detections by the binary choice of \texttt{FLAGS}$<8$ (i.e., no corruption of the {\sc Source Extractor} photometry), then ranking images with \texttt{IMG\_QUAL} of 1 or 2 higher than images with \texttt{IMG\_QUAL} of 3, and then sorting for the highest peak counts divided by the larger of 4~pixels or the measured object \texttt{FWHM}. 

The per-filter primary detections are collated into a single all-filter table which is then spatially cross-matched against itself. In this step, a linking distance of 1.5~arcsec is adopted and the same high-count-good-seeing metric is applied to identify the master source which seeds the eventual \texttt{master} table. Based on those master source positions, any object identifiers that can be retained from SMSS~DR3 are adopted (using a spatial cross-match distance of 2~arcsec, and keeping only the closest match if multiple new master sources were matched to a DR3 object). Objects newly identified in DR4 (nearly 122 million sources) are assigned \texttt{OBJECT\_ID} values beginning from 2E9 (while sources new to DR3 were assigned \texttt{OBJECT\_ID} values beginning from 1E9).

The list of individual detections are then associated with those master \texttt{OBJECT\_ID} identifiers by adopting the closest master source within 10~arcsec (allowing a larger distance to account for any cumulative offsets between links spanning the six filters). This procedure leaves only a very small number of detections unassociated with a \texttt{master} table entry: a total of 1376 detections, for a rate of 1 in 10 million. If, conversely, a single master object is associated with multiple detections in a given image, all of those detections have a bit value of 256 added to their \texttt{FLAGS} column in the \texttt{photometry} table.

\subsection{Mean properties per object}

To calculate the mean object properties in each filter, we then select those \texttt{OBJECT\_ID}-associated detections that appear to be of good quality. We first consider all detections for which \texttt{FLAGS} $< 4$ and \texttt{NIMAFLAGS} recorded $< 5$ masked pixels within the source's isophotal area. The number of good detections thus defined in the filter \texttt{\{f\}} is recorded in the column \texttt{\{f\}\_NGOOD}. We then omit all \texttt{IMG\_QUAL} $= 3$ detections from the mean properties calculation when quality level 1 or 2 data is available for a source and set the \texttt{USE\_IN\_CLIPPED} column of the low-quality entries in the \texttt{photometry} table to $-1$. We also add up the number of good detections over all six filters for the total number of good detections recorded in column \texttt{NGOOD}. 

\begin{figure*}[ht]
\begin{center}
\includegraphics[width=\columnwidth]{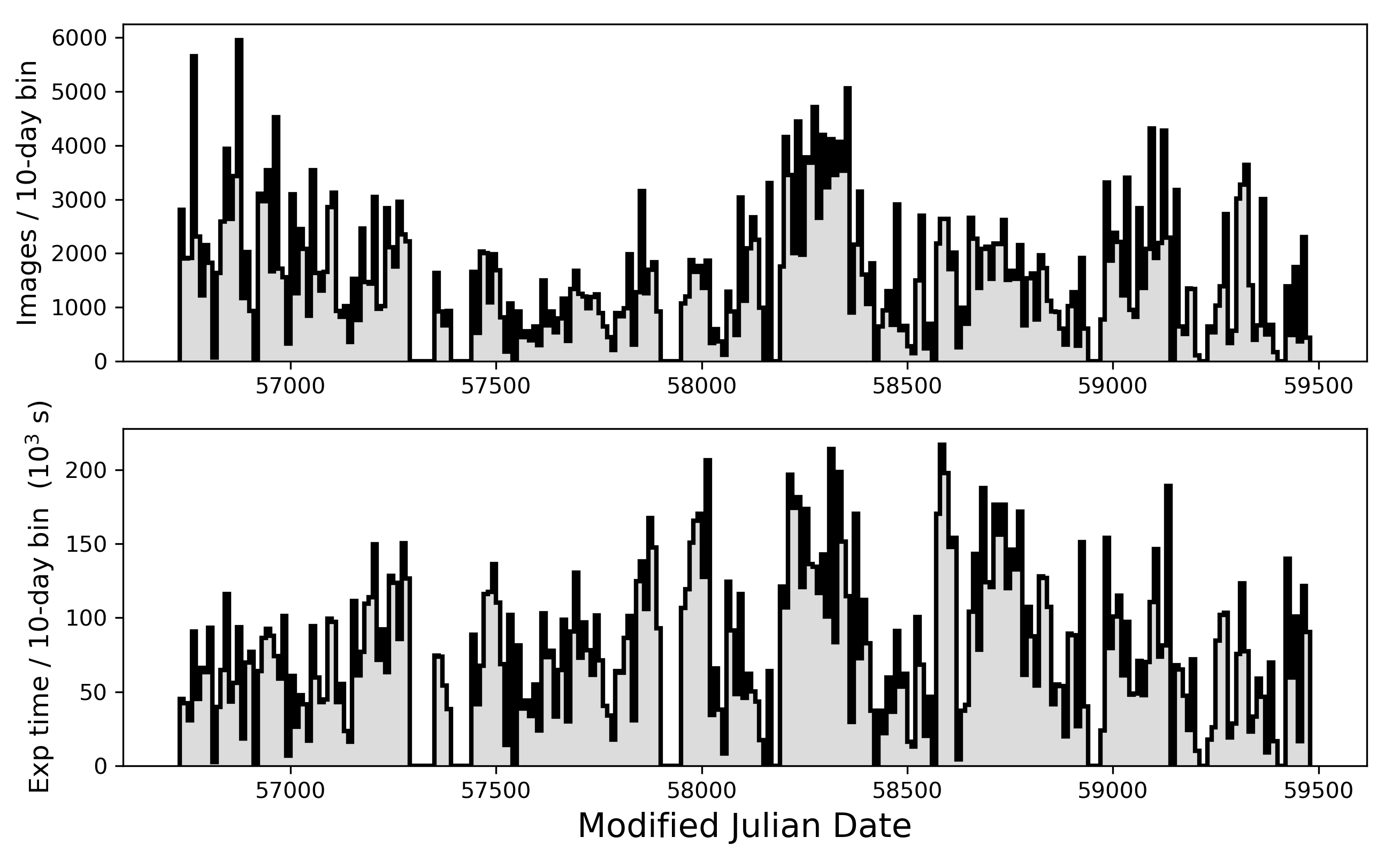}
\caption{For the images included in DR4, we show the number of images ({\it top}) and their total exposure time ({\it bottom}), in bins of 10 nights. Longer periods of downtime include three failures of the cooling system (around MJD 57300, 57400 and 57900).
}\label{fig:bydate}
\end{center}
\end{figure*}

\begin{figure*}
    \begin{center}
    \includegraphics[width=\columnwidth]{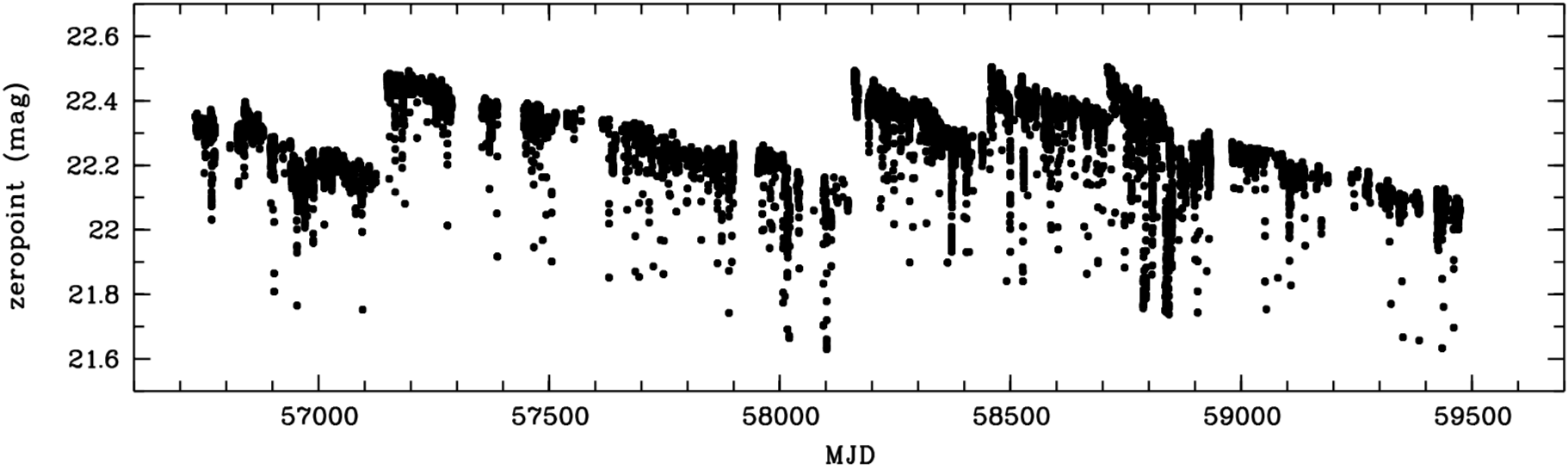}
    \caption{Time series of $g$-band image zeropoints in DR4, normalised by airmass and exposure time, from 15 March 2014 to 16 Sep 2021 (MJD 56733 to 59473). The gradual degradation of the ZP stems from loss of mirror reflectivity of $\sim 1$~per~cent per month. Sudden improvements appear after mirror cleaning (on MJD 57147, 58150, 58448 and 58710).}
    \label{fig:ZPtrends}
    \end{center}
\end{figure*}

Then we make a further distinction: for the determination of, e.g., mean magnitudes of an object, we assume {\it a priori} that the object is not variable. We enhance signal-to-noise by clipping potential outlier measurements, even if they have good-quality flags. Such outliers may be faulty measurements that are not recognised as such by our data processing, or they may be genuine astrophysical signals such as a rare binary eclipse affecting only one in several available magnitude measurements. Thus, the \texttt{USE\_IN\_CLIPPED} column uses three values to indicate data levels for the calculation of mean properties: $-1$ denotes bad data excluded from a distilled mean, $0$ denotes good data clipped to reduce scatter, and $1$ means data retained after clipping and used for the calculation of the final mean and error.

We separately average properties of sources for which {\it only} saturated detections or detections with otherwise bad flags exist in a given filter, to ensure they are not omitted from the \texttt{master} table by the restrictive quality selections above. The photometry of these objects may be useful but unreliable, and the errors on their photometry will be unknown; we indicate this further by listing their distilled magnitude errors as \texttt{NULL}. Their coordinates are listed with useful errors derived from the scatter among the detections, but they may still be on average less accurate than those of well-measured objects.

\subsection{Mean photometry per object}
\label{sec:mean_phot_per_obj}

For each object in each filter, we then compute the weighted median PSF magnitudes and their median absolute deviations (MADs, measured with respect to the weighted median). The \texttt{USE\_IN\_CLIPPED} column of the \texttt{photometry} table is set to 1 for magnitude values which are less than 3 MADs (or 3 times the square-added errors of the PSF magnitude and ZP RMS, if that is larger) from the weighted median PSF magnitude, and 0 otherwise.

From the detections that survive the clipping process, we calculate for each filter a summary of characteristics for each object: the mean properties (with the magnitude estimates weighted by their individual errors, but other entries unweighted), as well as the maximum \texttt{CLASS\_STAR} and\linebreak
\texttt{FLUX\_MAX} values, the OR-combined \texttt{FLAGS} bits, and the sum of the \texttt{NIMAFLAGS} entries (capped at 32767 to limit the column to a 16-bit representation). The weighting for the magnitude columns includes (by square-adding) the RMS of the photometric ZP fit to the image.

In contrast to previous DRs, which tabulated a $\chi^2$-like estimator of whether a source was consistent with being non-variable (\texttt{\{f\}\_RCHI2VAR} for filter \texttt{\{f\}}), for DR4 we adopt a simpler metric of the minimum-to-maximum range of PSF magnitudes (with no clipping considered): \texttt{\{f\}\_MMVAR}. This excludes detections in \texttt{IMG\_QUAL}$=3$ images when better images exist, as well as any that were flagged as bad.

The \texttt{master} table has also been expanded from previous DRs to record the weighted mean 5\arcsec\ aperture magnitude (corrected for aperture losses as described above). This small-aperture photometry, \texttt{\{f\}\_APC05}, is useful for galaxy colours as it focuses on well measured inner parts, i.e., core- or bulge-dominated regions in nearby galaxies and near-total colours in distant galaxies. As in past DRs, Petrosian magnitudes are included (using standard {\sc Source Extractor} parameters) that integrate out to poorly measured peripheral regions of galaxies and thus produce noisier colours.

From the per-filter summarised characteristics, we generate the final source descriptors for the \texttt{master} table: weighted-mean positions and uncertainties; mean and RMS MJD of the observations; OR-combined \texttt{FLAGS}; summed \texttt{NIMAFLAGS} and number of good detections (i.e., those with \texttt{USE\_IN\_CLIPPED}$ > -1$); mean \texttt{FWHM} (recorded in \texttt{MEAN\_FWHM}) and
\texttt{RADIUS\_PETRO} values; and maximum
\texttt{FLUX\_MAX} and
\texttt{CLASS\_STAR} values. 

A special case is the \texttt{CHI2\_PSF} parameter (determined from the light profile in a 15 arcsec aperture): the values in the \texttt{photometry} table are first averaged per filter and then the largest value among the filters is used in the \texttt{master} table, because blended sources of different colours may only be recognised in some bands but not others. However, the largest-value selection implies that the \texttt{CHI2\_PSF} values in the \texttt{master} table are expected to be larger than 1 on average, as they are sampled from the larger-than-average tail of the distribution.

A cleaning is then applied to sources close to the RA $= 0/360$~deg divide to ensure appropriate calculation of the mean position and uncertainty.

We finally create a \texttt{FLAGS\_PSF} entry for each object to indicate whether their PSF magnitude may be biased by close neighbours adding flux to their apertures. We construct a bitmask by testing whether any of the neighbours are closer in arcsec than \mbox{$5 + 2\times \left( m_{\rm object} - m_{\rm neighbour} \right )$}. Neighbours closer than this limit are expected to brighten the PSF magnitude by over 0.01~mag, with no upper limit. The flagging is performed independently for each filter, with $u$-band to $z$-band encoded as descending bit values from 32 to 1. The procedure differs from previous DRs, where only the single closest source was considered, while other, potentially brighter, sources had been ignored.

\subsection{Cross-matches to external catalogues}

With the final coordinates for the \texttt{master} table established, we cross-match SMSS~DR4 against the variety of other catalogues described in Section~\ref{sec:data}, as well as the next three closest sources within the \texttt{master} table itself.

One of the new cross-matches is the {\it Planck} satellite map of interstellar reddening, based on the generalised needlet internal linear combination (GNILC) method\footnote{We note a typo in Section~6 of \citet{2016A&A...596A.109P}, which reports the optical depth scale factor as 1.49$\times10^{-4}$ rather than 1.49$\times10^{4}$ \cite[cf.][]{2014A&A...571A..11P}.} \citep{2016A&A...596A.109P}. Comparisons of the GNILC reddening with the \citet{1998ApJ...500..525S} map from an earlier generation of infrared/microwave satellites ({\it COBE} and {\it IRAS}) have found that near-IR-bright galaxies show a smaller colour scatter with the GNILC maps, especially at low Galactic latitudes \citep{2021MNRAS.503.5351S}. Thus, we provide the \citet{1998ApJ...500..525S} $E(B-V)$ estimate alongside the GNILC $E(B-V)$ estimate and its error in the \texttt{master} table.

Finally, to make the data available to the community, we transfer the \texttt{master} table, the modified \texttt{photometry} tables (now containing the \texttt{OBJECT\_ID} labels, additional \texttt{FLAGS} bits, and \texttt{USE\_IN\_CLIPPED} indicators), the \texttt{images} and \texttt{ccds} tables to the PostreSQL database (version 9.5.25) underlying the SkyMapper node of the All-Sky Virtual Observatory (ASVO).

\section{SMSS DR4 Properties}
\label{sec:properties}

In this section, we describe the image selection criteria for inclusion in SMSS DR4, and the resulting distribution of image properties and sky coverage. We then assess the astrometric and photometric performance of the data.

\subsection{DR4 Input Images}
\label{sec:inputs}

SMSS DR2 and DR3 incorporated images taken up to March 2018 and October 2019, respectively. With DR4, we bring the processing of images up to September 2021. (Faults in the detector controller electronics on UT~2021-09-16 resulted in periods of time with incomplete mosaic operation, which will require pipeline modifications to properly calibrate, and thus defined a natural cutoff for DR4.)

In addition to the Survey image types released previously (Shallow Survey and Main Survey), DR4 has been allowed to consider the SMT survey and other non-Survey images acquired since the start of the SMSS, provided they meet the various quality metrics indicated in Section~\ref{sec:selection}.
Finally, the Standard fields containing the CALSPEC standard stars (see Sec.~\ref{sec:design}) have also been included in DR4.
The incorporation of these additional images means that the observational cadence is enhanced, even within the range of dates covered by previous DRs.

\begin{table}
\begin{threeparttable}
    \caption{Requirements for image quality levels: Constraints for quality 3 also apply to quality 2, and those of quality 2 also apply to quality 1. All numbers are upper limits except for the number of ZP stars.}
    \label{tab:image_quality}
    \setlength{\tabcolsep}{3pt}
    \centering
    \begin{tabular}{l|c|c|c|c|c|c|c}
\toprule
\headrow {\bf Property} & {\bf Units} & \multicolumn{5}{c}{{\bf Passband}} \\
\midrule 
\headrow &  & u & v & g & r & i & z \\
         \midrule \midrule
         {\bf Image quality 1}  \\
         \midrule 
         Zeropoint RMS          & mag & 0.05  & 0.033 & 0.015 & 0.015 & 0.015 & 0.02 \\
         \midrule
         Gradient ZP offset     & mag & \multicolumn{6}{c}{0.01} \\
         \midrule 
         Gradient ZP slope      & mag~px$^{-1}$ & 6e$-$6 & \multicolumn{4}{c}{3e$-$6} \\
         \midrule \midrule
         {\bf Image quality 2}  \\
         \midrule 
         Zeropoint RMS          & mag    & 0.07  & 0.06  & 0.04  & 0.04  & 0.055 & 0.055 \\
         PSF FWHM               & arsec  & 4.00  & 4.00  & 4.00  & 3.90  & 3.75  & 3.60 \\
         Background             & counts &  300  &  300  & 1000  & 1000  & 3000  & 3000 \\
         \midrule 
         Gradient ZP offset     & mag & \multicolumn{6}{c}{0.025} \\
         Gradient ZP slope      & mag~px$^{-1}$ & \multicolumn{6}{c}{1.2e$-$5} \\
         Airmass                & & \multicolumn{6}{c}{2.02} \\
         \midrule \midrule
         {\bf Image quality 3} \\
         \midrule 
         Zeropoint RMS          & mag & 0.12  &  0.10  & 0.08  & 0.08  & 0.08  & 0.08 \\
         \midrule 
         Gradient ZP slope      & mag~px$^{-1}$ & \multicolumn{6}{c}{2.1e$-$5} \\
         PSF FWHM               & arcsec & \multicolumn{6}{c}{5.0} \\
         PSF elongation         & &  \multicolumn{6}{c}{1.4} \\ 
         Background             & counts & \multicolumn{6}{c}{5000} \\
         Number of ZP stars    &       & \multicolumn{6}{c}{>20} \\
         Throughput loss        & mag   & \multicolumn{6}{c}{1.0} \\
         WCS RMS                & arcsec & \multicolumn{6}{c}{0.5} \\
         Compressor noise       & fraction & \multicolumn{6}{c}{0.2} \\
         \bottomrule
         \multicolumn{8}{l}{}
    \end{tabular}
\end{threeparttable}
\end{table}

\begin{table*}[ht]
\begin{threeparttable}
\caption{DR4 includes all images taken with SkyMapper that pass the quality criteria, irrespective of original purpose. Image properties are: final survey depth as peak in histogram of sources with $\sim 0.1$~mag errors in the PSF magnitude (source detection was made on individual images); saturation magnitude in the Shallow Survey; distribution of exposure times ($t_{\rm exp}$) across all images and for the Shallow and Main Survey specifically; fraction ($f_\mathrm{image}$) of Shallow and Main Survey images among the DR4 set; and image quality distributions. A small fraction ($<0.5$~per~cent) of $u$ and $v$ images have 300- or 400-sec durations.}
\label{tab:imcount}
\centering
\begin{tabular}{lcccccccrrr}
\toprule
\headrow Filter & 10$\sigma$ depth & Shallow Saturation & Min -- Max $t_{\rm exp}$ & Mean $t_{\rm exp}$ & Shallow / Main $t_{\rm exp}$ & Shallow / Main $f_\mathrm{image}$ & \multicolumn{3}{c}{Number of Images}  \\
\headrow & ABmag & ABmag & (s) -- (s) & (s) & (s) / (s) & (\%) / (\%) & Qual 1  & Qual 2  & Qual 3 \\
\midrule
$u$ & 18.6 & 8.9 & 15 -- 600 & 61.6 & 40 / 100 & 51 / 31 & 35300 &  7740 & 16750 \\
$v$ & 18.9 & 8.2 & 15 -- 600  & 52.5 & 20 / 100 & 49 / 30 & 23077 & 26779 & 13535 \\
$g$ & 20.5 & 9.4 & 0.1 -- 600 & 43.1 & \phantom{0}5 / 100 & 46 / 22 & 25616 & 28874 & 15478 \\
$r$ & 20.2 & 9.4 & 1 -- 400   & 40.2 & \phantom{0}5 / 100 & 48 / 19 & 20827 & 34590 & 13875 \\
$i$ & 19.4 & 9.5 & 1 -- 600   & 48.0 & 10 / 100 & 41 / 35 & 31584 & 39635 & 11385 \\
$z$ & 18.6 & 9.6 & 1 -- 300   & 50.2 & 20 / 100 & 47 / 38 & 28615 & 32472 & 11091 \\
\bottomrule
\end{tabular}
\end{threeparttable}
\end{table*}

\begin{table}
\begin{threeparttable}
\caption{Galactic Plane Coverage: DR4 contains overall $\sim 3.5\times$ as many images as DR2, but at low Galactic latitudes DR4 has $\sim 7\times$ and $\sim 12\times$ as much data in the $u$ and $v$ bands, respectively. Astrometry methods used in earlier releases often failed at calibrating $uv$ data in crowded fields.}
\label{tab:galimcount}
\centering
\begin{tabular}{lcccc}
\toprule
\headrow Filter & \multicolumn{2}{c}{DR4 CCDs} & \multicolumn{2}{c}{DR2 CCDs} \\
\headrow        &  $|b|<5\deg$ & $|b|=5-10\deg$ & $|b|<5\deg$  & $|b|=5-10\deg$ \\
\midrule
$u$ & 137911 & 183870 & 20559 & 40895 \\
$v$ & 141604 & 189137 & 11647 & 26622 \\
$g$ & 137738 & 177893 & 42467 & 46338 \\
$r$ & 143541 & 181825 & 42796 & 46911 \\
$i$ & 171576 & 208026 & 54018 & 58697 \\
$z$ & 174562 & 215288 & 55254 & 61200 \\
\bottomrule
\end{tabular}
\end{threeparttable}
\end{table}

\subsection{Image Selection Criteria}
\label{sec:selection}

We use three image quality categories with requirements that depend on the filter, and reject all images that do not conform with minimum requirements. The requirements consider the RMS ($\sigma$) of the ZP determination that is used for the frame (either flat or gradient) and also the gradient fit parameters themselves when the ZP gradient solution is adopted for an image. Furthermore, high airmass and PSF FWHM beyond a threshold affect the image quality level irrespective of the ZP quality, and are also constrained (see Table~\ref{tab:image_quality}).

Images that do not meet the criteria for quality level 3 are not included in the data release. When a unique object in a unique filter has images from both quality level 3 and the better quality levels (1 and 2), then the images of quality level 3 are not included for distilling quantities in the \texttt{master} table of mean object properties (further details are given in Section~\ref{sec:distill}), although the data is still present in the \texttt{photometry} table and useful for probing behaviour of variable sources at additional epochs. Thus, in the interest of better characterisation of non-variable sources, data at quality level 3 only work their way into the \texttt{master} table if no better-quality images exist in that filter for a given object.

In Table~\ref{tab:imcount}, we present the distribution of DR4 images by filter and image quality. Overall, more than 80~per~cent of the DR4 images have image quality 1 or 2.

The distribution of images over time is shown in the two panels of Figure~\ref{fig:bydate}, which present the number of images as well as the total exposure time, in intervals of 10 nights. A number of periods of technical downtime are evident as gaps in the time coverage.

\subsection{Sky Coverage and Completeness}

The sky area covered in DR4 is essentially the complete sky South of declination $+16\deg$ plus a few special fields North of that line. The most intense coverage in terms of depth and repeat observations is South of the celestial equator. However, it is not guaranteed that every position in the southern sky has fallen onto good pixels in all filters, because the bad pixel patterns, the telescope dither pattern, the pointing accuracy of the telescope, and the de-selection of bad images for the release all affect the coverage of the data set.

Compared to earlier releases, DR4 has much improved coverage of  well calibrated data near the Galactic Plane. Such data was previously missing, especially in the $u$ and $v$ filters, where the number of images in DR4 has increased by a factor of $\sim 7$ and $\sim 12$, respectively  (counted as individual CCDs). This is partly due to improved astrometry methods in DR4, where previously several CCDs in an image were lost for lack of reliable per-CCD astrometry; the mosaic-global astrometry in DR4 recovers those CCDs as embedded in a whole-mosaic solution. Another important reason is the new zeropoint star catalogue, which provides a higher density of sources, especially in the $u$ and $v$ band in regions with higher foreground reddening or source density.

\begin{figure}[ht]
\begin{center}
\includegraphics[width=\columnwidth]{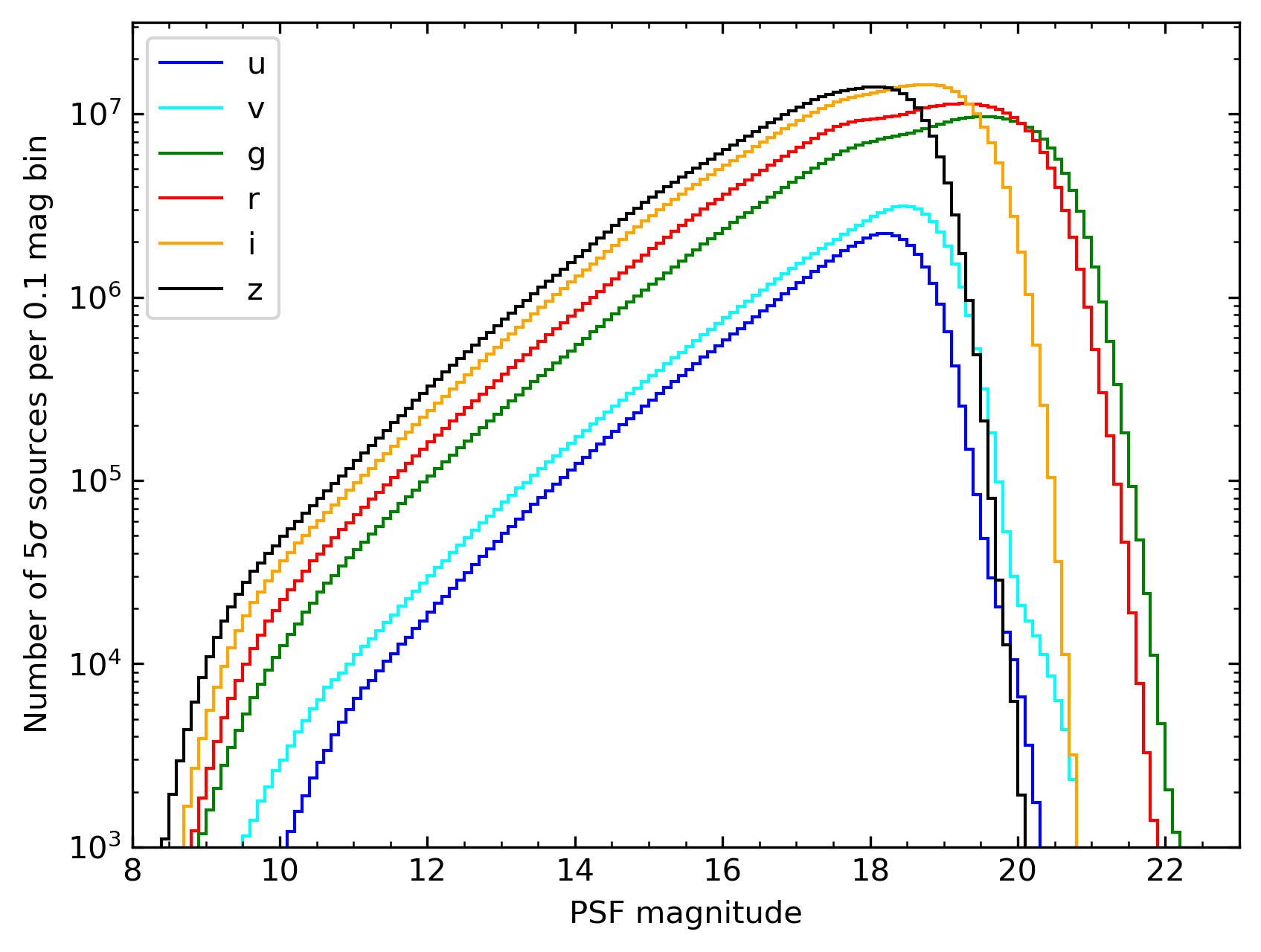}
\caption{Distribution of mean magnitudes in each SMSS filter for sources below $0^\circ$ Dec having 5+$\sigma$ photometry in that filter.
}\label{fig:phot_hist}
\end{center}
\end{figure}

Among the 700 million distinct objects contained in DR4's \texttt{master} table, we have magnitude estimates in the six SMSS filters for 9, 13, 59, 73, 85, and 72 percent of the objects in $u, v, g, r, i$, and $z$, respectively. The histogram of source magnitudes (5$\sigma$ or higher) is shown in Figure~\ref{fig:phot_hist}.

\subsubsection{Completeness versus DES}

We assess the completeness of the DR4 dataset by comparison to the deeper DES DR2 catalogue \citep{2021ApJS..255...20A}, focusing on the $r$-band. In practice, the fraction of faint sources detected by SMSS will depend upon the particular mix of image seeing and photometric zeropoint for any given sky position, but we illustrate the typical behaviour.

To avoid any biases arising from sky location, we sample the DES data in a series of 30-arcmin-radius circles around every other SMSS field centre (which form a regular grid on the sky) in alternating Declination stripes, thus giving a spacing of about 4~deg in each direction. The SMSS field centres considered were also limited to those with at least three 100~s Main Survey exposures in DR4, to avoid artificially underestimating the completeness from those rare sky areas only observed with short exposures. From the DES catalogue, we select those objects identified as unsaturated point sources \cite[\texttt{EXTENDED\_CLASS\_COADD}=0 and \texttt{FLAGS\_R}<4; see][]{2021ApJS..255...20A}. Corresponding SMSS sources were required to be less than 1~arcsec from the DES positions.

\begin{figure}[!ht]
\begin{center}
\includegraphics[width=\columnwidth]{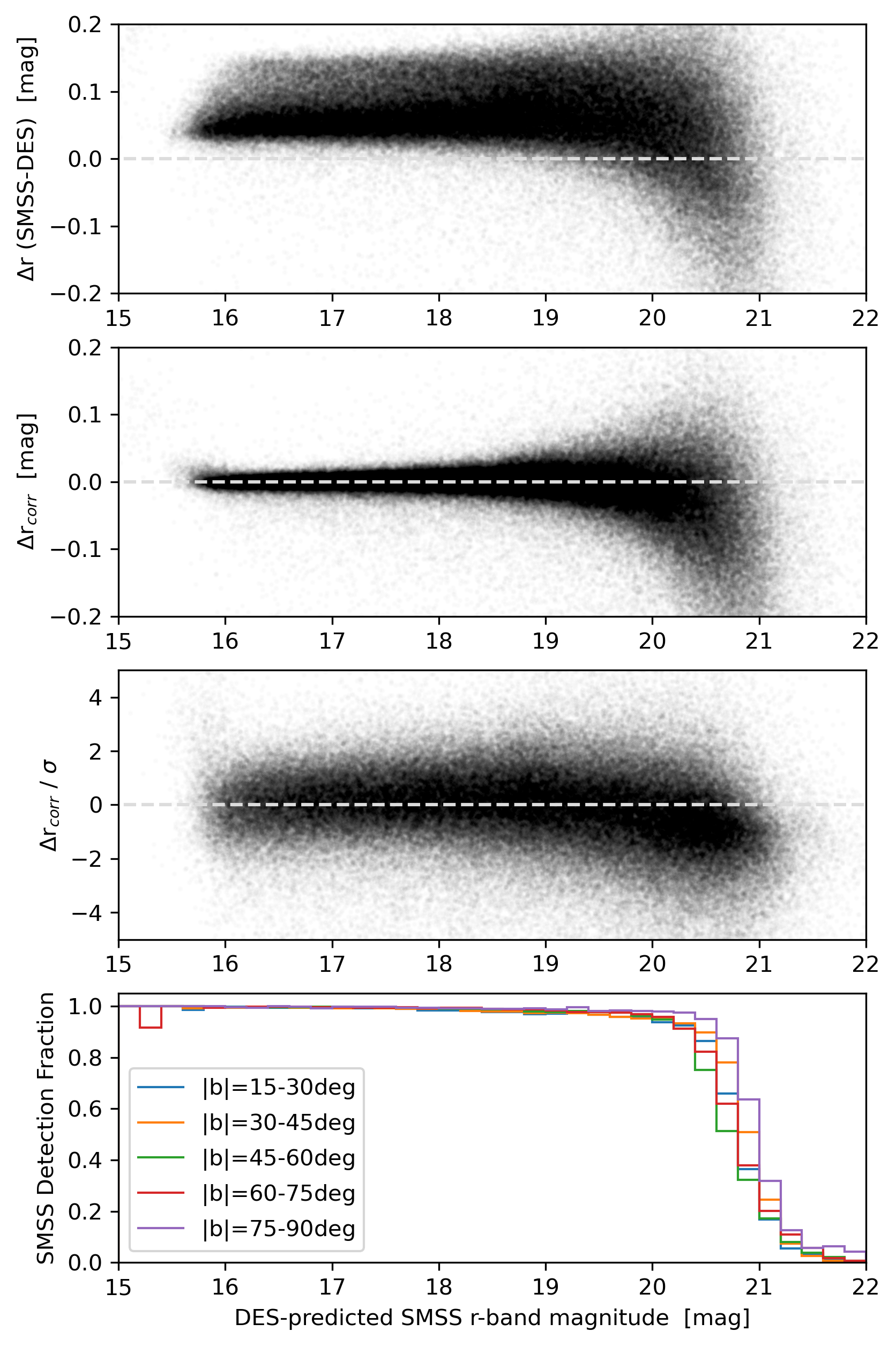}
\caption{Difference in $r$-band photometry between SMSS DR4 and DES DR2 as a function of DES-predicted magnitude in the SMSS system. Grey dashed lines in the first three panels are drawn at a magnitude difference of zero. ({\it Top panel}) Raw photometric difference prior to colour correction (in this panel only, the x-axis shows the original DES $r$-band magnitudes). ({\it Second panel}) Photometric difference after application of the colour correction, significantly tightening the relation. ({\it Third panel}) Colour-corrected photometric difference in units of photometric error (square-adding the DES and SMSS errors). The SMSS errors are consistently representing the photometric differences until magnitudes fainter than $r\sim20$~mag, beyond which the SMSS photometry exhibits a slight bias to brighter measurements. ({\it Bottom panel}) SMSS detection fraction as a function of magnitude, plotted in different ranges of Galactic latitude, showing high completeness to $r\sim20$~mag.
}\label{fig:des_delta}
\end{center}
\end{figure}

Because of differences in the DECam and SkyMapper $r$-band filter curves, there are colour terms when comparing the DES and SMSS photometry. We correct for these by restricting the DES-measured colour range to $0<(g-r)<1.3$~mag and fitting the SMSS-minus-DES magnitude difference, $\Delta r$. We adopt a colour correction relation of $(-0.007 + 0.118\times(g-r))$ that is applied to the DES $r$-band photometry to predict corresponding SMSS $r$-band magnitudes. In Figure~\ref{fig:des_delta}, we show the improvement in the photometric comparison before ($\Delta r$, top panel) and after ($\Delta r_{corr}$, second panel) applying the colour correction.

The third panel of Figure~\ref{fig:des_delta} shows the colour-corrected magnitude difference normalised by the photometric errors (with the SMSS and DES errors being square-added). Until $r\sim20$~mag, the $\Delta r_{corr}$ distribution is consistently distributed around zero, with fainter sources showing a slight bias towards brighter magnitudes in DR4. At $r=20.7$~mag, the median offset to brighter magnitudes begins to exceed $1\sigma$. The bottom panel of Figure~\ref{fig:des_delta} shows the fraction of DES sources having corresponding SMSS matches as a function of $r$-band magnitude, and divided into different bins of Galactic latitude, |$b$|. The variations as a function of |$b$| are small, but the high-latitude, low-extinction region of the sky is the most complete in DR4. The 50~per~cent completeness level is typically around $r = 20.9$~mag in these fields, a brightness at which the mean DR4 magnitude error for sources with just a single detection (which cannot take advantage of multiple measurements to reduce the uncertainty) is $\sim0.1$~mag.

\begin{figure}[!ht]
\begin{center}
\includegraphics[width=\columnwidth]{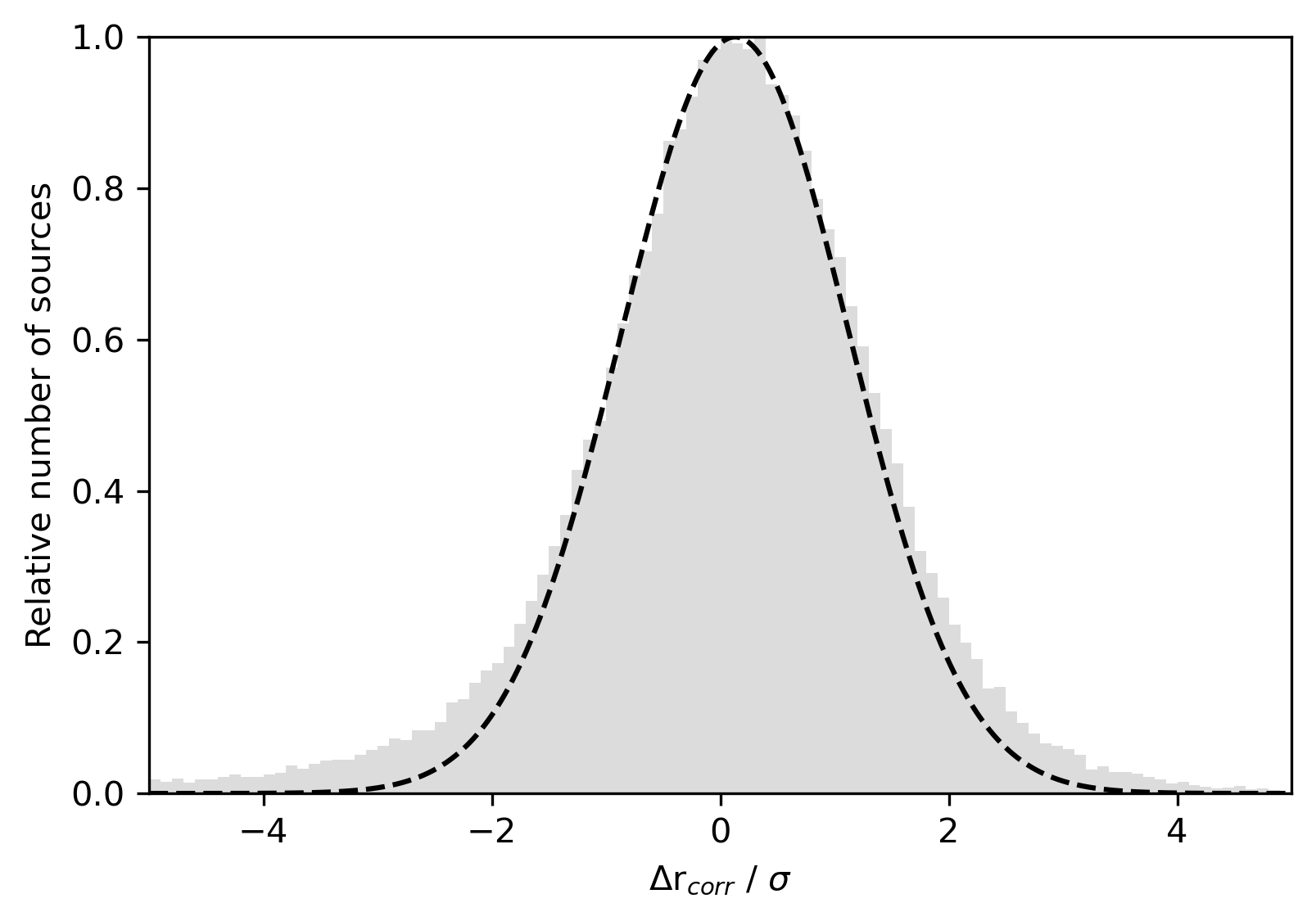}
\caption{Distribution of differences in colour-corrected $r$-band photometry between SMSS DR4 and DES DR2 in units of photometric error (square-adding the DES and SMSS errors). The sample of point sources was limited to DES $r$-band magnitudes between 16 and 19. The dashed line indicates a Gaussian of standard deviation equal to one, which fits the bulk of the population, indicating the reliability of the SMSS photometric errors. The sources at larger magnitude differences include variable sources.
}\label{fig:des_hist}
\end{center}
\end{figure}

Finally, Figure~\ref{fig:des_hist} shows distribution of error-normalised, colour-corrected magnitude differences for sources with DES $r$-band magnitudes between 16 and 19. The dashed line overplotted is a Gaussian of unit standard deviation, which is an excellent representation of the distribution. The long tails in the histogram likely reflect the brightness changes of variable objects. Approximately 13~per~cent of the sources lie above the Gaussian shown in the Figure. This illustrates that the SMSS photometric errors are reliable indications of the magnitude uncertainty.

\subsection{Astrometric Performance}
\label{sec:prop_astrom}

The WCS solutions derived for the SMSS~DR4 images typically involved 800 stars from across the mosaic, with quartiles of $\sim500$ and $\sim1500$ stars. 
While no significant differences between filters are seen in the number of stars fit, the resulting RMS in the WCS fit\footnote{Catalogued in the \texttt{WCS\_RMS} column of the \texttt{images} table.} did vary from median values of 0.155~arcsec in $u$-band to 0.110~arcsec in $z$-band, likely due to the wavelength-dependent trends in seeing. 

We also see a date-related trend in the WCS RMS, worsening by $\approx50$~per~cent over the MJD range included in DR4. This effect may be driven by underlying trends in the median seeing, which also increased by $\sim1$~arcsec between 2014 and 2021 across all 6 filters. In addition, stellar proper motions since the {\it Gaia} DR2 epoch of 2015.5 (which anchor our WCS solutions) will contribute a small level of scatter for more recent observations.

\begin{figure*}
\begin{center}
\includegraphics[width=\columnwidth]{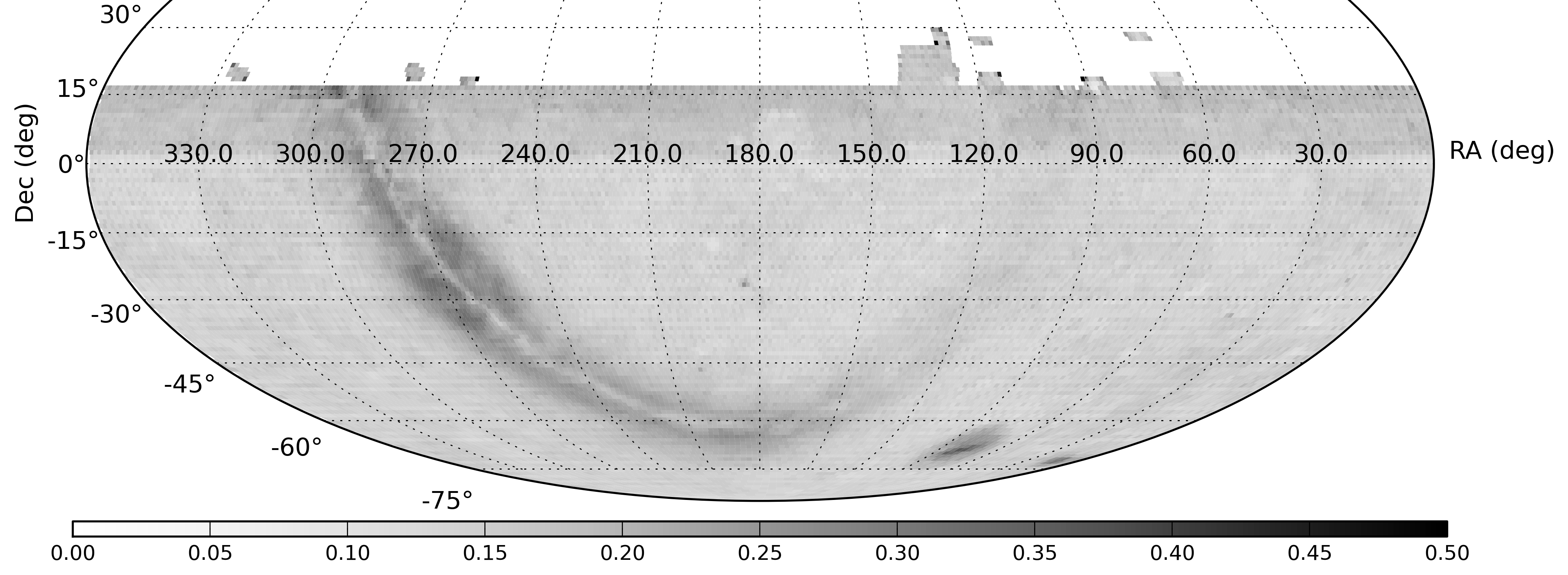}
\caption{Sky map of the median astrometric offset between SMSS DR4 and {\it Gaia} DR3; the colourbar indicates the offset in units of arcsec. The median offset per square degree ranges between 0.04 and 0.54~arcsec, with all values larger than 0.25~arcsec (0.5~pixels) occurring at the edges of SMSS DR4's sky coverage or in regions of high density and source blending. 
}\label{fig:gaia_offsets}
\end{center}
\end{figure*}

Overall, amongst the \texttt{master} table sources with {\it Gaia} DR3 matches within 1~arcsec, the median astrometric offset was 0.13~arcsec, or 0.26~pixels. The sky map of median {\it Gaia} offset distance per square degree is shown in Figure~\ref{fig:gaia_offsets}. The only regions in which the median offsets exceed 0.5 pixels (0.25\arcsec) are some highly crowded regions (e.g., the centre of the Large Magellanic Cloud) and at the edges of the SMSS DR4 sky coverage (where the numbers of sources in the square degree is small and the WCS solution in a given image is least constrained).

\begin{figure}
\begin{center}
\includegraphics[width=\columnwidth]{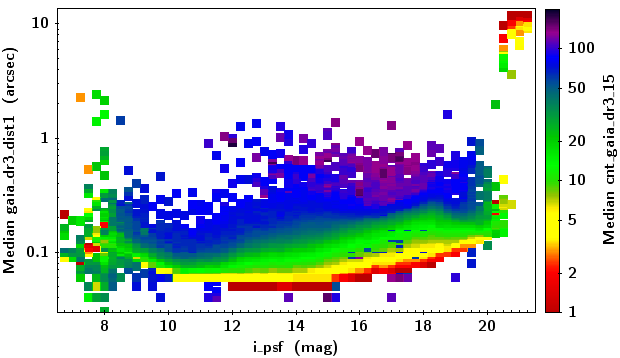}
\caption{Median astrometric offset between SMSS DR4 and {\it Gaia} DR3 as a function of SMSS $i$-band magnitude; the colourbar indicates the median number of {\it Gaia} sources within 15~arcsec. Very bright sources (many of which have significant proper motion) and very faint sources (for which the cross-match may be spurious) deviate from the expected trends of smoothly, but mildly increasing offset at both fainter source magnitude and increasing source density (where SMSS is subject to blending).
}\label{fig:gaia_offset_mag_density}
\end{center}
\end{figure}

We show in Figure~\ref{fig:gaia_offset_mag_density} how the offset from {\it Gaia} positions depends upon both the local source density (number of {\it Gaia} sources within 15~arcsec) and the source magnitude (SMSS DR4 $i$-band). Aside from the extremes in $i$-band magnitude, the offsets are well behaved, with mild increases at both fainter magnitudes and denser fields.

\subsection{Photometric Performance}
\label{sec:prop_phot}

\begin{figure*}
\begin{center}
\includegraphics[width=0.495\columnwidth]{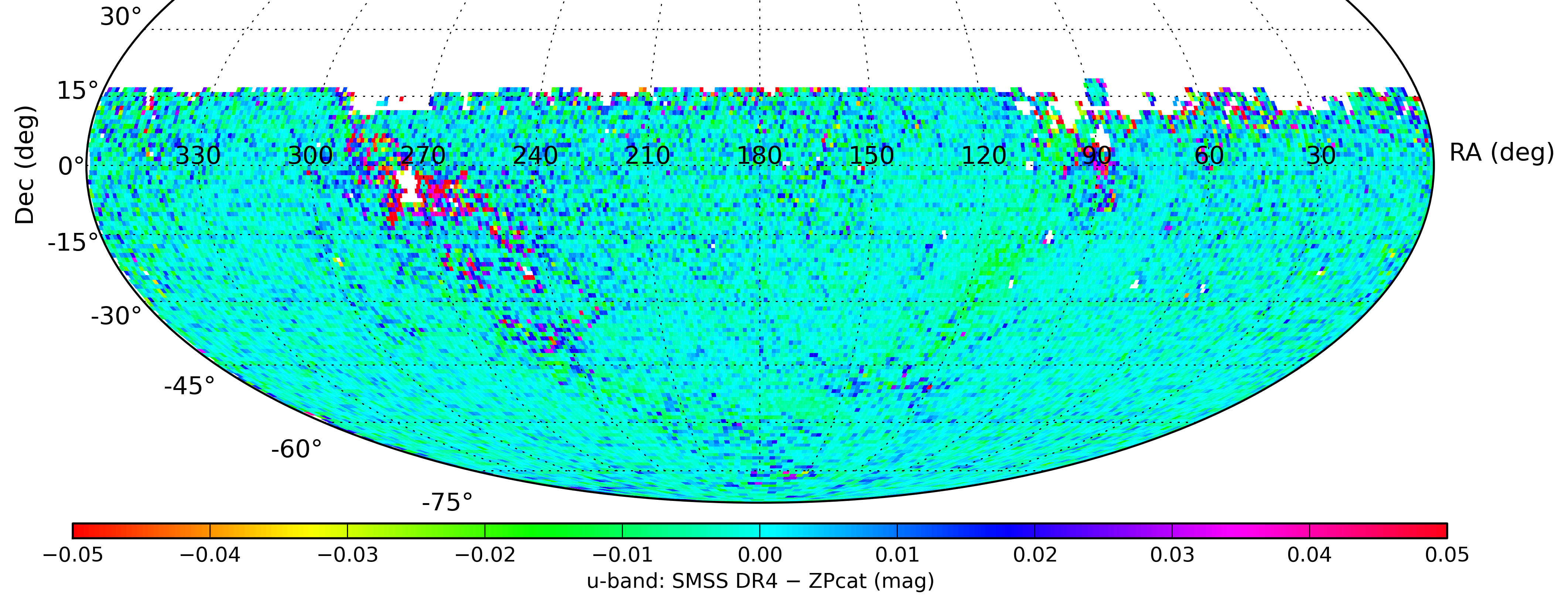}
\includegraphics[width=0.495\columnwidth]{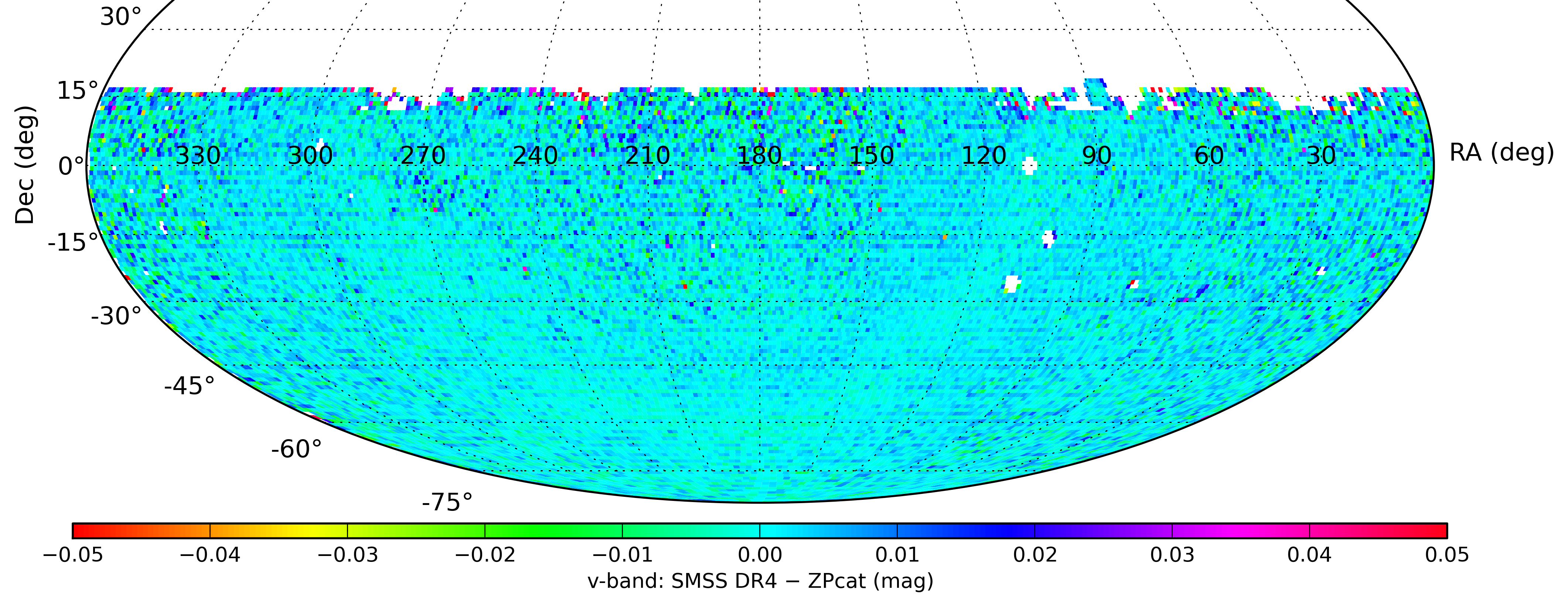}\\
\includegraphics[width=0.495\columnwidth]{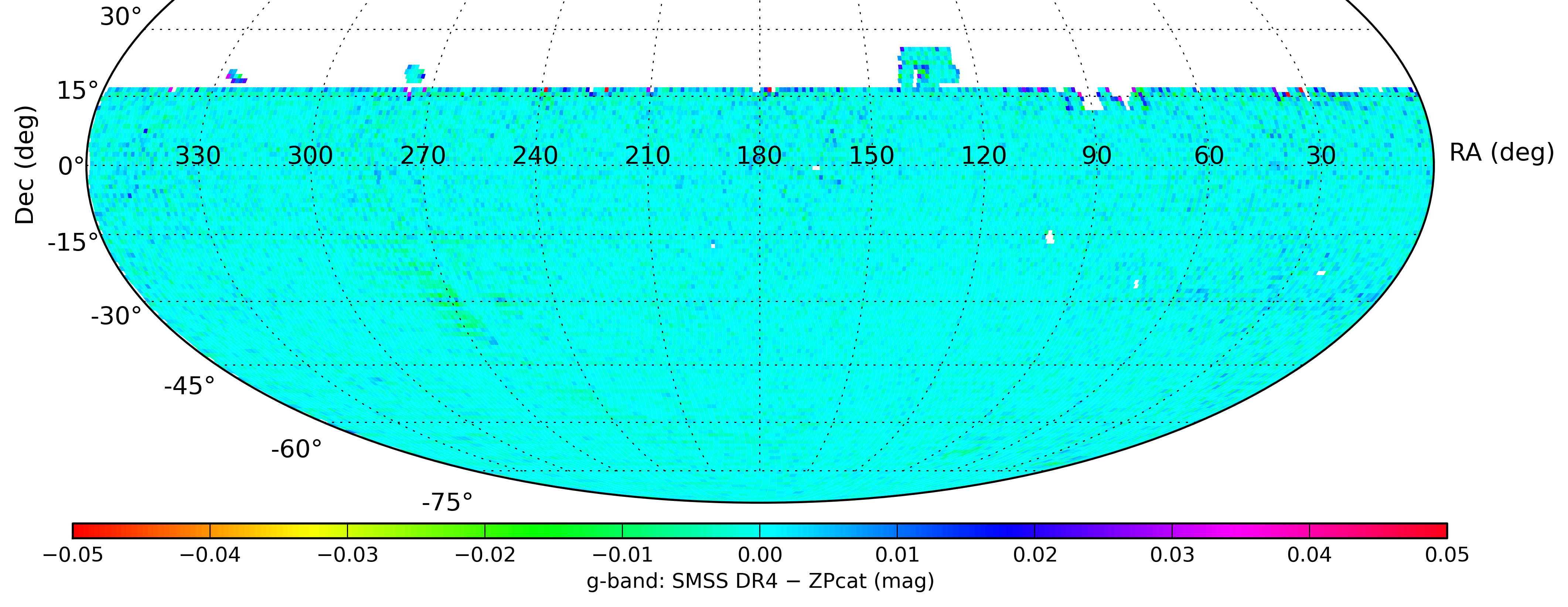}
\includegraphics[width=0.495\columnwidth]{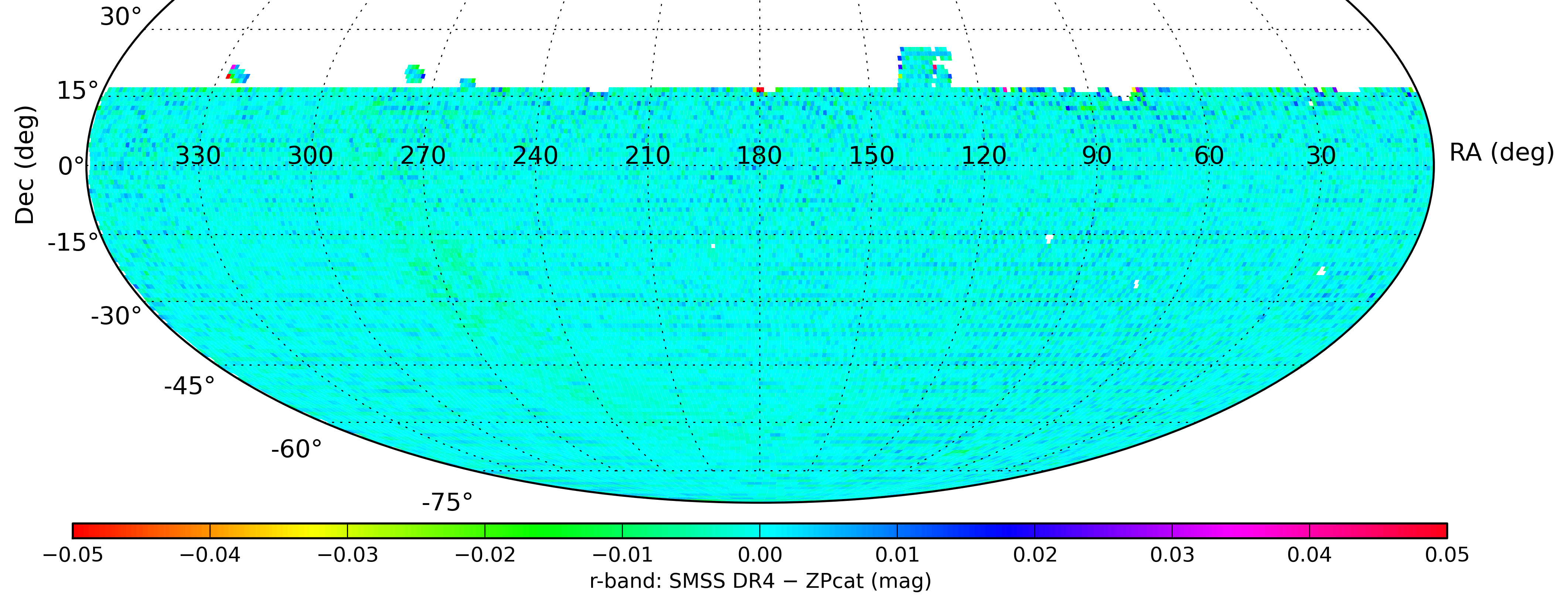}\\
\includegraphics[width=0.495\columnwidth]{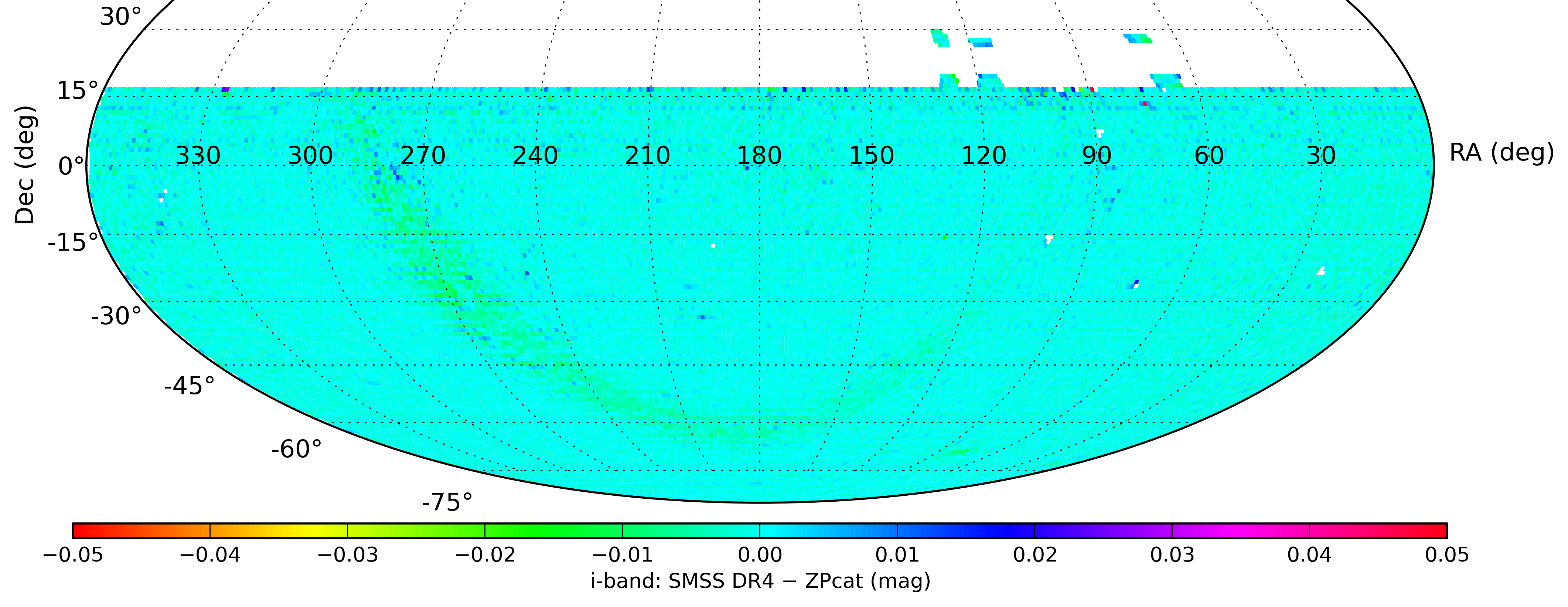}
\includegraphics[width=0.495\columnwidth]{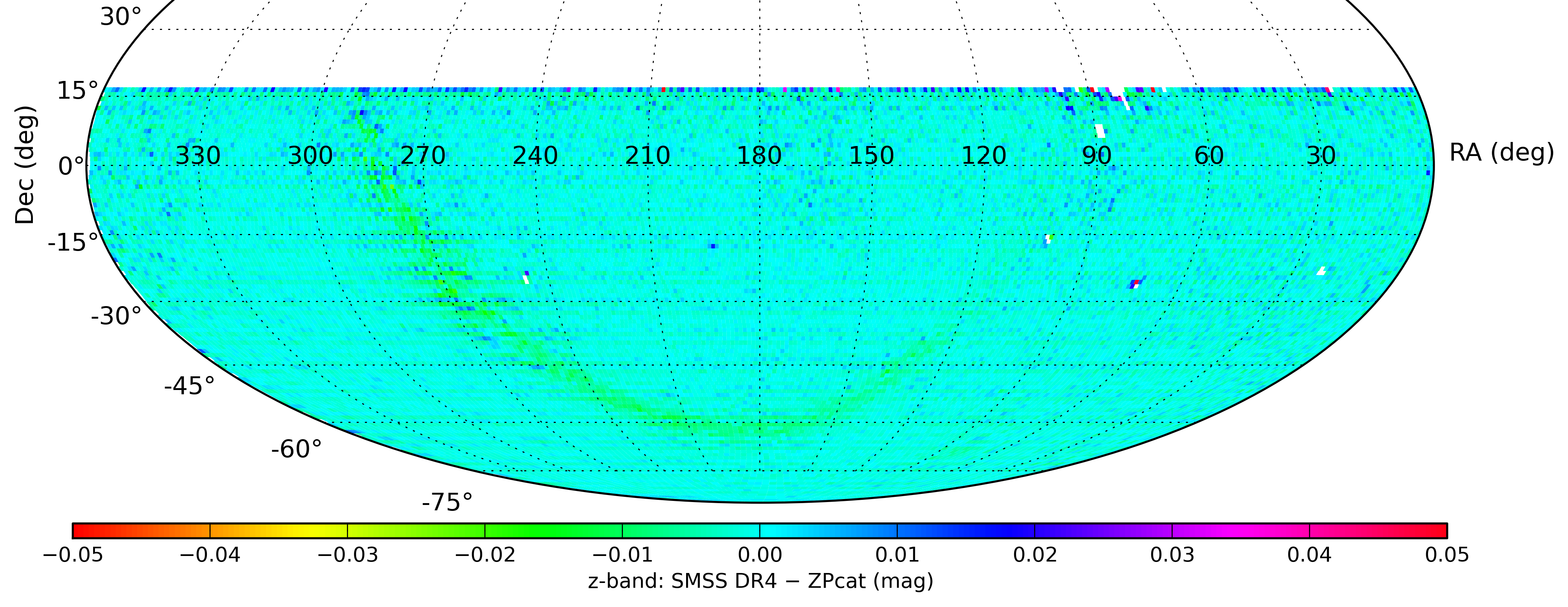}
\caption{Median photometric differences (DR4-observed minus ZP-catalogue-predicted) per square degree for each of the 6 SMSS filters ($uvgriz$). In contrast to Figure~\ref{fig:dr4_dr2}, the colourbars now cover a uniform range of $\pm0.05$~mag. 
}\label{fig:zpcat_dr4}
\end{center}
\end{figure*}

We first confirm that the calibration of the \texttt{master} table photometry yields results that match those of the ZP catalogue. Figure~\ref{fig:zpcat_dr4} shows that the DR4 mean photometry of the ZP stars is well matched to the {\it Gaia} synthetic photometry. The small residuals in regions of high density arise because of the effects of crowding on the SkyMapper photometry.

In contrast to previous DRs --- where the per-image ZP fits had a median RMS values in $u$- and $v$-band of $\approx0.1$~mag --- with the revised ZP catalogue in DR4, more than 99~per~cent of $u$ and $v$ images have ZP RMS values smaller than 0.1~mag. The median RMS values in DR4 are 0.03~mag for $u$ and $v$, 0.02~mag in $z$, and 0.01~mag in the other filters.

\begin{figure*}
\begin{center}
\includegraphics[width=0.495\columnwidth]{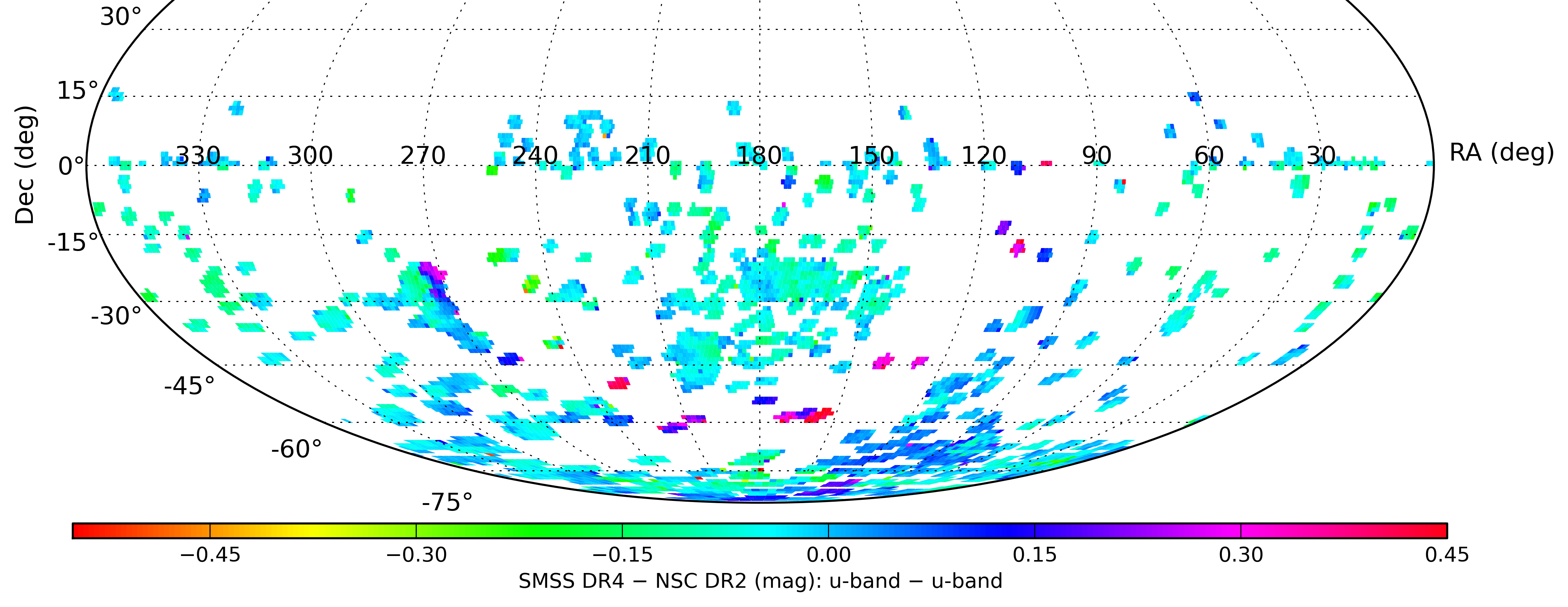}
\includegraphics[width=0.495\columnwidth]{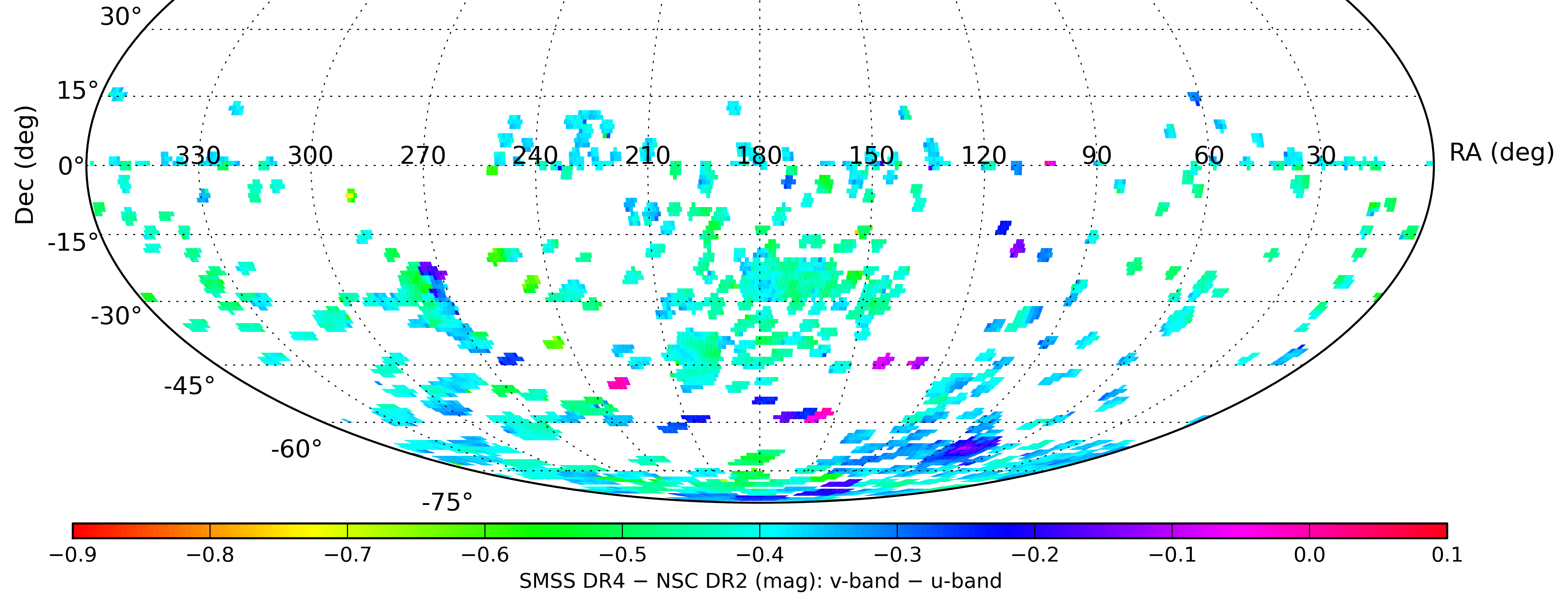}\\
\includegraphics[width=0.495\columnwidth]{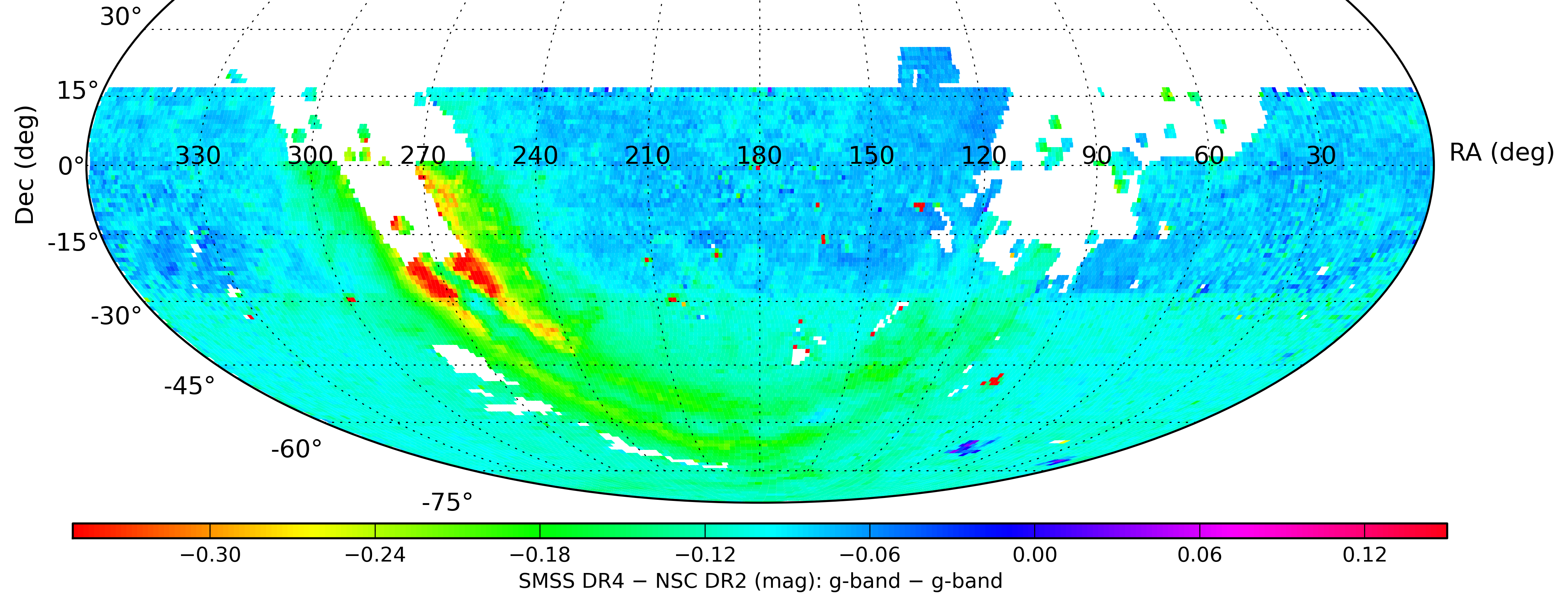}
\includegraphics[width=0.495\columnwidth]{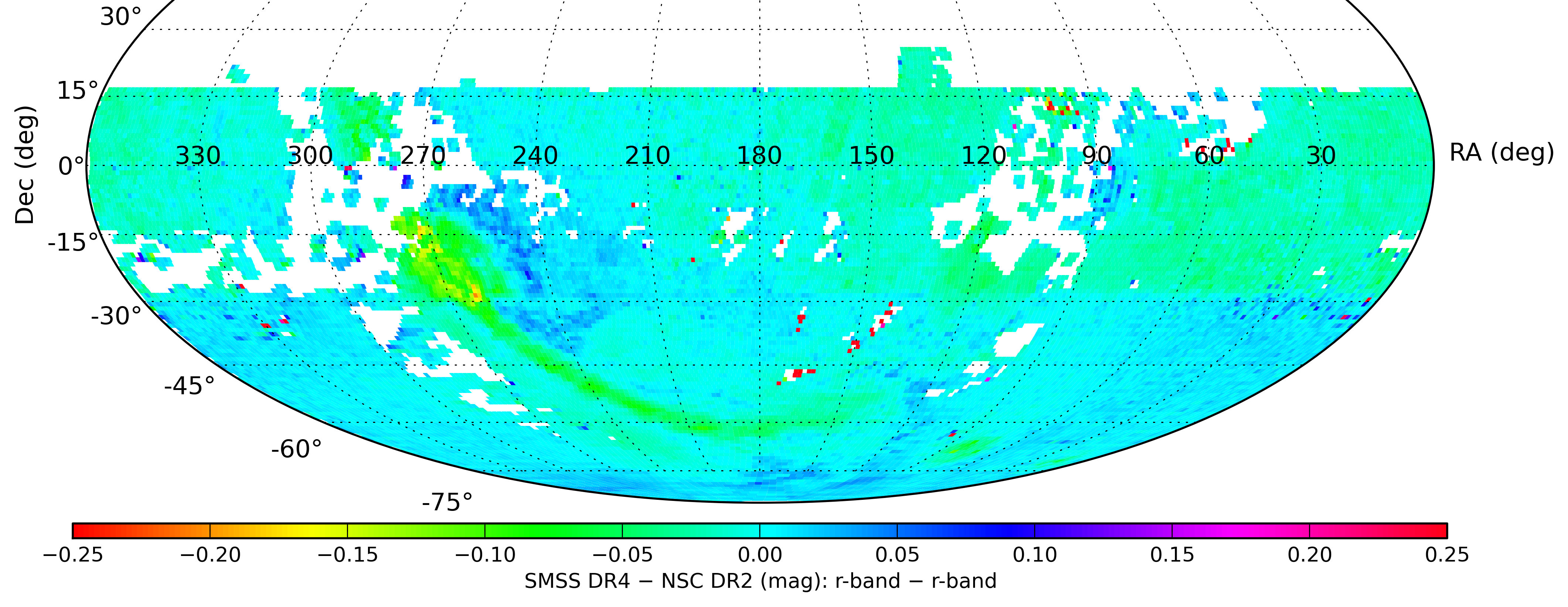}\\
\includegraphics[width=0.495\columnwidth]{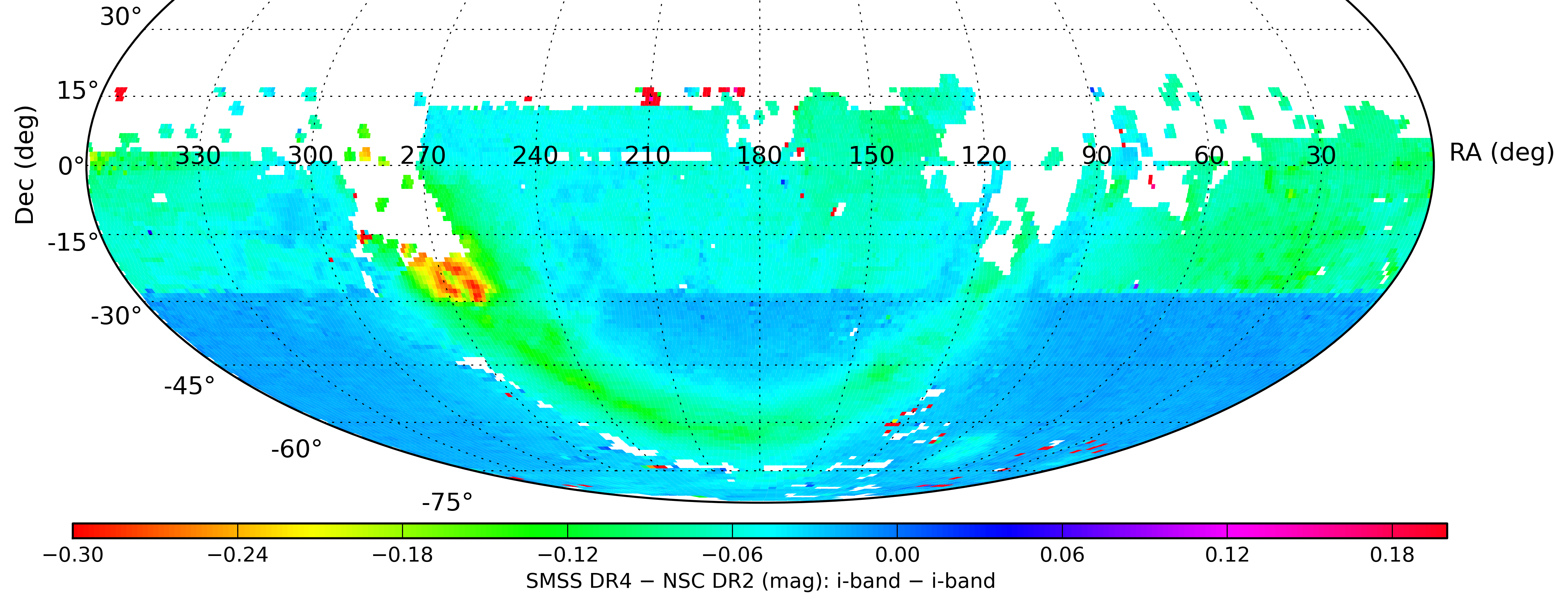}
\includegraphics[width=0.495\columnwidth]{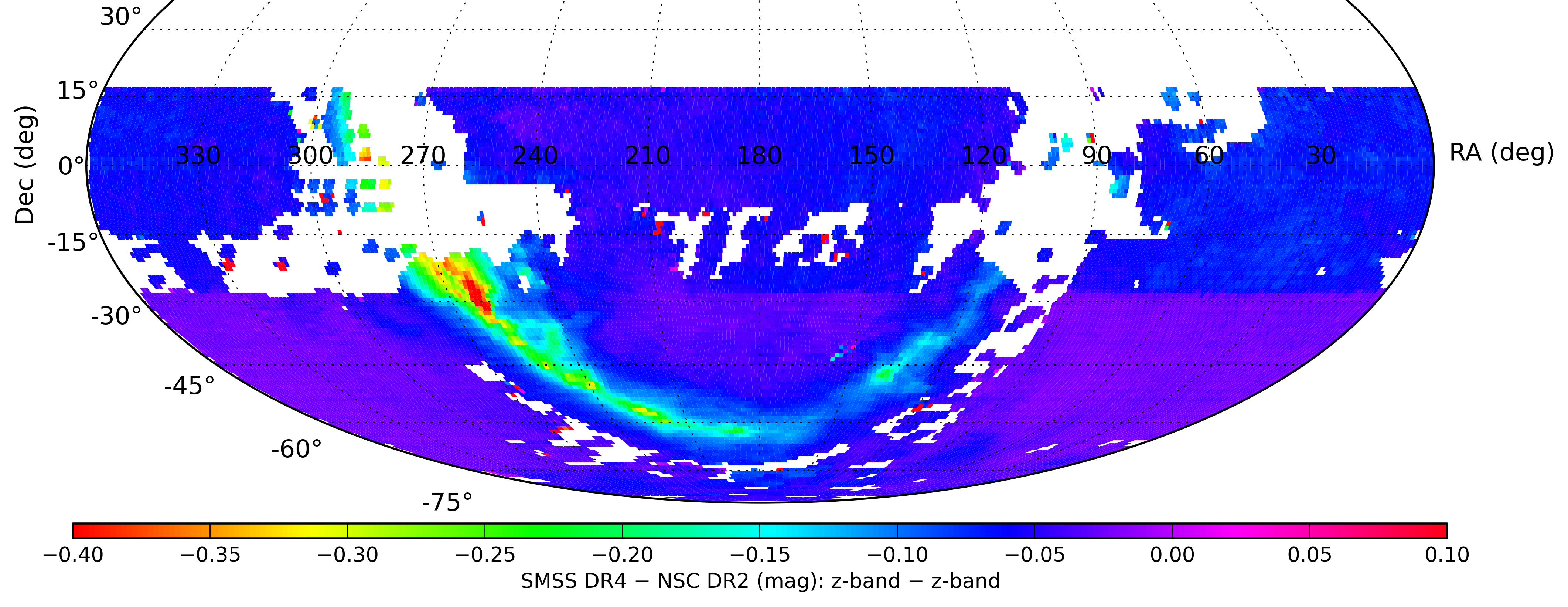}
\caption{Median photometric differences (DR4-observed minus NSC DR2) per square degree for each of the 6 SMSS filters ($uvgriz$) compared to the $ugriz$ of NSC DR2. The SMSS $u$- and $v$-band maps are both made in comparison to the NSC $u$-band, and have larger colourbar ranges than $griz$. 
}\label{fig:dr4_nsc}
\end{center}
\end{figure*}

In terms of hemispheric coverage in the South, one of the largest and deepest optical datasets available for comparison to SMSS DR4 is the NOIRLab Source Catalog (NSC) DR2 \citep{2021AJ....161..192N}. NSC has compiled a broad set of images, which, for the southern hemisphere, primarily come from the NOIRLab Blanco 4m telescope, and processed them in a homogeneous way. We show the sky maps of the median magnitude differences per square degree in Figure~\ref{fig:dr4_nsc}. The maps are driven by two factors: 
\begin{enumerate}
    \item Primarily, they illustrate bandpass differences between SkyMapper and DECam. The SkyMapper $u$ and $v$ filters have no equivalent bandpass in the NSC dataset and are both compared with the NSC $u$-band. The SkyMapper $griz$ filters have eponymous NSC bands with different transmission curves. The AB colours for an unreddened M0V star, e.g., are $(g-r,r-i,i-z)=(0.83,0.76,0.31)$ for SkyMapper passbands, while they are $(g-r,r-i,i-z)=(1.28,0.62,0.31)$ for the DECam passbands in NSC. 
    
    \item A secondary factor is actual calibration differences between the two data sets. The sparse NSC DR2 data in $u$-band limits the visual comparison between the photometry of the two surveys in Figure~\ref{fig:dr4_nsc}. However, because SMSS DR1.1 provided the $u$-band photometric calibration of NSC, one can anticipate significant spatial residuals that reflect the improvement of the SMSS photometric zeropoints. 
    For the $griz$ sky maps, we see a clear change in behaviour at Dec $\sim -30^{\circ}$, which marks the southern limit of Pan-STARRS DR1's coverage and the transition to where the NSC calibration source, ATLAS Refcat2 \citep{2018ApJ...867..105T}, more heavily relies on SkyMapper DR1.1 (in conjunction with other surveys).
\end{enumerate}

\begin{figure*}
\begin{center}
\includegraphics[width=0.33\columnwidth]{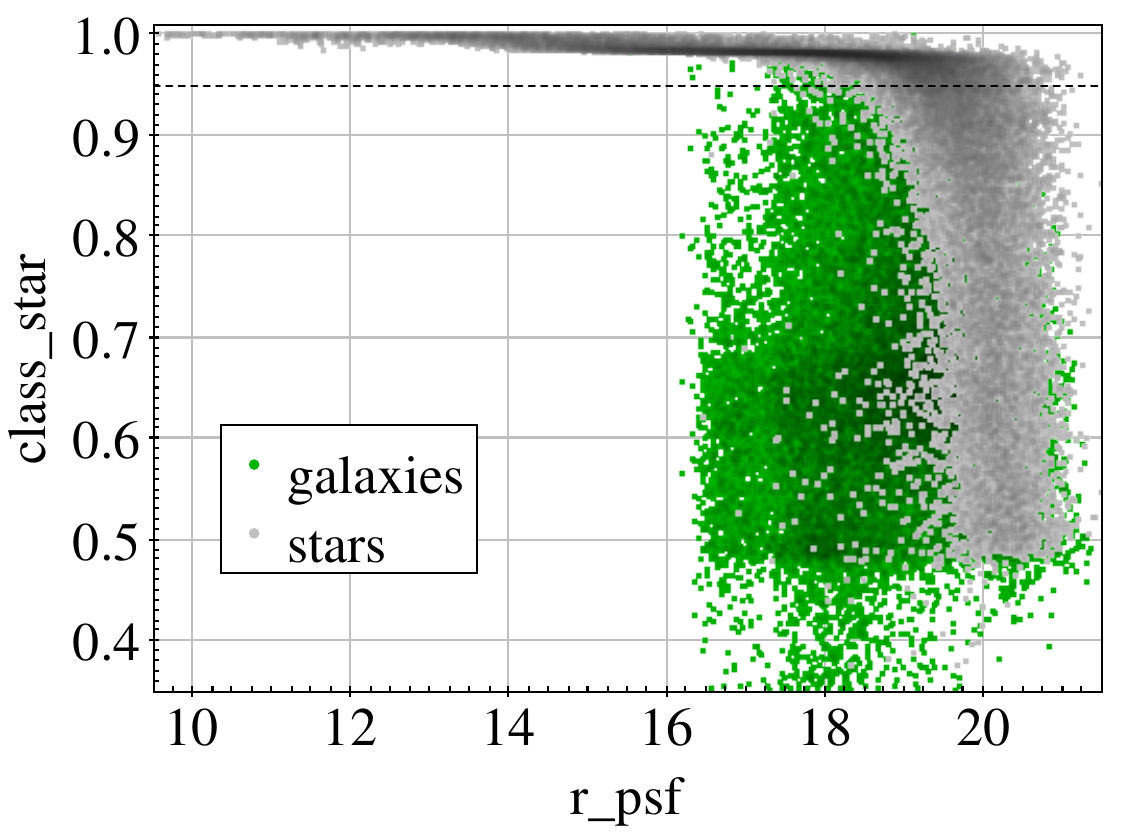}
\includegraphics[width=0.33\columnwidth]{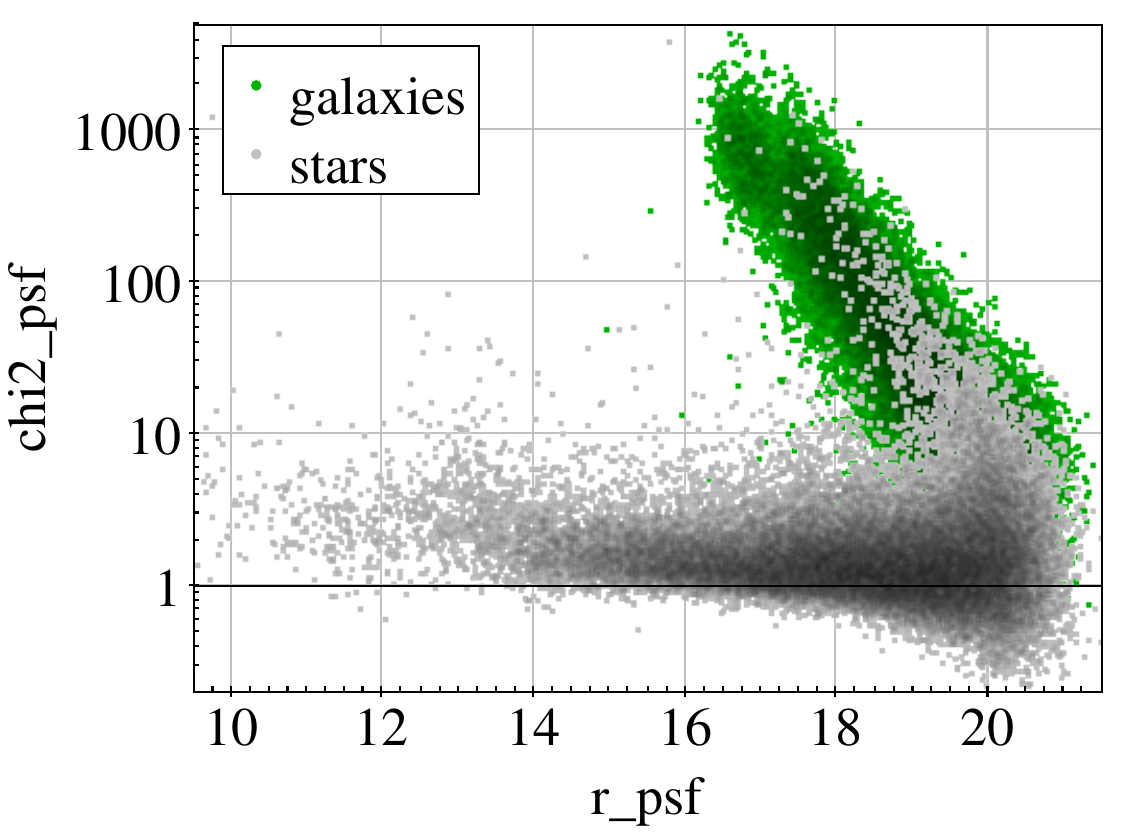}
\includegraphics[width=0.33\columnwidth]{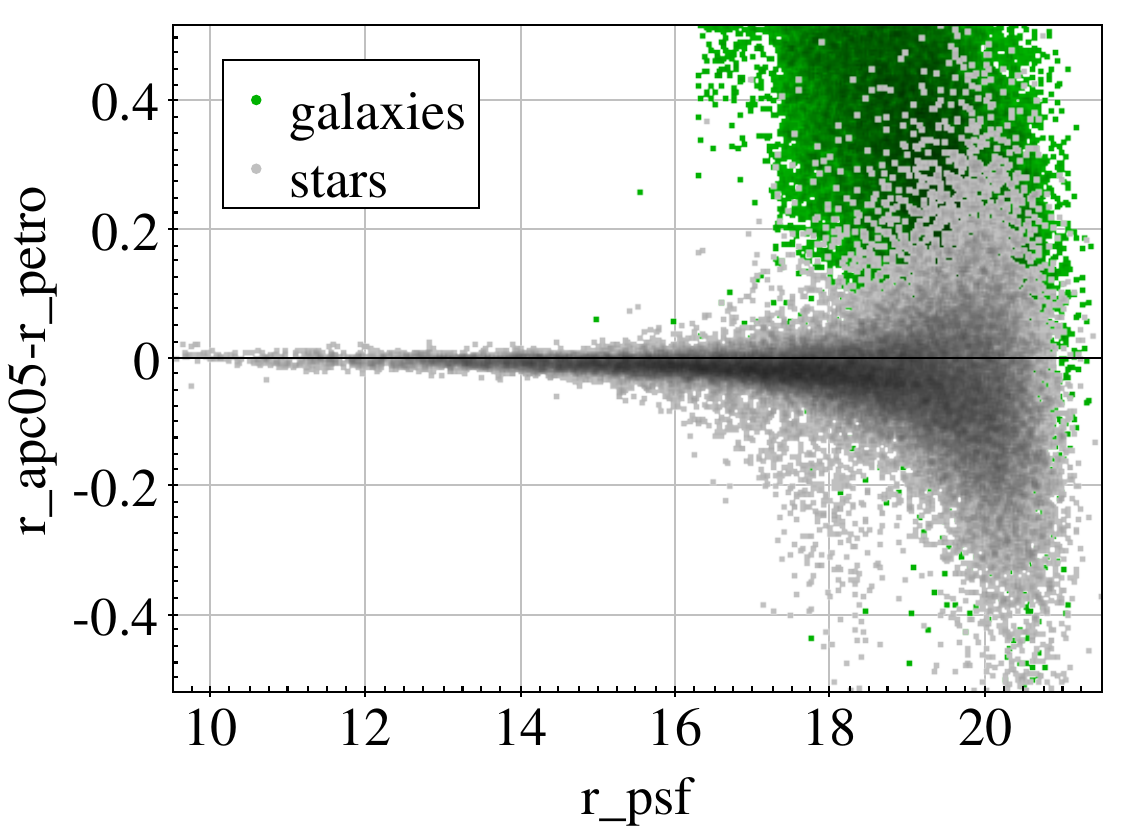}
\caption{Star-galaxy separation as probed by three indicators: the {\sc Source Extractor} \texttt{CLASS\_STAR} (left) is $>0.95$ for point sources; the \texttt{CHI2\_PSF} (centre) ranges mostly between 1 and 3 for point sources; and the difference between the APC05 magnitude and the Petrosian magnitude (right) is close to zero for point sources. 
Stars are a subset of {\it Gaia} sources with parallax and proper motion signals of at least $5\sigma$ significance and no neighbours seen within 15 arcsec radius, and compares them to galaxies from the 2dFLenS Survey \citep{2016MNRAS.462.4240B}. The double fan seen among stars in the top right panel results from objects with only Shallow Survey imagery vs the majority that has also Main Survey data. 
}\label{fig:sg_separation}
\end{center}
\end{figure*}

\subsection{Revisiting the passband transmission curves}
\label{calspec_comp}

After the production of the full DR4 data products, we compared the measurements of CALSPEC stars against the prediction from synthetic photometry and noticed imperfections hinting at inaccurate bandpass definitions. Already before the start of the survey, \citet{2011PASP..123..789B} argued that empirical data will eventually facilitate improved bandpass definitions. 

\begin{figure}
\begin{center}
\includegraphics[angle=270,width=\columnwidth]{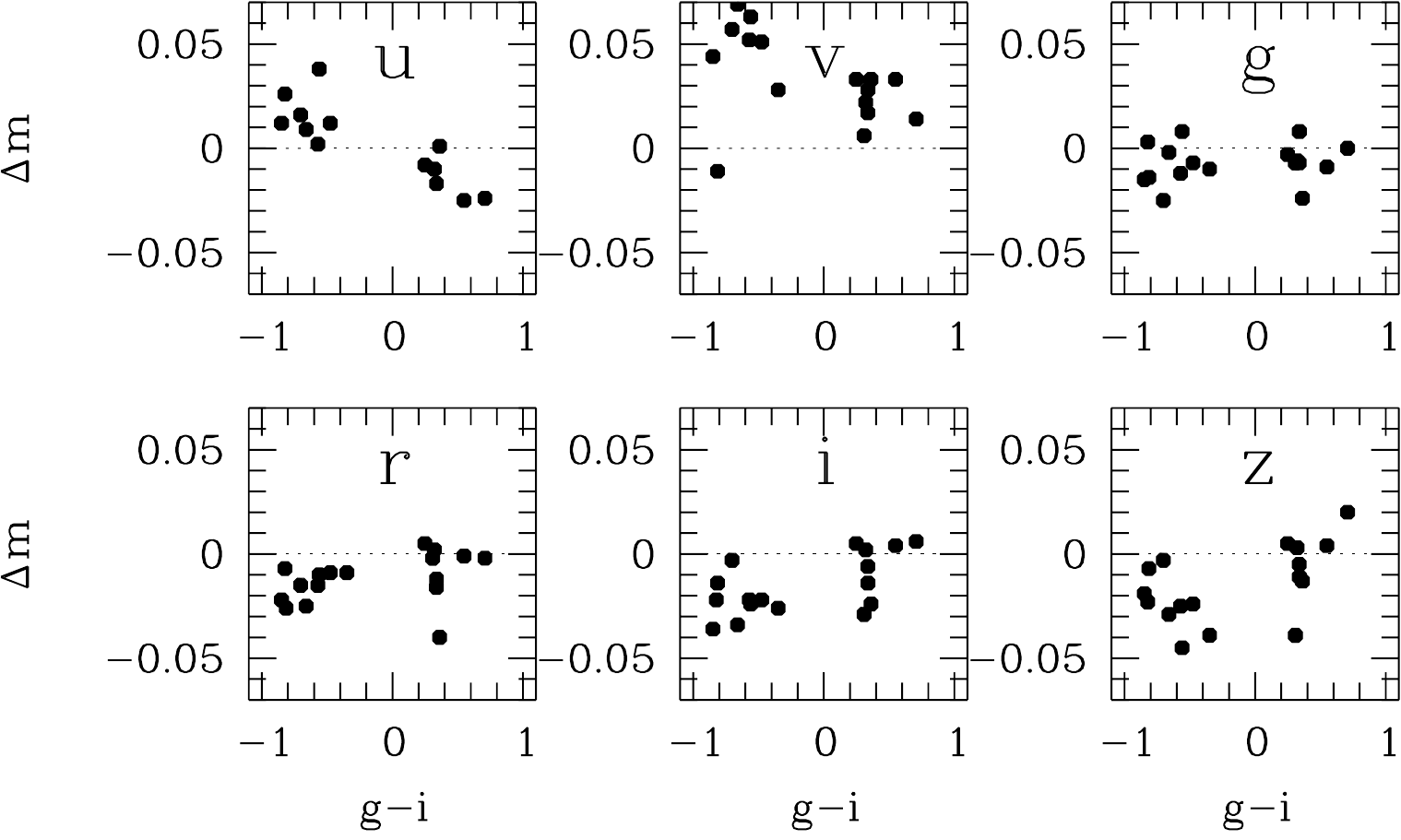}
\caption{Magnitude differences, DR4 (measured) minus predicted, of CALSPEC spectrophotometric standard stars. Two of the hot stars and two of the cool ones do not appear in the $u$-band panel, where they have $\Delta m\approx -0.1$ for reasons that remain under investigation.
}\label{fig:calspec}
\end{center}
\end{figure}

For CALSPEC stars with AB colours in the range of $g-i=[0,1]$, the offsets between measured and predicted magnitudes are mostly less than 0.02~mag (see Figure~\ref{fig:calspec}). The strongest offsets are seen in hot blue stars of $g-i<-0.4$, which can reach 0.05~mag. The current evidence suggests that these colour terms could be explained by shifting the mean wavelength of the passbands by between 1 and 5~nm.

A full explanation of these effects is beyond the scope of this paper, however, we briefly discuss the empirical evidence.
First, existing records revealed an ambiguity around the mean CCD quantum efficiency curve. Using the alternative CCD curve may nearly remove the colour trends in the $u$, $g$ and $z$ bandpasses, which are in areas where the CCD efficiency is not flat across the width of the filter. Shifts in the $r$ and $i$ filter might be explained by the differences between convergent beams in the telescope and a parallel beam in the 2010 laboratory measurements that produced the existing bandpass definitions. A 0.05~mag offset seen in the $v$ band for hot stars might be explained by a bandpass shift, although that is not expected in a glass filter. It may also reflect systematic uncertainties in the {\it Gaia} spectral reconstructions at the wavelengths of $v$-band.

In future work, we will revisit these questions by exploiting more evidence from DR4 data, especially considering many more objects with well-known spectra. There may also be an opportunity to repeat laboratory measurements of the filters to determine whether their transmission has changed due to ageing.

\subsection{Notes on Source Types}
\label{sec:additional_notes}

In this section, we describe some important features of the DR4 dataset, broken down by source type.

\subsubsection{Moving sources}

Sources with significant sky motion, principally Solar System objects, will show up with a range of properties depending on their apparent speed:
\begin{enumerate}
    \item Slow-moving objects such as stars with moderate proper motion will show large position uncertainties associated with a single \texttt{OBJECT\_ID} that captures all detections of the object over the years of the survey.
    \item Stars with high proper motions and very distant Solar System objects will see their detections broken up into more than one \texttt{OBJECT\_ID}. The example of Pluto was presented in Sect. 3.7 of \citet{2019PASA...36...33O}.
    \item Distant Solar System objects are bound to show a separate \texttt{OBJECT\_ID} for each night of their observation, and may also show large position uncertainties when a sequence of images was observed. 
    \item Typical Main-Belt asteroids may appear broken up into several \texttt{OBJECT\_ID} entries already during a Shallow Survey (4 min duration) or Main Survey (20 min duration) colour sequence.
    \item Near-Earth asteroids and artificial satellites will produce streaks covering one or more CCDs (see, e.g., Figure~\ref{fig:trails}), which could each seed their own \texttt{OBJECT\_ID} (or multiple, if split into separate {\sc Source Extractor} detections).
\end{enumerate}
In any case, there is a risk of such objects being associated with the \texttt{OBJECT\_ID} of a persistent source from beyond the Solar System and being blended with it; one such case is discussed in detail in Sect. 3.7 of \citet{2019PASA...36...33O}. In this case, also the \texttt{\{f\}\_MMVAR} column for the filter \texttt{\{f\}} observed during and affected by the transient blending may show an inflated value above a true variability level of the persistent source, although the rare outlier measurements may be clipped from the distilled magnitude estimates. Finally, the fastest and faintest moving objects may pass so quickly that the flux is too diffuse to appear in the \texttt{photometry} table at all.

The main challenge for selecting moving or transient sources from the \texttt{master} table is to differentiate between genuine objects and spurious detections. In persistent astrophysical sources, the multiple detections indicated by \texttt{NGOOD} values above 1 tend to weed out spurious imposters. But among single detections, a variety of indicators need to be consulted for a pure sample of real objects: good \texttt{FLAGS} ($<4$) and sensible \texttt{MEAN\_FWHM} values should be required.

\subsubsection{Small-separation sources}

Some objects, such as those in crowded areas or binary/triple stars or merging galaxies, require special attention: it is possible that two objects are physically present, but three unique objects appear in the \texttt{master} table. This happens when, in some images, the two are separately detected, and in others, a blended object is detected at an intermediate position. The two sets would appear in mutually exclusive image sets. In such cases, the spatially resolved photometry remains useful for the two genuine sources, and the blended detection should be ignored. Recognising which is which requires consulting the images or the \texttt{photometry} table with an eye on positions and \texttt{FWHM} values of the detections.

Similarly, at intermediate separations, no third object may appear with its own \texttt{OBJECT\_ID}, but the blended version may be associated with one of two true sources, bringing brighter photometry into the detection list; the \texttt{\{f\}\_MMVAR} column for each filter \texttt{\{f\}} may show an inflated value far above the true variability; to what extent measurements might be deemed outliers and thus be clipped from the distilled magnitude estimates depends on the individual measurement statistics.

Finally, tight binary objects will usually not be separated, and thus they will have blended photometry in most or all images, which then dominates the available photometry and the clipping process retains the blended measurement.

\subsubsection{Variable sources}

Variable objects are easily found in the \texttt{master} table by looking for objects with a \texttt{\{f\}\_MMVAR} value that is atypically high for the object's magnitude, which points to excess variability above the expected noise variation. Chance projections with asteroids, however, can enhance the brightness of otherwise non-variable objects in a small number of detections and thus make them appear variable. 

As discussed in Sec.~\ref{sec:history}, the SMSS has not followed a strict cadence pattern, but collected the observations over time to accrue the desired depth. One specific cadence that was retained, however, is that of the Main Survey Colour Sequence, which completed a $uvgruvizuv$ pattern on one field within a span of 20 minutes. Such data have proven particularly useful for measuring short-term variability in the $uv$ bands \citep[see e.g.][]{2022MNRAS.515.3370L}.

\begin{figure}
\begin{center}
\includegraphics[width=\columnwidth]{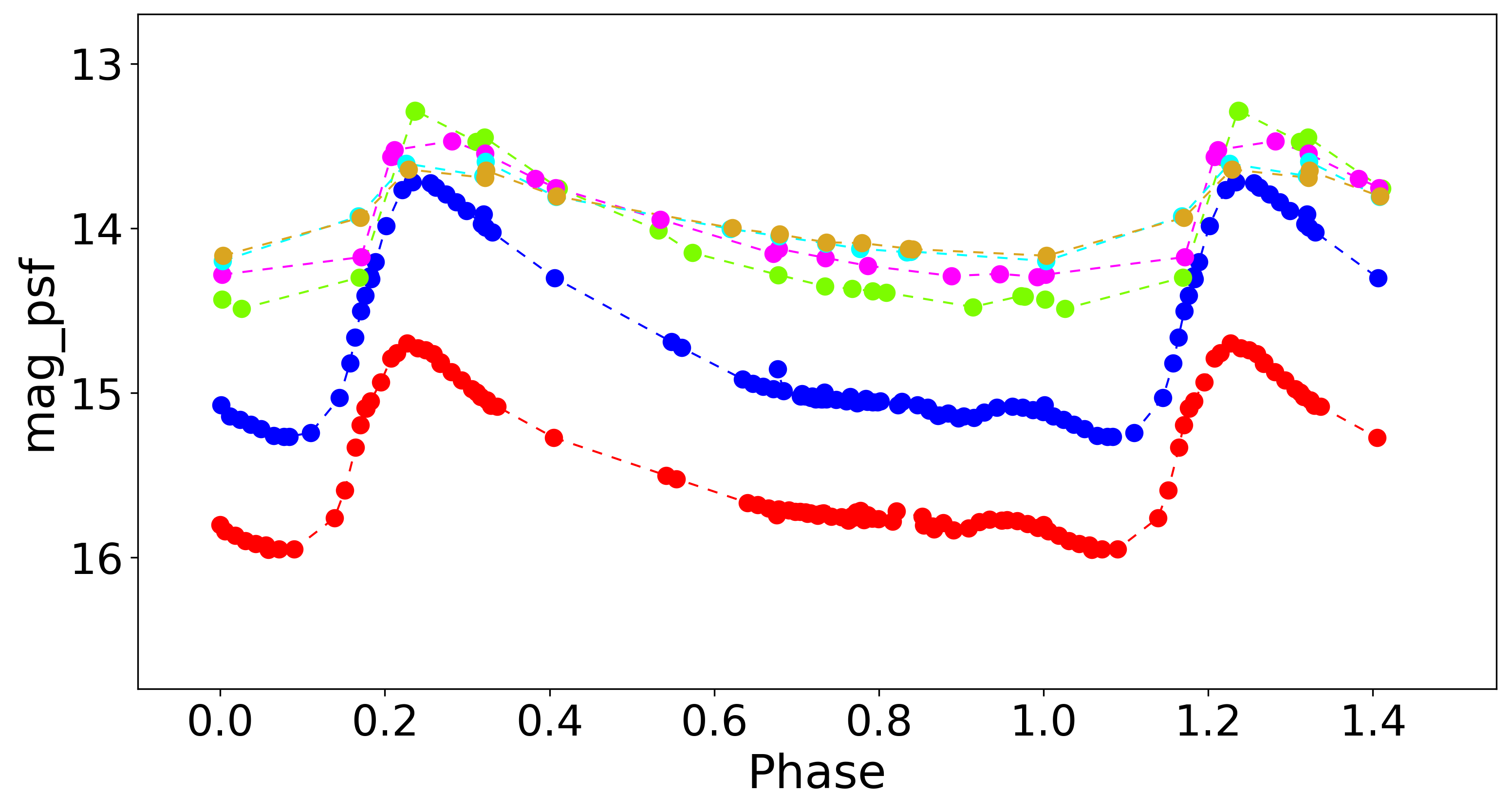}
\includegraphics[width=\columnwidth]{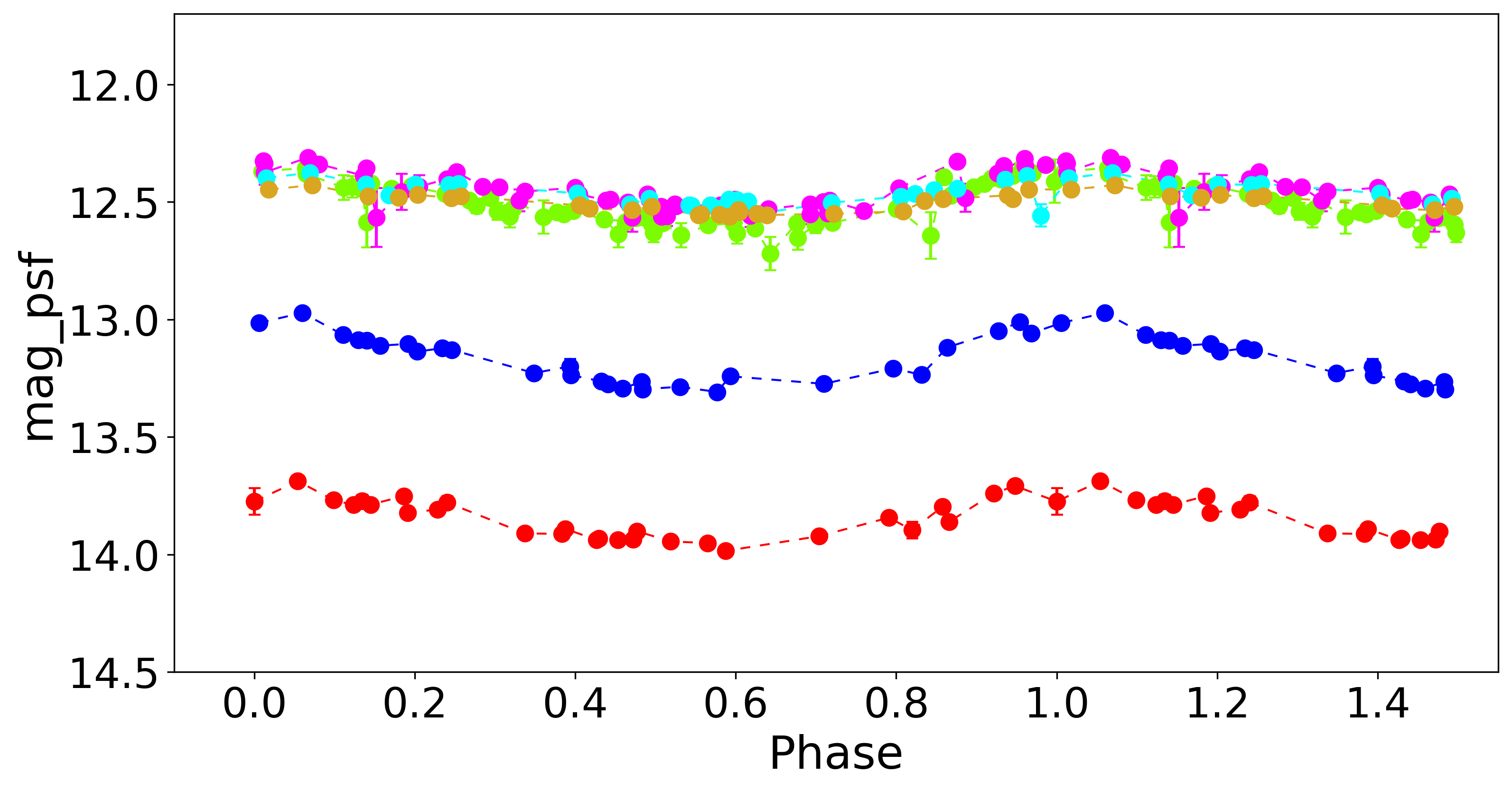}
\caption{SMSS DR4 phase-folded lightcurves of two known variable stars: ({\it top}) the fundamental-mode RR~Lyr (RRab) star, SX~Gru (\texttt{OBJECT\_ID} 489799903), with a period of 0.5934~d; ({\it bottom}) the $\delta$~Scuti pulsating variable, ATO~J136.5509-00.8024 (\texttt{OBJECT\_ID} 89456216), with a period of 0.1206~d. In each panel, the points are colour-coded by filter: $u$=red, $v$=blue, $g$=green, $r$=magenta, $i$=cyan, $z$=gold. The phase coverage is more complete for these two stars than the average DR4 source (even the redder four filters in the top panel have $2-3\times$ more measurements than typical), but a small fraction of sources will have even more detections in DR4 than the $\sim80$ for $u$- and $v$-band in the top panel.
}\label{fig:var}
\end{center}
\end{figure}

Around 90~per~cent of the moderately bright objects in DR4, with $g_\mathrm{PSF}\sim 17$, have less than ten observations per band with good flags. While the total number of visits by the telescope has been larger, that includes observations with bad quality flags and images not even included in DR4. The median number of good detections in the $g$ and $r$ filters is \texttt{\{f\}\_NGOOD}$=6$. As DR4 includes all data taken with the SkyMapper Telescope, irrespective of the original purpose, some areas in the sky have many more visits, including the area of the SkyMapper Transient Survey \citep{2017PASA...34...30S}, the Standard Fields, and smaller non-Survey programs. As a result, about 10~per~cent of the objects have 10 or more good detections, 1~per~cent have 25 or more, 0.3~per~cent have over 100 detections, and finally 0.1~per~cent of the objects have over 500 to 600 $g$- and $r$-band detections. Therefore, some specific sky areas have a strong time-domain coverage, while most of the hemisphere has few visits. In Fig.~\ref{fig:var} we show the phase-folded light curves of two known variable stars.

\subsubsection{Extended and point-like sources}
\label{sec:extended}

Marginally extended sources can be distinguished from point-like sources using one of three indicators (see Figure~\ref{fig:sg_separation}):
\begin{enumerate}
    \item The \texttt{CLASS\_STAR} parameter measured by {\sc Source Extractor} on individual images is recorded in the \texttt{photometry} table, and the \texttt{master} table records the largest value detected for any distilled object. Down to a magnitude of $r\approx 19$, nearly all point sources show \texttt{CLASS\_STAR} values $>0.95$.
    
    \item The \texttt{CHI2\_PSF} parameter is determined from light profiles in 15 arcsec apertures; the values of individual detections in the \texttt{photometry} table get averaged per filter and then the largest value among the filters is listed in the \texttt{master} table (and hence the average value in the \texttt{master} table will be larger than 1 even if the mean value was $\sim$1 among the detections). For point sources, typical values of \texttt{CHI2\_PSF} range from 1 to 3, but they show little correlation to the \texttt{CLASS\_STAR} parameter. Towards faint point sources, the \texttt{CLASS\_STAR} values decrease, while the \texttt{CHI2\_PSF} values remain at the same level just with increased scatter. Some galaxies with bright nuclei show high, point-like \texttt{CLASS\_STAR} values although their \texttt{CHI2\_PSF} values are large, indicating a significant envelope of light.
    
    \item The difference, \texttt{\{f\}\_APC05} $-$ \texttt{\{f\}\_PETRO}, between the PSF-corrected 5\arcsec\ aperture magnitude and the Petrosian magnitude in filter \texttt{\{f\}} can be calculated for individual detections in the \texttt{photometry} table but will have the least noisy value from distilled values in the \texttt{master} table. It is well-correlated with the \texttt{CHI2\_PSF} parameter.
\end{enumerate}

Of course, this is no reliable indication for the physical nature of an object: AGN and QSOs may show point-like appearance especially when the nucleus is significantly brighter than the host, and stars could appear extended when they are in projected or real binary systems that are only marginally resolved by SkyMapper.

\subsubsection{Colours of extended sources}
\label{sec:extended_colour}

Integrated colours of extended objects depend on the choice of aperture due to potential colour gradients. However, the light profiles of galaxies imply that the photometry of outer regions is noisy. Hence, colours show lower scatter when calculated from the PSF-corrected 5\arcsec\ aperture magnitude rather than the Petrosian magnitude; the aperture colours are also corrected for PSF variation between and within images. For example, the bright red-sequence galaxies at $z=0.10\pm 0.01$ in 2dFLenS (with \texttt{R\_PETRO}$=[15.9,16.7]$) show observed-frame APC05 colours of $v-r=2.41\pm 0.115$ (mean and RMS scatter) but Petrosian colours of $v-r=2.89\pm 0.315$. Propagating formal magnitude errors, which are dominated by the $v$-band errors, suggests typical errors for the APC05 colour of $0.086$~mag and for the Petrosian colour of $0.13$~mag. After square-subtracting these from the colour scatter, we derive an intrinsic colour scatter (not corrected for the slope of the colour-magnitude relation) of $0.075$~mag from APC05 colours and $0.29$~mag from Petrosian colours, i.e., four times larger. Although the two results come from two different-sized physical footprints on the galaxies, and the red sequence is contaminated with red spirals with modest colour gradients, the inflated scatter in the Petrosian $v-r$ colour is likely due to mostly larger noise that is not captured in the formal errors of the Petrosian magnitude.

\subsubsection{Large extended sources}
\label{sec:extended_large}

One known deficiency in the current bias-PCA algorithm (Sec.~\ref{sec:bias}) is inadequate masking around the edges of large extended objects. This particularly affects the $u$- and $v$-band images, which are naturally faint in the outskirts of galaxies. The consequence is that the extended emission is treated as part of the bias level, and so is fit and subtracted by the PCA. Furthermore, because the PCs tend to be slowly varying functions with pixel position, the enhanced count level in the outer regions can force the PC fit higher than appropriate, resulting in dark wings on either side of the galaxy.

\subsection{Other Known issues}

\begin{itemize}

\item Around 50 million sources have \texttt{NGOOD}=0 in the \texttt{master} table, which are a combination of spurious and unreliable sources. 95~per~cent of them have \texttt{FLAGS}$>=$4. The remainder have \texttt{NIMAFLAGS}$>=$5, which excluded them from providing magnitude data to the distilling process. The spurious sources are most often ($\sim$50~per~cent) single spurious detections close to CCD borders and not real objects (with bit value 16 set in \texttt{FLAGS}). Nearly as common, comprising over 40~per~cent of the sources, is the case of real objects in the vicinity of a nearby bright star, and so having bit value 1024 set in \texttt{FLAGS} because their photometry could be affected by scattered light.

\item Slightly over 6 million \texttt{MEAN\_FWHM} values are less than or equal to zero. In $\sim$95~per~cent of the cases these are objects with single spurious detections: in $\sim$70~per~cent of cases, they reside in the vicinity of bright stars (bit value 1024 in \texttt{FLAGS}), and in $\sim $25~per~cent of cases they are detected at the edges of a CCD (bit value 16 in \texttt{FLAGS}). Even when they have good flags, they have at most one detection and usually no counterparts in the {\it Gaia} DR3 catalogue; they are often image artefacts that have not been flagged or masked.

\item Some detections in the \texttt{photometry} table show identical PSF magnitudes for all apertures, and thus \texttt{CHI2\_PSF} values equal to 0. This phenomenon occurs on CCDs where five or fewer stars were available to determine the PSF map; in these cases, the adopted PSF aperture corrections is chosen to be the median of the available stars, and thus, unintentionally, when the number of PSF stars is odd, one of them ends up with a seemingly perfect PSF shape by construction. In most cases, this problem is irrelevant, as multiple detections will exist with proper \texttt{CHI2\_PSF} values, the largest of which gets distilled into the entry for the \texttt{master} table. However, the \texttt{master} table does contain 48 objects with good flags and a \texttt{CHI2\_PSF} value of 0; all of these objects have only a single detection in one single filter. Some of these objects appear genuine in the images, but as they are not associated with a source known from another survey such as {\it Gaia} DR3, they may be asteroids. However, the majority appear to be unflagged cosmic rays and image artefacts near the central amplifier boundaries of the CCDs. 

\item Extended sources in shallow images may have poorly determined central coordinates. In certain circumstances, this can result in multiple \texttt{OBJECT\_ID} values being created for the same astrophysical object, which were unintentionally left unmerged during the calculation of the mean object properties (Sec.~\ref{sec:distill}).
Among the consequences of this proliferation of sources in the \texttt{master} table, we find 1592 sources that have \texttt{SELF\_DIST1} < 0.1~arcsec, four that have \texttt{SELF\_DIST2} < 0.2~arcsec, and one that has \texttt{SELF\_DIST3} < 0.3~arcsec. More than half of these cases are clustered around low-redshift galaxies in the 6dF Galaxy Survey \citep{2009MNRAS.399..683J}, especially those that fall within the Standard fields (Sec.~\ref{sec:design} and Tab.~\ref{tab:calspec}), which were visited thousands of times with short exposures. As indicated in Section~\ref{sec:flags}, most of these cases arose from the Standard Field images, which were subsequently assigned a \texttt{FLAGS} value of 16384.

\item When aperture corrections for a given CCD did not yield valid results for a given aperture size, then certain corrected aperture magnitudes will be missing in the \texttt{photometry} table, although the uncorrected aperture fluxes remain. As a result, no determination of the \texttt{MAG\_PSF} could be made for the detections in that CCD. This implies that the detection is not considered for distilling and \texttt{USE\_IN\_CLIPPED}$=-1$. However, the other magnitude columns --- the larger-aperture (15, 20, and 30~arcsec) magnitudes that do not receive aperture corrections, and those of \citet{1976ApJ...209L...1P} and  \citet{1980ApJS...43..305K} ---  will exist in the \texttt{photometry} table.

\end{itemize}

\section{SMSS DR4 Data Access and Format}
\label{sec:data}

In this section, we describe how to access the DR4 data, and the format of the available data.

\subsection{Data Access}
\label{sec:access}

The SMSS DR4 dataset is available through both the SkyMapper website (\url{https://skymapper.anu.edu.au/}) and the SkyMapper node of the All-Sky Virtual Observatory (ASVO). The website provides documentation for DR4 and previous releases; interfaces for cone-search, image cutout, and catalogue queries; summary pages for each SMSS~DR4 astrophysical object; and a User Forum to obtain additional information from the SMSS user community or the SkyMapper Team.

The SkyMapper ASVO node provides API access to the cone-search, image cutout, and catalogue query (Astrophysical Data Query Language [ADQL]) interfaces. Further details are available here: \url{https://skymapper.anu.edu.au/how-to-access/}.

Additional data access pathways for DR4 are being established through the Astro Data Lab (\url{https://datalab.noirlab.edu}) at NOIRLab's Community Science and Data Center (CSDC), and the VizieR catalogue service (\url{https://vizier.cds.unistra.fr/viz-bin/VizieR}) at the Centre de Donn\'{e}es astronomiques de Strasbourg (CDS).

\subsection{DR4 Data Format}
\label{sec:format}

We describe here the format of the information released in DR4, both the table data and the image data.

\subsubsection{Table Data}

The DR4 catalogue consist of five distinct tables:
\begin{itemize}
    \item \texttt{dr4.master} - main catalogue, listing the distinct astrophysical sources with their mean properties;
    \item \texttt{dr4.photometry} - per-image detection properties, which can be joined to the \texttt{master} table with the\linebreak
        \texttt{OBJECT\_ID} column;
    \item \texttt{dr4.images} - image-level properties, which can be joined to the \texttt{photometry} table with the \texttt{IMAGE\_ID} column;
    \item \texttt{dr4.ccds} - CCD-level properties, which can be joined to the \texttt{photometry} table with the (\texttt{IMAGE\_ID}, \texttt{CCD}) columns;
    \item \texttt{dr4.mosaic} - mapping between a detection's CCD and ($x_{\rm CCD}$, $y_{\rm CCD}$) position to the ($x_{\rm mosaic}$, $y_{\rm mosaic}$) position.
\end{itemize}

The \texttt{master} table contains a number of cross-match identifiers and spatial offset distances (in arcseconds) for other major catalogues. These external catalogues are stored in the \texttt{ext} schema
and can be utilised to extract the other data by matching on the unique identifier relevant for each table. Matches are provided with offset distances up to 15~arcsec. The cross-matched tables include:

\begin{itemize}
    \item \texttt{ext.allwise} - AllWISE Source Catalog, \citet{2013wise.rept....1C}
    \item \texttt{ext.catwise2020}$^*$ - CatWISE2020 Catalog, \citet{2021ApJS..253....8M}
    \item \texttt{ext.des\_dr2}$^*$  - Dark Energy Survey (DES) DR2 Catalog, \citet{2021ApJS..255...20A}
    \item \texttt{ext.refcat2} - ATLAS All-Sky Stellar Reference Catalog, \citet{2018ApJ...867..105T}
    \item \texttt{ext.gaia\_dr3} - {\it Gaia} DR3 Source Catalogue, \citet{2023A&A...674A...1G}
    \item \texttt{ext.galex\_guvcat\_ais} - {\it GALEX} Catalog of UV sources from the All-sky Imaging Survey (\texttt{GUVcat\_AIS\_fov055}), \citet{2017ApJS..230...24B}
    \item \texttt{ext.ls\_dr9}$^*$ - DESI Legacy Imaging Surveys (LS) DR9 Sweep Catalogs, \citet{2019AJ....157..168D}
    \item \texttt{ext.nsc\_dr2}$^*$ - NOIRLab Source Catalog (NSC) DR2, \citet{2021AJ....161..192N}
    \item \texttt{ext.ps1\_dr1}$^*$ - Pan-STARRS1 (PS1) DR1 Catalog, \citet{2016arXiv161205560C} (Note: only selected columns, restricted to decMean<30deg and nDetections>1)
    \item \texttt{ext.splus\_dr3}$^*$ - Southern Photometric Local Universe Survey (S-PLUS) DR3 Catalog, \citet{2019MNRAS.489..241M}
    \item \texttt{ext.twomass\_psc} - Two Micron All Sky Survey (2MASS) Point Source Catalog (PSC), \citet{2006AJ....131.1163S}
    \item \texttt{ext.vhs\_dr6} - VISTA Hemisphere Survey (VHS) DR6 Source Table, \citet{2013Msngr.154...35M} (Note: VISTA Science Archive copy of DR6, includes data through UT 2017-04-01)
\end{itemize}
where the asterisk indicates tables that have been newly added or expanded for SMSS~DR4.

In addition, we store SMSS~DR4 cross-match information in a number of smaller (primarily spectroscopic) tables in the \texttt{DR4\_ID} and \texttt{DR4\_DIST} columns, which record the \texttt{OBJECT\_ID} from the \texttt{master} table and the offset distance in arcseconds, respectively. Since these object lists are orders of magnitudes smaller than the DR4 \texttt{master} table, it is more efficient to find and store suitable DR4 matches in those tables. However, these ``reverse matches'' are affected by the issue that some unique astrophysical objects appear as two separate and non-identical entries in the \texttt{master} table, typically with one of the entries distilling most detections and the other one distilling only few. The best match is thus determined not only based on proximity but also on the volume of good-quality information. Hence, we allow reverse-matching only to sources with \texttt{FLAGS} $<16384$ and choose the match as the DR4 \texttt{master} source that minimises the value of (0.5\arcsec$+d$)/(1$+$\texttt{NGOOD}), for distance $d$ in arcsec.   

As above, these cross-matches extend up to 15~arcsec. This set of tables includes:
\begin{itemize}
    \item \texttt{ext.kids\_dr4p1}$^*$ - Kilo-Degree Survey (KiDS) DR4.1, \citet{2019A&A...625A...2K}
    \item \texttt{ext.milliquas\_v8}$^*$ - Million Quasars Catalog (Milliquas), version 8 (2023), \citet{2023arxiv230801505F}
    \item \texttt{ext.spec\_2dfgrs} - 2dF Galaxy Redshift Survey (2dFGRS) Final Data Release, \citet{2003astro.ph..6581C}
    \item \texttt{ext.spec\_2dflens} - 2dF Gravitational Lens Survey\linebreak
        (2dFLenS), \citet{2016MNRAS.462.4240B}
    \item \texttt{ext.spec\_2qz6qz} - 2dF and 6dF QSO Redshift Surveys (2QZ/6QZ) Final Catalogue, \citet{2004MNRAS.349.1397C}
    \item \texttt{ext.spec\_6dfgs} - 6dF Galaxy Survey (6dFGS) DR3, \citet{2009MNRAS.399..683J}
    \item \texttt{ext.spec\_anu2p3} - spectroscopic classifications from the ANU 2.3m telescope for quasar candidates, extremely metal poor stars, and other programs \citet{2022MNRAS.511..572O, 2023PASA...40...10O, 2019MNRAS.489.5900D}
    \item \texttt{ext.spec\_galah\_dr3}$^*$ - GALAH+ DR3, \citet{2021MNRAS.506..150B}
    \item \texttt{ext.spec\_gama\_dr3} - Galaxy Mass and Assembly\linebreak
        (GAMA) Survey DR3, \citet{2018MNRAS.474.3875B}
    \item \texttt{ext.spec\_hesqso} - Hamburg/ESO survey for bright QSOs, \citet{2000A&A...358...77W}
    \item \texttt{ext.spec\_ozdes\_dr2}$^*$ - Australian Dark Energy Survey (OzDES) DR2, \citet{2020MNRAS.496...19L}
    \item \texttt{ext.spec\_rave\_dr6} - Radial Velocity Experiment\linebreak
        (RAVE) DR6, combining the \texttt{sparv}, \texttt{classification}, and \texttt{obsdata} tables, \citet{2020AJ....160...82S}
    \item \texttt{ext.spec\_twomrs} - 2MASS Redshift Survey (2MRS), \citet{2012ApJS..199...26H}
    \item \texttt{ext.viking\_dr5} - VISTA Kilo-degree Infrared Galaxy (VIKING) Survey DR5 Source Table, \citet{2013Msngr.154...32E} (Note: VISTA Science Archive copy of DR5, includes data through UT 2018-02-15)
    \item \texttt{ext.vsx\_20230626}$^*$ - AAVSO International Variable Star Index, version 2023-06-26, \citet{2006SASS...25...47W}
\end{itemize}
where the asterisk again indicates tables that have been newly added or expanded for SMSS~DR4.

\subsubsection{Image Data}

The pipeline-processed image data and associated pixel masks that underlie the photometry in the tables are also made available to users. The images and masks are available in FITS or PNG formats via the Image Cutout service of the ASVO. At the initial release of DR4, each individual CCD is presented independently, with no stitching together of the separate CCDs in each mosaic image or coadding of multiple exposures. Such processes can be undertaken by the user, tuned to the needs of their own scientific analysis. 

The images provided by the Image Cutout service have been bias-subtracted, flat-fielded, and defringed (where appropriate), but the sky level has not been subtracted. 
In the header of each FITS image, the \texttt{ZPAPPROX} keyword reports the centre-of-mosaic ZP, but users should be aware that the 2D ZP gradient applied to photometry in the tables has not been applied to the image cuouts (neither to renormalise the ZP to the centre of the cutout nor to apply a gradient in the count levels). However, because of the limits placed on ZP gradients in order for images to be included in DR4, application of the simple relation,
\begin{equation}
    mag = \texttt{ZPAPPROX} - 2.5 \log_{10}({\rm counts}),
\end{equation}
is likely to provide a calibration better than 0.1~mag for photometry measured by users directly from the image data.

\section{Future Activities}
\label{sec:future}

The wealth of DR4 data will provide the opportunity to update the passband definitions from on-sky data, and thus tidy up the overall calibration of the catalogue. We anticipate such an update as part of a refined DR4.1 release.

We intend to produce co-added images in large sky tiles, taking advantage of dithered pointings to cover the gaps in the CCD mosaic, bad pixels, and pixels masked by cross-talk. Different image depths allow us to ignore pixel values that are saturated or masked for other reasons, including pixel overflow. The result will be deeper images with wider dynamic range. These will be produced with native coadded PSFs but also after convolution to a common PSF FWHM, where the latter will also allow the creation of robust colour maps. This imagery will then be used to update the {\sc Aladin Lite}-based SkyViewer on the SkyMapper website\footnote{The SkyViewer utilises {\sc Aladin Lite} version 2 \citep{2014ASPC..485..277B}.}.

Future work may explore additional methods of correcting for the high-frequency noise (Section~\ref{sec:ingest}) and slower bias level fluctuations (Section~\ref{sec:bias}), 
which would improve the photometry in and around extended sources.
To address the WCS solution biases in the mosaic corners (see Section~\ref{sec:distill}), DR4 images with high source density could be used to create a fundamental high-fidelity, high-order WCS mapping, which could then be used for improved corner astrometry while allowing just subtle adjustments based on the observed sources in each image.

\section{Summary}
\label{sec:summary}

We present the 4th data release of the SkyMapper Southern Survey. The data set is nearly complete relative to the original public Survey plans; for the first time, the release includes data taken with SkyMapper for other science programmes, such as the SkyMapper Transient Survey \citep[SMT; ][]{2017PASA...34...30S}, and $\sim$6\,000 very short exposures on each of seven standard fields. Improvements in sky coverage and data processing should enhance its usefulness compared to previous releases. 

In summary, we highlight the following properties:
\begin{enumerate}
    \item Compared to the previous public data release \citep[DR2;][]{2019PASA...36...33O}, SMSS DR4 more than triples the number of images, adds three years of additional time baseline (now 2014 to 2021), and 2,000~deg$^2$ of additional sky coverage.
    \item WCS solutions are now based on {\it Gaia} astrometry, and mosaic-wide solutions have improved the sky coverage near the Galactic Plane, especially in the short-wavelength $u$ and $v$ filters. 
    \item A new photometric zeropoint catalogue uses synthetic photometry derived from {\it Gaia} DR3 low-resolution spectra. The precision in the $u$ and $v$ filters is much improved over previous releases and naturally obviates the need for the \citet{2021ApJ...907...68H} corrections. The calibration quality now appears limited by our knowledge of the quantum efficiency of the CCDs and filter throughput curves; for the latter, we have characterised the total system throughput. 
    \item Calibration comparisons with other surveys are now also limited by knowledge of the exact colour terms that follow from the system efficiency curves.
    \item Source completeness at the faint end is limited by object detection on individual images and ranges from $\sim$18~mag (AB) in $u$- and $z$-band to nearly 20~mag in $g$- and $r$-band. However, photometric errors for non-variable objects are relatively small due to repeat measurements: the $10\sigma$ depth ranges from $18.6$~mag in $u$ and $z$ to $20.5$~mag in $g$ band.
    \item Photometric errors are improved, especially for the PSF magnitudes, where they were previously underestimated.
    \item \texttt{OBJECT\_ID}s from DR2/3 have been kept where appropriate, while new sources have been given new IDs starting with from a value of 2E9.
    \item In total, the catalogues include measurements for nearly 13~billion detections from $\sim700$~million astrophysical sources over 26,000~deg$^2$ of sky; these were derived from over 417\,000 on-sky images with exposures ranging from 0.1~sec to 600~sec.
    \item In some specific sky areas, hundreds of epochs are available in the $g$ and $r$ filters of the SkyMapper Transient Survey.
    \item The average PSF FWHM of the images is 2.7~arcsec, but has steadily degraded from 2018 to the end of the survey period.
    \item Updated cross-matches with additional multi-wavelength catalogues have been included.
\end{enumerate}

We look forward to utilisation of DR4 by the astronomical community via the tools of the All-Sky Virtual Observatory and the SkyMapper website (\url{https://skymapper.anu.edu.au}), or through partner data hosts.

\begin{acknowledgement}
We acknowledge the Gamilaroi people as the traditional owners of the land on which the SkyMapper Telescope stands. We are indebted to the original PIs of the S4 and SMSS projects: Brian P.\ Schmidt, Paul J.\ Francis, and MSB; as well as to the numerous students, postdocs, and technical staff who devoted themselves to the success of SkyMapper.
We thank Ian Adams and the other hard-working staff of Siding Spring Observatory, whose diligent efforts have supported SkyMapper's ongoing productivity. We thank Annino Vaccarella and Mike Ellis for their tireless dedication to resolving technical challenges. We thank Peter Onaka, Sidik Isani, and the STARGRASP team for critical assistance with the CCD controller system. 
We thank Andrew Robinson, Robert Cohen, Andrew Howard, James Fitzsimmons, and the rest of the staff from the National Computational Infrastructure (NCI) for their long-standing support of the SkyMapper node of the All-Sky Virtual Observatory (ASVO) and the SkyMapper project. We thank Chris Ramage for retrieving the AAT seeing records.

We thank the {\it Gaia} team for their assistance with the generation of synthetic photometry used in our photometric calibration, particularly Francesca De Angeli and Nicholas Walton from the Institute of Astronomy at the University of Cambridge, and Paolo Montegriffo from the Osservatorio Astronomico di Bologna of the Italian National Institute for Astrophysics.

We thank the team from the NOIRLab Astro Data Lab, particularly Robert Nikutta, David Nidever, Adam Scott, Mike Fitzpatrick, and Benjamin Weaver, for making the DES DR2 and NSC DR2 datasets available for cross-matching to this data release.

The national facility capability for SkyMapper has been funded through ARC LIEF grant LE130100104 from the Australian Research Council (ARC), awarded to the University of Sydney, the Australian National University, Swinburne University of Technology, the University of Queensland, the University of Western Australia, the University of Melbourne, Curtin University of Technology, Monash University and the Australian Astronomical Observatory. Parts of this project were conducted by the Australian Research Council Centre of Excellence for All-sky Astrophysics (CAASTRO), through project number CE110001020. 
We acknowledge support from the ARC Discovery Projects program, most recently through DP190100252.

SWC acknowledges support from the National Research Foundation of Korea (NRF) grants, No. 2020R1A2C3011091 and No. 2021M3F7A1084525 funded by the Ministry of Science and ICT (MSIT). This research was also supported by Basic Science Research Program through the NRF funded by the Ministry of Education (RS-2023-00245013).

Development and support for the SkyMapper node of the ASVO has been funded in part by Astronomy Australia Limited (AAL) and the Australian Government through the Commonwealth's Education Investment Fund (EIF) and National Collaborative Research Infrastructure Strategy (NCRIS), particularly the National eResearch Collaboration Tools and Resources (NeCTAR) and the Australian National Data Service Projects (ANDS). The NCI, which is supported by the Australian Government, has contributed resources and services to this project and hosts the SkyMapper node of the ASVO. We also thank the ANU Major Equipment Committee (via grant 14MEC25), the Research School of Astronomy \& Astrophysics, and CAASTRO for financial contributions toward the purchase of the SkyMapper server on which the mean object properties were distilled. The NCI resources used for the image processing were granted through the ANU Merit Allocation Scheme.

This work has made use of data from the European Space Agency (ESA) mission
{\it Gaia} (\url{https://www.cosmos.esa.int/gaia}), processed by the {\it Gaia}
Data Processing and Analysis Consortium (DPAC,
\url{https://www.cosmos.esa.int/web/gaia/dpac/consortium}). Funding for the DPAC
has been provided by national institutions, in particular the institutions
participating in the {\it Gaia} Multilateral Agreement. This work has made use of the {\sc Python} package GaiaXPy, developed and maintained by members of the {\it Gaia} Data Processing and Analysis Consortium (DPAC), and in particular, Coordination Unit 5 (CU5), and the Data Processing Centre located at the Institute of Astronomy, Cambridge, UK (DPCI).

This publication makes use of data products from the Wide-field Infrared Survey Explorer, which is a joint project of the University of California, Los Angeles, and the Jet Propulsion Laboratory/California Institute of Technology, and NEOWISE, which is a project of the Jet Propulsion Laboratory/California Institute of Technology. WISE and NEOWISE are funded by the National Aeronautics and Space Administration.

This research uses services or data provided by the Astro Data Lab at National Science Foundation's National Optical-Infrared Astronomy Research Laboratory. NOIRLab is operated by the Association of Universities for Research in Astronomy (AURA), Inc. under a cooperative agreement with the NSF.

This research has made use of "Aladin sky atlas" developed at CDS, Strasbourg Observatory, France.

This research made use of {\sc Astropy}, a community-developed core {\sc Python} package for Astronomy \citep{2013A&A...558A..33A,2018AJ....156..123A}. 

\end{acknowledgement}

\printendnotes

\bibliography{bib} 

\appendix

\onecolumn
\begin{center}
\begin{longtable}{lp{7cm}crr}
\caption{Name, description, units, minimum value, and maximum value for each column in the SMSS DR4 \texttt{master} table.}\\
\label{tab:master}
Column Name & Description & Units & Minimum & Maximum\\
\hline \hline
\endfirsthead

Column Name & Description & Units & Minimum & Maximum\\
\hline \hline
\endhead

\hline \multicolumn{2}{l}{Continued on next page}\\
\hline
\endfoot

\hline \hline
\endlastfoot

object\_id & Global unique object ID in the master table. & --- & 1 & 2724329343\\
raj2000 & Mean ICRS Right Ascension of the object & deg & 0 & 359.999998\\
dej2000 & Mean ICRS Declination of the object & deg & -89.992883 & 29.182411\\
e\_raj2000 & RMS variation around the mean Right Ascension in milliarcseconds & mas & 36 & 7311\\
e\_dej2000 & RMS variation around the mean Declination in milliarcseconds & mas & 36 & 7562\\
smss\_j & SkyMapper Southern Survey designation of the form SMSS Jhhmmss.ss$\pm$ddmmss.s derived from mean ICRS coordinates & --- & 000000.00+095415.5 & 235959.99-755208.7\\
mean\_epoch & Mean MJD epoch of the observations & d & 56730.6305 & 59473.7663\\
rms\_epoch & RMS variation around the mean epoch & d & 0 & 1934.46\\
glon & Galactic longitude derived from ICRS coordinates. Not to be used as primary astrometric reference. & deg & 6.71978E-7 & 360\\
glat & Galactic latitude derived from ICRS coordinates. Not to be used as primary astrometric reference. & deg & -89.9936 & 79.3879\\
flags & Bitwise OR of Source Extractor flags across all observations & --- & 0 & 16384\\
nimaflags & Total number of flagged pixels from bad or saturated or crosstalk-affected pixel masks across all observations & --- & 0 & 23069556\\
ngood & Number of observations used across all filters & --- & 0 & 11306\\
u\_flags & Bitwise OR of Source Extractor flags from u-band measurements in photometry table & --- & 0 & 26136\\
u\_nimaflags & Number of flagged pixels from bad or saturated or crosstalk-affected pixels masks from u-band measurements in photometry table & --- & 0 & 32767\\
u\_ngood & Number of u-band observations used & --- & 0 & 1604\\
u\_nclip & Number of u-band observations with magnitudes clipped from the final PSF magnitude estimate & --- & 0 & 252\\
v\_flags & Bitwise OR of Source Extractor flags from v-band measurements in photometry table & --- & 0 & 26136\\
v\_nimaflags & Number of flagged pixels from bad or saturated or crosstalk-affected pixels masks from v-band measurements in photometry table & --- & 0 & 32767\\
v\_ngood & Number of v-band observations used & --- & 0 & 1739\\
v\_nclip & Number of v-band observations with magnitudes clipped from the final PSF magnitude estimate & --- & 0 & 362\\
g\_flags & Bitwise OR of Source Extractor flags from g-band measurements in photometry table & --- & 0 & 28160\\
g\_nimaflags & Number of flagged pixels from bad or saturated or crosstalk-affected pixels masks from g-band measurements in photometry table & --- & 0 & 32767\\
g\_ngood & Number of g-band observations used & --- & 0 & 1914\\
g\_nclip & Number of g-band observations with magnitudes clipped from the final PSF magnitude estimate & --- & 0 & 846\\
r\_flags & Bitwise OR of Source Extractor flags from r-band measurements in photometry table & --- & 0 & 28176\\
r\_nimaflags & Number of flagged pixels from bad or saturated or crosstalk-affected pixels masks from r-band measurements in photometry table & --- & 0 & 32767\\
r\_ngood & Number of r-band observations used & --- & 0 & 1968\\
r\_nclip & Number of r-band observations with magnitudes clipped from the final PSF magnitude estimate & --- & 0 & 833\\
i\_flags & Bitwise OR of Source Extractor flags from i-band measurements in photometry table & --- & 0 & 26139\\
i\_nimaflags & Number of flagged pixels from bad or saturated or crosstalk-affected pixels masks from i-band measurements in photometry table & --- & 0 & 32767\\
i\_ngood & Number of i-band observations used & --- & 0 & 2381\\
i\_nclip & Number of i-band observations with magnitudes clipped from the final PSF magnitude estimate & --- & 0 & 859\\
z\_flags & Bitwise OR of Source Extractor flags from z-band measurements in photometry table & --- & 0 & 26136\\
z\_nimaflags & Number of flagged pixels from bad or saturated or crosstalk-affected pixels masks from z-band measurements in photometry table & --- & 0 & 32767\\
z\_ngood & Number of z-band observations used & --- & 0 & 2175\\
z\_nclip & Number of z-band observations with magnitudes clipped from the final PSF magnitude estimate & --- & 0 & 854\\
class\_star & Maximum stellarity index from photometry table (between 0=no star and 1=star) & --- & 0 & 1\\
chi2\_psf & Maximum chi-squared from photometry table & --- & 0 & 3.2558e+06\\
flags\_psf & Bitmask indicating whether photometry is likely biased by neighbours at >1\%; bits 0-5 correspond to filters z-u & --- & 0 & 63\\
radius\_petro & Mean r-band Petrosian radius & pix & 3.5 & 10.56\\
mean\_fwhm & Mean FWHM of detections & pix & -2.27989e+08 & 3696.12\\
u\_psf & Weighted mean u-band PSF magnitude & mag & 3.5168 & 22.4221\\
e\_u\_psf & Error in weighted mean u-band PSF magnitude & mag & 0.0014 & 3.0543\\
u\_petro & Weighted mean u-band Petrosian magnitude & mag & 0.3224 & 31.7358\\
e\_u\_petro & Error in weighted mean u-band Petrosian magnitude & mag & 0.0014 & 2939.69\\
u\_apc05 & Weighted mean u-band aperture-corrected magnitude from flux in 5 arcsec diameter & mag & 6.1501 & 31.6061\\
e\_u\_apc05 & Error in weighted mean u-band aperture-corrected magnitude from flux in 5 arcsec diameter & mag & 0.0015 & 2.4124\\
u\_mmvar & Magnitude range between minimum and maximum unclipped u-band PSF magnitudes & mag & 0 & 7.3489\\
v\_psf & Weighted mean v-band PSF magnitude & mag & 3.1073 & 22.2866\\
e\_v\_psf & Error in weighted mean v-band PSF magnitude & mag & 0.0009 & 2.5687\\
v\_petro & Weighted mean v-band Petrosian magnitude & mag & 0.1233 & 35.1441\\
e\_v\_petro & Error in weighted mean v-band Petrosian magnitude & mag & 0.0009 & 108244\\
v\_apc05 & Weighted mean v-band aperture-corrected magnitude from flux in 5 arcsec diameter & mag & 6.0923 & 30.1388\\
e\_v\_apc05 & Error in weighted mean v-band aperture-corrected magnitude from flux in 5 arcsec diameter & mag & 0.0009 & 2.3113\\
v\_mmvar & Magnitude range between minimum and maximum unclipped v-band PSF magnitudes & mag & 0 & 7.3253\\
g\_psf & Weighted mean g-band PSF magnitude & mag & 2.8979 & 24.0783\\
e\_g\_psf & Error in weighted mean g-band PSF magnitude & mag & 0.0006 & 4.1083\\
g\_petro & Weighted mean g-band Petrosian magnitude & mag & 0.4787 & 36.1903\\
e\_g\_petro & Error in weighted mean g-band Petrosian magnitude & mag & 0.0006 & 82992.5\\
g\_apc05 & Weighted mean g-band aperture-corrected magnitude from flux in 5 arcsec diameter & mag & 6.0203 & 30.9735\\
e\_g\_apc05 & Error in weighted mean g-band aperture-corrected magnitude from flux in 5 arcsec diameter & mag & 0.0007 & 3.7124\\
g\_mmvar & Magnitude range between minimum and maximum unclipped g-band PSF magnitudes & mag & 0 & 11.3157\\
r\_psf & Weighted mean r-band PSF magnitude & mag & 2.5215 & 25.3334\\
e\_r\_psf & Error in weighted mean r-band PSF magnitude & mag & 0.0006 & 4.2441\\
r\_petro & Weighted mean r-band Petrosian magnitude & mag & 0.5543 & 35.4118\\
e\_r\_petro & Error in weighted mean r-band Petrosian magnitude & mag & 0.0006 & 23988.6\\
r\_apc05 & Weighted mean r-band aperture-corrected magnitude from flux in 5 arcsec diameter & mag & 5.8349 & 34.6417\\
e\_r\_apc05 & Error in weighted mean r-band aperture-corrected magnitude from flux in 5 arcsec diameter & mag & 0.0006 & 3.1773\\
r\_mmvar & Magnitude range between minimum and maximum unclipped -band PSF magnitudes & mag & 0 & 9.7395\\
i\_psf & Weighted mean i-band PSF magnitude & mag & 2.4086 & 25.446\\
e\_i\_psf & Error in weighted mean i-band PSF magnitude & mag & 0.0003 & 3.9294\\
i\_petro & Weighted mean i-band Petrosian magnitude & mag & 0.586 & 34.3901\\
e\_i\_petro & Error in weighted mean i-band Petrosian magnitude & mag & 0.0003 & 186518\\
i\_apc05 & Weighted mean i-band aperture-corrected magnitude from flux in 5 arcsec diameter & mag & 5.4024 & 32.7377\\
e\_i\_apc05 & Error in weighted mean i-band aperture-corrected magnitude from flux in 5 arcsec diameter & mag & 0.0004 & 2.8266\\
i\_mmvar & Magnitude range between minimum and maximum unclipped i-band PSF magnitudes & mag & 0 & 9.1216\\
z\_psf & Weighted mean z-band PSF magnitude & mag & 1.3949 & 22.2704\\
e\_z\_psf & Error in weighted mean z-band PSF magnitude & mag & 0.0007 & 3.8631\\
z\_petro & Weighted mean z-band Petrosian magnitude & mag & 0.6096 & 33.6571\\
e\_z\_petro & Error in weighted mean z-band Petrosian magnitude & mag & 0.0007 & 359127\\
z\_apc05 & Weighted mean z-band aperture-corrected magnitude from flux in 5 arcsec diameter & mag & 4.7756 & 32.5408\\
e\_z\_apc05 & Error in weighted mean z-band aperture-corrected magnitude from flux in 5 arcsec diameter & mag & 0.0008 & 2.9392\\
z\_mmvar & Magnitude range between minimum and maximum unclipped z-band PSF magnitudes & mag & 0 & 9.0278\\
self\_id1 & Unique identifier (object\_id) of closest SMSS DR4 source & --- & 1 & 2724329340\\
self\_dist1 & Distance on sky to closest SMSS DR4 source & arcsec & 0.03 & 15\\
self\_id2 & Unique identifier (object\_id) of second-closest SMSS DR4 source & --- & 2 & 2724329343\\
self\_dist2 & Distance on sky to second-closest SMSS DR4 source & arcsec & 0.18 & 15\\
self\_id3 & Unique identifier (object\_id) of third-closest SMSS DR4 source & --- & 40 & 2724327091\\
self\_dist3 & Distance on sky to third-closest SMSS DR4 source & arcsec & 0.26 & 15\\
cnt\_self\_15 & Number of SMSS DR4 sources within 15 arcsec & --- & 1 & 107\\
ebmv\_sfd & E(B-V) from Schlegel+1998 extinction maps at the ICRS coordinates & mag & 0.0006 & 152.148\\
ebmv\_gnilc & E(B-V) from Planck satellite GNILC reddening maps at the ICRS coordinates & mag & 0 & 164.918\\
ebmv\_g\_err & Error in the E(B-V) from GNILC & mag & 0 & 52.6726\\
gaia\_dr3\_id1 & Unique identifier (source\_id) of closest Gaia DR3 source & --- & 4295806720 & 6917528997577384320\\
gaia\_dr3\_dist1 & Distance on sky to closest Gaia DR3 source & arcsec & 0 & 15\\
gaia\_dr3\_id2 & Unique identifier (source\_id) of second-closest Gaia DR3 source & --- & 34361129088 & 6917528997577384320\\
gaia\_dr3\_dist2 & Distance on sky to second-closest Gaia DR3 source & arcsec & 0.09 & 15\\
cnt\_gaia\_dr3\_15 & Number of Gaia DR3 sources within 15 arcsec & --- & 1 & 333\\
twomass\_key & Unique identifier (pts\_key) of closest 2MASS PSC source & --- & 40 & 1339356373\\
twomass\_dist & Distance on sky to closest 2MASS PSC source & arcsec & 0 & 15\\
allwise\_cntr & Unique identifier (cntr) of closest AllWISE source & --- & 1601351000001 & 3584116601351035683\\
allwise\_dist & Distance on sky to closest AllWISE source & arcsec & 0 & 15\\
catwise\_id & Unique identifier (id) of closest CatWISE2020 source & --- & 0000m016\_b0-000001 & 3584p166\_b0-080605\\
catwise\_dist & Distance on sky to closest CatWISE2020 source & arcsec & 0 & 15\\
refcat2\_id & Unique identifier (objid) of closest ATLAS Refcat2 source (only useful for this copy of Refcat2) & --- & 1 & 992637833\\
refcat2\_dist & Distance on sky to closest ATLAS Refcat2 source & arcsec & 0 & 15\\
ps1\_dr1\_id & Unique identifier (objid) of closest Pan-STARRS1 DR1 source & --- & 65810810454945744 & 143011300804053051\\
ps1\_dr1\_dist & Distance on sky to closest Pan-STARRS1 DR1 source & arcsec & 0 & 15\\
galex\_guv\_id & Unique identifier (objid) of closest GALEX GUVcat AIS source (Bianchi et al. 2017) & --- & 6374012077942505627 & 6388191387412599267\\
galex\_guv\_dist & Distance on sky to closest GALEX GUVcat AIS source & arcsec & 0 & 15\\
vhs\_dr6\_id & Unique identifier (objid) of closest Pan-STARRS1 DR1 source & --- & 472446402562 & 473820609749\\
vhs\_dr6\_dist & Distance on sky to closest Pan-STARRS1 DR1 source & arcsec & 0 & 15\\
ls\_dr9\_id & Unique identifier (unique\_id) of closest Legacy Survey DR9 source & --- & 100000236480000014 & 120004860270005003\\
ls\_dr9\_dist & Distance on sky to closest Legacy Survey DR9 source & arcsec & 0 & 15\\
des\_dr2\_id & Unique identifier (coadd\_object\_id) of closest DES DR2 source & --- & 870245760 & 1700560060\\
des\_dr2\_dist & Distance on sky to closest DES DR2 source & arcsec & 0 & 15\\
nsc\_dr2\_id & Unique identifier (id) of closest NSC DR2 source & --- & 100000\_1000 & 99999\_9999\\
nsc\_dr2\_dist & Distance on sky to closest NSC DR2 source & arcsec & 0 & 15\\
splus\_dr3\_id & Unique identifier of closest S-PLUS DR3 source & --- & DR3.HYDRA-0011.000000 & DR3.STRIPE82-0170.061552\\
splus\_dr3\_dist & Distance on sky to closest S-PLUS DR3 source & arcsec & 0 & 15\\
\end{longtable}
\end{center}

\clearpage
\twocolumn

\end{document}